\documentclass[aps,prd,preprint,showpacs,superscriptaddress,linenumbers]{revtex4} 

\usepackage{amsmath}   
\usepackage{graphicx}  
\usepackage{xspace}
\usepackage{units}
\usepackage{lscape}

\usepackage{hyperref}

\newcommand{\sizecheck}{0} 

\newif\ifpdf
\ifx\pdfoutput\undefined
   \pdffalse
\else
   \pdfoutput=1
   \pdftrue
\fi
\ifpdf
   \usepackage{graphicx}
   \usepackage{epstopdf}
\else
   \usepackage{graphicx}
\fi

\begin{document}

\title{Elastic hadron-nucleus scattering in neutrino-nucleus reactions and transverse kinematics measurements}


\newcommand{\UMD}{Department of Physics, University of Minnesota -- Duluth, Duluth, Minnesota 55812, USA}
\newcommand{\Rochester}{University of Rochester, Rochester, New York 14627 USA}
\newcommand{\oxford}{Oxford University, Department of Physics, Oxford, United Kingdom}

\author{L.~A.~Harewood}  \affiliation{\UMD}
\author{R.~Gran}                           \affiliation{\UMD}

%

\date{\today}

\begin{abstract}

Rescattering following a neutrino-nucleus reaction changes the number, energy, and direction of detectable hadrons.  In turn, this affects the selection and kinematic distributions of subsamples of neutrino events used for interaction or oscillation analysis.
This technical note focuses on three forms of two-body  rescattering.  Elastic hadron+nucleus scattering primarily changes the direction of the hadron with very little energy transfer.  Secondly, a hadron+nucleon quasi-elastic process leads to the knockout of a single struck nucleon, possibly with charge exchange between the two hadrons.  Also, a pion can be absorbed leading to the ejection of two nucleons.  There was an error in the code of the {\small GENIE} neutrino event generator that affects these processes.    We present examples of the change with the fixed version of the scattering process, but also compare these specifically to turning off elastic scattering completely, which is similar to other neutrino event generator configurations or a potential Equick-fix to already generated samples.   Three examples are taken from current topics of interest:  transverse kinematics observables in quasielastic neutrino reactions, the pion angle with respect to the incoming and outgoing lepton for $\Delta$ reactions with a charged pion in the final state, and the angle between two protons in reactions with no pions present.  Elastic hadron+nucleus scattering in its unfixed form makes a large distortion in distributions of transverse kinematic imbalances scattering, but only mild distortion in other observables.  The distortion of the other two processes is also mild for all distributions considered.   The correct form of hadron+nucleus scattering process could play a role in describing the width and center of the sharp peak in the inferred Fermi-motion of the struck nucleon or be benchmarked using (e,e'p) data.

\end{abstract}
\ifnum\sizecheck=0
\maketitle
\fi


\section{introduction}

Final state (re-)interactions (FSI) are an important part of modeling neutrino reactions in nuclei.  They change the direction, energy, species, and number of hadrons as they exit the nucleus.   Neutrino reaction measurements require subsets of interactions with specific hadron content that are favorable to probe weak interaction parameters and/or the the environment of the nucleus.  Neutrino oscillation spectrum measurements likely benefit from using sub-samples with the highest resolution and least biased reconstructed neutrino energy.   Uncertainties in the strength and details of FSI feed into the efficiency of these selections as well as the predicted distributions of signal and background after selection.

This note explores the effect of FSI on three special final states of current experimental and theoretical interest.   The first is the selection of events which are enriched in charged-current quasielastic reaction (CCQE) and therefore allow for the best energy reconstruction.  They are ideal for evaluating transverse kinematic imbalances that arise from Fermi motion and FSI.  Also included separately are contributions to the same distributions from resonance production and two particle reactions leading to two-nucleon knockout (2p2h).   Second are the angle of a pion with respect to the initial incoming neutrino or the outgoing muon, especially for resonance interactions.  The third are events with exactly two protons in the final state.  The latter samples also lead to a set of transverse kinematic imbalance distributions.

From versions 2.6 to 2.12 and continuing into version 3.0, the {\small GENIE} neutrino event generator \cite{Andreopoulos:2015wxa} has a flaw in the core FSI routine ``TwoBodyKinematics'' .   This includes hadron + nucleus elastic scatters, single nucleon knockout reactions, and pion absorption on two nucleons.   In this work, a candidate replacement for this code is compared to the original flawed code.   

Additional comparisons illustrate proposals to eliminate the hadron + nucleus scattering process completely, either in already generated Monte Carlo samples or in future {\small GENIE} samples.  Beyond the flawed code, these comparisons provide interesting guidance on the role of FSI for modern precision measurements.  

This document offers a roadmap for users of recent versions of {\small GENIE} to determine how their analyses and interpretations may be affected. Supplementary material includes code patches for the {\small GENIE} 2.12.10 hA (shown here) and hA2014 models widely used by current experiments for their recent and upcoming publications.  Modifying GENIE with the patches privided with this technical note constitutes a private version of GENIE with fixed FSI (hereafter {\small GENIE}v2.12.10+FSIfix) and are not currently available in an official GENIE release.  For convenience, the patches offer configurations from old behavior to fully fixed behavior driven by a run-time option.  These are to enable others' studies beyond those presented here, such as different energies or fluxes, nuclei, or observable and diagnostic distributions.

\subsection{contemporary measurements affected by FSI}

Some of the foundational explorations of the role of FSI for CCQE events were in support of analyses by K2K, MiniBooNE, and NOMAD \cite{Gran:2006jn,AguilarArevalo:2007ab,Lyubushkin:2008pe}.   These analyses used three different event generators' models for FSI:  {\small NEUT} uses Oset's approach \cite{Oset:1986sy}, {\small NUANCE} \cite{Casper:2002sd} inherited code from IMB \cite{MudanThesis}, and NOMAD used the cosmic ray code {\small DPMJET} \cite{Battistoni:1998hh,Ranft:1994fd,Roesler:2000he}.  To select a quasielastic-rich sample, requirements on angle and/or missing momentum were placed on the events where a muon and proton and no other particles were reconstructed.    Among other criteria, events where the observed proton was coplanar with the muon were likely accepted and other muon+proton events were placed in a non-QE sideband sample.  The discrepancies between data and model for these two samples were partially attributed to uncertainties in the FSI models.

As neutrino and anti-neutrino datasets have become more sophisticated and higher in statistics, analysis has moved toward more complex observables that can differentiate between FSI, Fermi-motion, and binding energy/shell-model effects.  Experiments are chasing the equivalents to electron (e,e') and (e,e'p) measurements, even without the advantage of knowing the beam energy on an event-by-event basis.  

When a muon and a proton can be fully reconstructed, and the beam direction is known, the billiard ball (quasi-)elastic process of neutrino + nucleon is overconstrained.  In this case, experimenters have access to additional information about the details of the struck nucleon and pursue analysis of ``transverse kinematics imbalance'' observables.   The simplest of these is the coplanarity of the muon and proton directions.  Using the {\small GENIE} 2.8 event generator and its Fermi-motion and FSI models, the intention was to show whether the coplanarity angle was well reproduced  \cite{Walton:2014esl,Betancourt:2017uso}.  It was not; a major distortion of the {\small GENIE} prediction near perfect coplanarity is observed, which is fixed with these patches.

Current effort separates the Fermi-motion and FSI effects with a richer set of observables.    Theoretical considerations were outlined by a number of people \cite{Lu:2015hea,Lu:2015tcr,Furmanski:2016wqo,Lu:2019nmf} followed by measurements from both MINERvA and T2K data \cite{Lu:2018stk,Abe:2018pwo,Dolan:2018zye}.   The first model comparisons in this paper address four of these observables from \cite{Lu:2018stk}.

Additional progress has been made analyzing pion production.  The CC $\Delta$ production process is of particular interest when it is likely all hadrons have been measured. It should yield a high resolution neutrino energy measurement, and also provide its own set of transverse kinematics quantities \cite{Lu:2019nmf}.  A simpler set of comparisons show the effects on the observed pion angle, with respect to both the incident neutrino direction and the outgoing muon.

A two-nucleon knockout process, once seen as a background to the QE reaction, produces a significant rate and leads to an uncertain fraction of the hadronic energy going to neutrons.   Models provide a direct 2p2h reaction and also an indirect component when a single pion reaction loses its pion through FSI absorption.  The third model comparison looks at the predicted opening angle between two ejected protons and the coplanarity of the protons and muon.

This progress has been made using outstanding statistical power plus continuous improvement in detector systematic and flux uncertainties, which are under 10\% for MINERvA for most quantities of interest here.   Just as important are the community's improvements to the input modeling of the cross sections and event rates in these kinematics.  Aggressive strategies to reduce dependence on the input model (external constraints, sideband tuning, warping studies, iterative unfolding) always leave some model dependence when extracting cross sections, and better inputs always produce better results.


\subsection{Outline of FSI code}

This subsection outlines the basic components of a generic cascade implementation of FSI and details specific to the {\small GENIE} hA and hN models.   Once a hadron has momentum and energy transfer from the lepton, the following analysis is most concerned with the mix of fates (including no rescattering)  that the hadron might experience on its way out of the nucleus.  The overall FSI cascade strategy is:

\begin{itemize}
\item generate a neutrino reaction, including its hadronic final state
\item place each hadron in the nucleus and make each step its way out
\item according to a mean free path, determine if FSI should happen
\item modify the final state according to the chosen hadron fate
\item if it is a full cascade ({\small GENIE} hN but not hA), repeat these steps
\item place the resulting hadron(s) outside the nucleus and give them to Geant4.
\end{itemize}

{\small GENIE} uses energy-dependent hadron-nucleon mean free path from total cross sections taken from the SAID \cite{SAID} database maintained by and for \cite{Arndt:2006bf,Arndt:2007qn} among others.  If stepping through the nucleus leads to a fate, the energy dependent fates for proton and neutron scattering are obtained from a one-time run of the intranuclear cascade model CEM03 by S.G. Mashnik and collaborators \cite{Mashnik:2005ay,Mashnik:2000up} evaluated for proton+$^{56}$Fe.   A failure generating FSI because something is unphysical usually means to retry with a different struck-nucleon, start with a new fate, or totally give up on the event, depending on the severity of the failure.   


For nucleons in the hA model, every nucleon experiences exactly one of the following fates:
\begin{itemize}
\item fate 1 no FSI at all (the stepper's random number never passed to the FSI fates)
\item fate 2 charge exchange with single nucleon knockout
\item fate 3 elastic hadron+nucleus scattering
\item fate 4 ``inelastic'' single nucleon knockout
\item fate 5 multi-nucleon knockout (including pion absorption)
\item fate 8 pion production
\end{itemize}
This was designed to be an adequate approximation to a full cascade and allows for relatively easy reweighting fate by fate.   The utility is fast evaluation of systematic uncertainties for fully-generated Monte Carlo samples used by experiments.  A full cascade such as the hN model or even CEM03 has too many combinatorics to isolate any one fate without regenerating large alternate samples and/or a complex reweighting scheme.  For our purposes the simple combinatorics has additional benefit of showing how specific FSI mechanisms populate distributions differently.


\subsection{Elastic is in the eye of the beholder}

The terminology gets confusing because we draw data and models from different corners of nuclear and particle physics.   Here is a summary of the more common jargon. 

A neutrino+nucleus scattering person says ``coherent'' is elastic scattering where the nucleus remains in its ground state or possibly a low-lying excited state with no nucleon knockout.  Single nucleon knockout via a 1p1h process (prior to FSI) is quasielastic.   The ``quasi'' refers to the energy cost to remove the nucleon from the nucleus, same as for the (e,e') folks, and to produce the mass of the outgoing charged lepton.   Resonance production and quark-level deeply inelastic scattering (DIS) are all inelastic processes, and are often but not always accompanied by pion or heavier meson production.

A hadron+nucleon scattering person also says ``inelastic'' when resonance, exotic baryons, or DIS interactions occur, which are often accompanied by pion production.  All other reactions are elastic, with the special case of charge exchange between ground state nucleons.  These may include ``diffractive'' effects.   These folks include the partial-wave analysis work and the SAID database.

Finally, hadron+nucleus scattering folks would say ``elastic'' if the nucleus remains in its ground state after the reaction.   Experimentally this is observed as a change in momentum with little or no change in energy.  The nucleus can also be put into its lowest excited states, or single and multiple nucleon knockout can occur.   The elastic-nucleus scattering process has a diffraction character, described after the results in Sec.~\ref{whyelastic}.   This is the outlook of many nucleus cascade models, including CEM03.

For completeness, it is worth mentioning that FSI has two meanings in the literature.   In this work we mean a possibly off-shell hadron, regardless of its origin, is being transported through a nucleus where it encounters other hadrons and interacts.   In other theoretical calculations it means diagram-level exchanges of momentum and energy including the interferences among amplitudes, in addition to the W or Z boson.

Users of {\small GENIE} may note that the hN scheme is different in multiple ways that affect truth-level analysis code.   Every nucleon can experience more than one fate, based on the mean free path.  Secondly, the elastic hadron+nucleus fate does not exist in hN.   Both the fate numbering and the C++ enum names are different.  Single nucleon knockout in hN is called elastic (i.e. equivalent to neutrino+nucleus quasi-elastic or free nucleon scattering) and uses fate = 3. The inelastic knockout fate = 4 is the multi-nucleon knockout process for hN.  Charge exchange with single nucleon knockout remains fate = 2 for both hN and hA and is handled by the same TwoBodyKinematics routine that is so far fixed only for hA.

Also, the reader may be interested in Steve Dytman's review \cite{Dytman:2009zza} of approaches to intranuclear rescattering by several neutrino event generator authors.   It was written contemporaneously with the {\small GENIE} 2.6 updated model from the version originally developed within the NEUGEN \cite{Gallagher:2002sf} neutrino event generator used by Soudan2 and MINOS. 

\subsection{Summary of fixes to the code}

The changes to the code fix a mistaken calculation in the routine TwoBodyKinematics used by elastic scattering, single-nucleon knockout, and absorption of pions (and photons) on two nucleons.  The nature of the mistake was to use the boost direction in an unnatural way, thus incorporating lab frame information when doing the center of momentum scattering calculation.   It also updates the code that pulls a center of momentum angle from an empirical distribution for the elastic scattering process.
 
An alternative approach is to turn off elastic hadron + nucleus scattering instead of fixing it; {\small GENIE}'s hA2015 and hN models already do this.  In the following, we compare to a second method accessible through the modified codes or by reweighting already generated Monte Carlo.

The different behavior is selected by a new user option with the following configurations
\begin{itemize}
\item ElasticConfig = 0;  // the old behavior
\item ElasticConfig = 1;   // only elastic scattering uses new code
\item ElasticConfig = 2;   // elastic scattering $\theta_{CM}$ = 0 = no-FSI using new code
\item ElasticConfig = 3;   // full fix, elastic and inelastic two-body reactions use new code
\item ElasticConfig = 4;   // elastic $\theta_{CM} = 0$, inelastic reactions also use new code  
\item use hA2015 // elastic scattering is turned off, inelastic scattering is increased
\end{itemize}

In ElasticConfig versions 0, 1, and 2, the inelastic and pion two-nucleon absorption process are not using the fixed code.   Comparing ElasticConfig 1 and 3 will show the effects of the inelastic processes only.  Comparing ElasticConfig 2 and 3 show the remaining distortion after a proposed reweight of fully generated {\small GENIE} 2.12.10 Monte Carlo to no-FSI.

Modifications are also made to the code that generates the center of momentum angle $\theta_{CM}$ for the hadron+nucleus elastic scatter.     However, the code that picked an angle from the data also did not behave as intended.   The data distributions are now converted to d$\sigma$/d$\theta$ and turned into a cumulative distribution function to be sampled.  More discussion in Sec.~\ref{whyelastic} follows the results of the main study.  The option to force it to zero produces events that are equivalent to no scattering at all, but keeping the FSI = 3 fate code in the output.

\subsection{Simulation setup}

The simulations here are monoenergetic 3 GeV neutrinos interacting with carbon nuclei.   Only charged current interactions are considered.  They are presented as binned histograms whose vertical axis is events from a sample of 200,000 drawn by {\small GENIE}, though the statistics used are actually 1000x to 3000x higher.  Many of the ratio plots have a smoothing function applied and are drawn as curves through bin center.  This is usually more clear but statistical fluctuations are evident in the most rare components of each process:

\begin{itemize}
\item Quasielstic reactions:  true CCQE events using MINERvA selection of \cite{Lu:2018stk}.
\item Pion production reactions:  require a charged pion above 75 MeV in the final state. 
\item Two-proton reactions:  all processes, exactly two protons above 50 MeV, no pions.
\end{itemize}

These selections are naturally centered on data samples like those from MINERvA's ``low energy'' dataset.  The {\small GENIE} simulation presented in this paper is modified following the MINERvA tune Mnv{\small GENIE}-v1.   This tune modifies the CCQE process to include an RPA screening effect \cite{Nieves:2004wx,Gran:2017psn} .   The 2p2h process is from the Valencia group \cite{Nieves:2011pp,Gran:2013kda,Schwehr:2016pvn} with the enhancement based on the measurements in \cite{Rodrigues:2015hik}.   A suppression of pion production from the {\small GENIE} DIS model is also applied \cite{Wilkinson:2014yfa, Rodrigues:2016xjj}.  The CC coherent pion production component is excluded for the convenience of coding these studies, and anyway would not experience FSI.

The conclusions will apply to other experiments (possibly even more so) at different energies, with oxygen and argon, with higher resolution detectors, and using different configurations of {\small GENIE}'s neutrino interaction model.   Theconclusions in this note are about the predicted effects of the underlying processes and not only about the distortions caused by the flawed code.

\section{Quasielastic reactions}

The quasielastic reaction has the least-random final state kinematics.   This allows for the extraction of the properties of the nuclear environment such as Fermi motion and also obtain the most detail of the weak interaction process.    However, the nature of the bug in the TwoBodyKinematics function  is to inappropriately use the CM boost direction, and therefore lab frame information, during the center of momentum frame scattering process.  The most prominent effect is that scattered hadron small angles distribute around perfect coplanarity (and by extension low transverse momentum imbalance), rather than Fermi-motion smeared like the no-FSI distribution.   

Charged-current quasielastic interactions are also special because with the hA model either zero or one FSI interaction happens, and it is exactly one proton that experiences the fate.  Resonance and 2p2h reactions have two hadrons prior to FSI which may be pion or nucleon and may separately experience different fates.   In the hN model, each particle may experience multiple fates on its own.   Thus from both physics, code, and an interpretation  perspective, the CCQE process is a simple place to start.

This section goes through four cases illustrating the distortion caused by the elastic and inelastic aspects of the bug, a fix that would reweight elastic to be no FSI ($\theta_{CM}=0$), and the fix that is actually implemented as {\small GENIE}'s hA2015.

\subsection{selection of CCQE sample}
For the 3 GeV neutrino + carbon reaction comparisons in this paper we have reproduced the selection of the MINERvA measurement \cite{Lu:2018stk}.   This involves standard MINERvA muon acceptance limit of $\theta_\mu < 20$ degrees (in some analyses it is 25 degrees).   For the low energy beam (data prior to 2012) the analysis required $1.5 < p_{\mu} < 10 $ GeV, which is almost always satisfied for the monoenergetic 3 GeV neutrinos in this study.   In order to select well-reconstructed protons, additional selections are applied.   Protons with angle $\theta_{\mathrm{Lab}} < 70$ degrees are accepted because the MINERvA detector's planar construction makes high angle tracks challenging to reconstruct.   Protons with momentum $p > 0.450$ GeV are selected, lower momenta are inefficiently reconstructed because of the minimum five-plane tracking threshold.  Protons must have $p < 1.2$ GeV because more energetic protons are poorly identified because it is likely the proton undergoes a hadronic inelastic interaction in the scintillator detector, masking both the proton nature and the energy of the particle.    Finally, no pions are allowed in the final state.

\begin{table}[h]
\begin{tabular}{c|ccccccc}
selection stage & total & no-FSI & CEX & elastic & single & multi & pi \\ \hline 
Before any cuts & 100 & 36.0 & 6.8 & 22.8 & 14.3 & 17.1 & 3.1 \\
pre-FSI energy & 58.8 & 24.5 & 3.7 & 12.5 & 7.6 & 10.2 & 0.5 \\
After all cuts new sim & 36.2 & 18.1 & 1.3 & 9.5 & 3.5 & 3.9 & 0 \\
After all cuts original sim & 38.4 & 18.1 & 1.3 & 11.7 & 3.4 & 3.9 & 0 \\
\end{tabular}
\caption{Percent of the total sample remaining, by fate, after two stages of selection.   The CCQE process produces a wide range of energies and the fates themselves are energy dependent.   To separate these, pre-FSI energy line requires $0.1<$ true kinetic energy $<0.6$ GeV, which are protons that would be in the right range to pass the selection if no FSI happened.   The final line shows the original simulation has 25\% more elastic+nucleus scattered protons to passing the angle selection.
\label{tab:neutroncounts}}
\end{table}

Table~\ref{tab:neutroncounts} shows how many protons of each fate survive the selection.  The first line shows that stepping through carbon, 36\% of the protons do not experience FSI at all, which is the combination of the stepper and the mean free path from SAID.   As nuclei get larger, this number goes down significantly, though there are always some reactions that take place at the edge of the nucleus.  Beyond the 36\% for carbon in the top line of Table~\ref{tab:neutroncounts}, the generated no-FSI fate happens 32\%, 18\%, 17\%, and 10\% of the time for $^{16}$O, $^{40}$Ar, $^{56}$Fe, and $^{208}$Pb, respectively.

Elastic, single-nucleon knockout, and multi-nucleon knockout have similar probabilities when averaged across all CCQE proton kinetic energies at the start.   This is less true for protons that are in the right energy range to be selected.   Events where pions are produced are higher energy and are rare for CCQE reactions, then they are explicitly cut from the signal definition.   The no-FSI and elastic fates preferentially survive the energy and angle cuts; the other fates lead to protons below threshold and are reduced by about one third.    

\begin{figure*}[tbh!]
\begin{center}
\includegraphics[width=5.4cm]{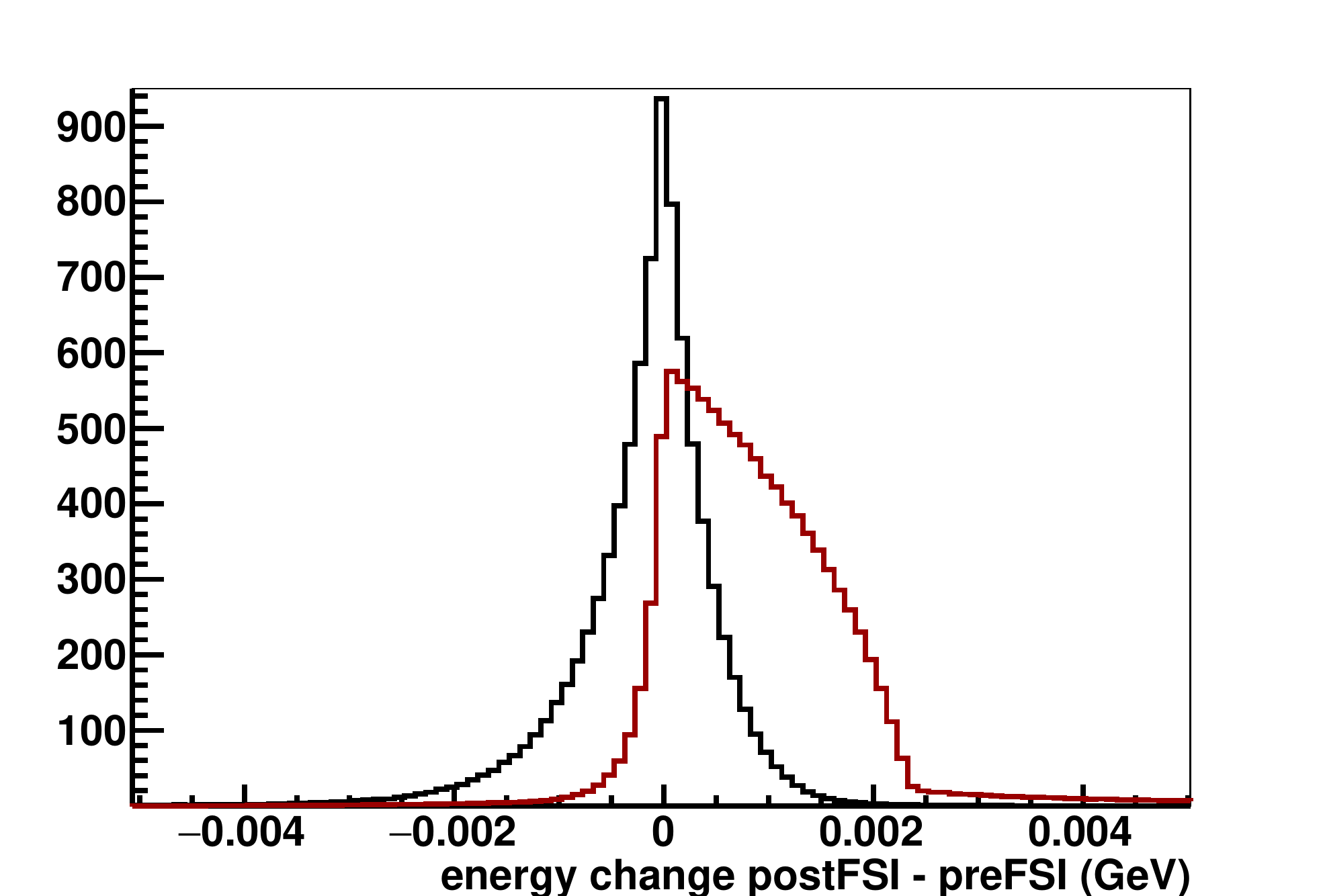}
\includegraphics[width=5.4cm]{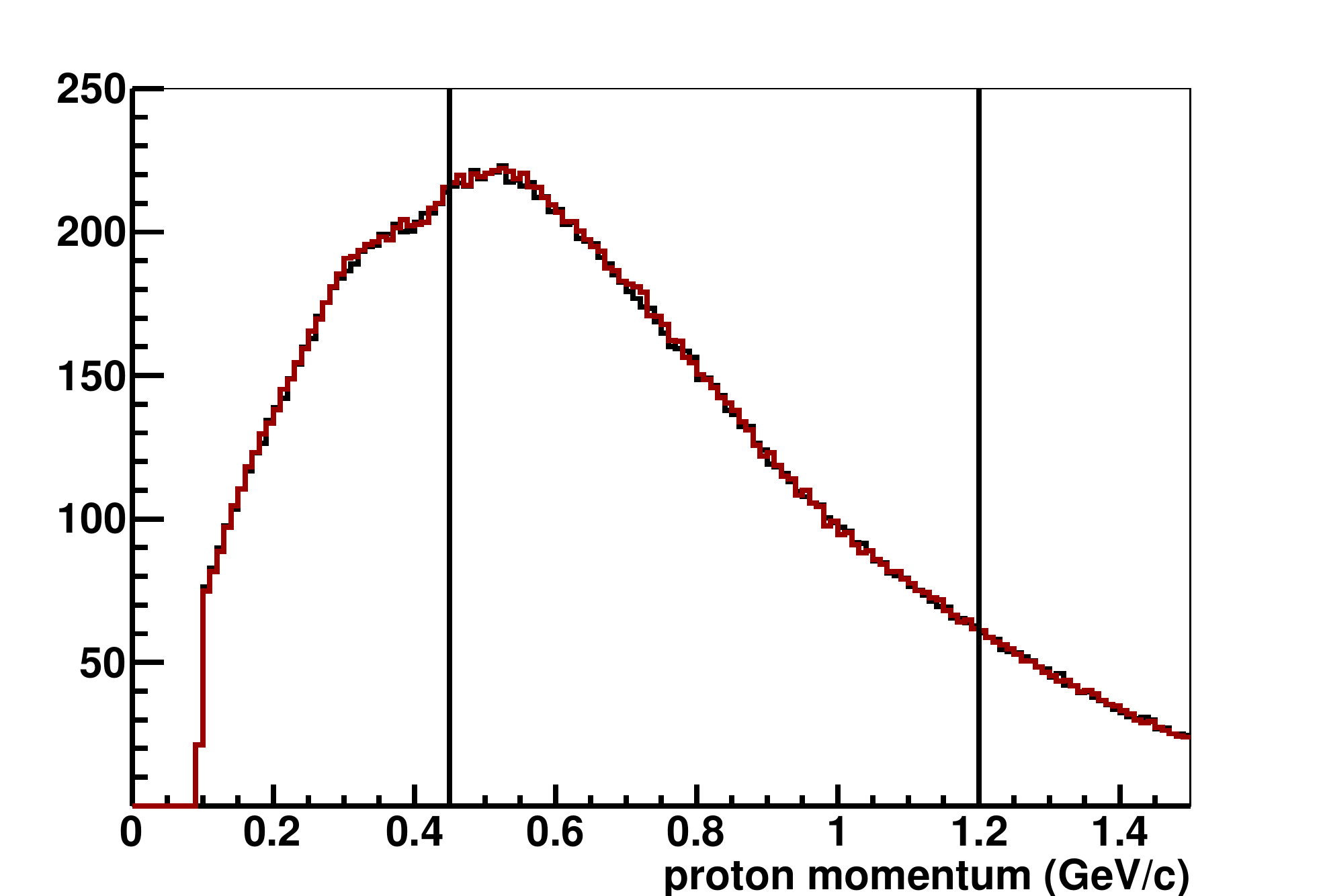}
\includegraphics[width=5.4cm]{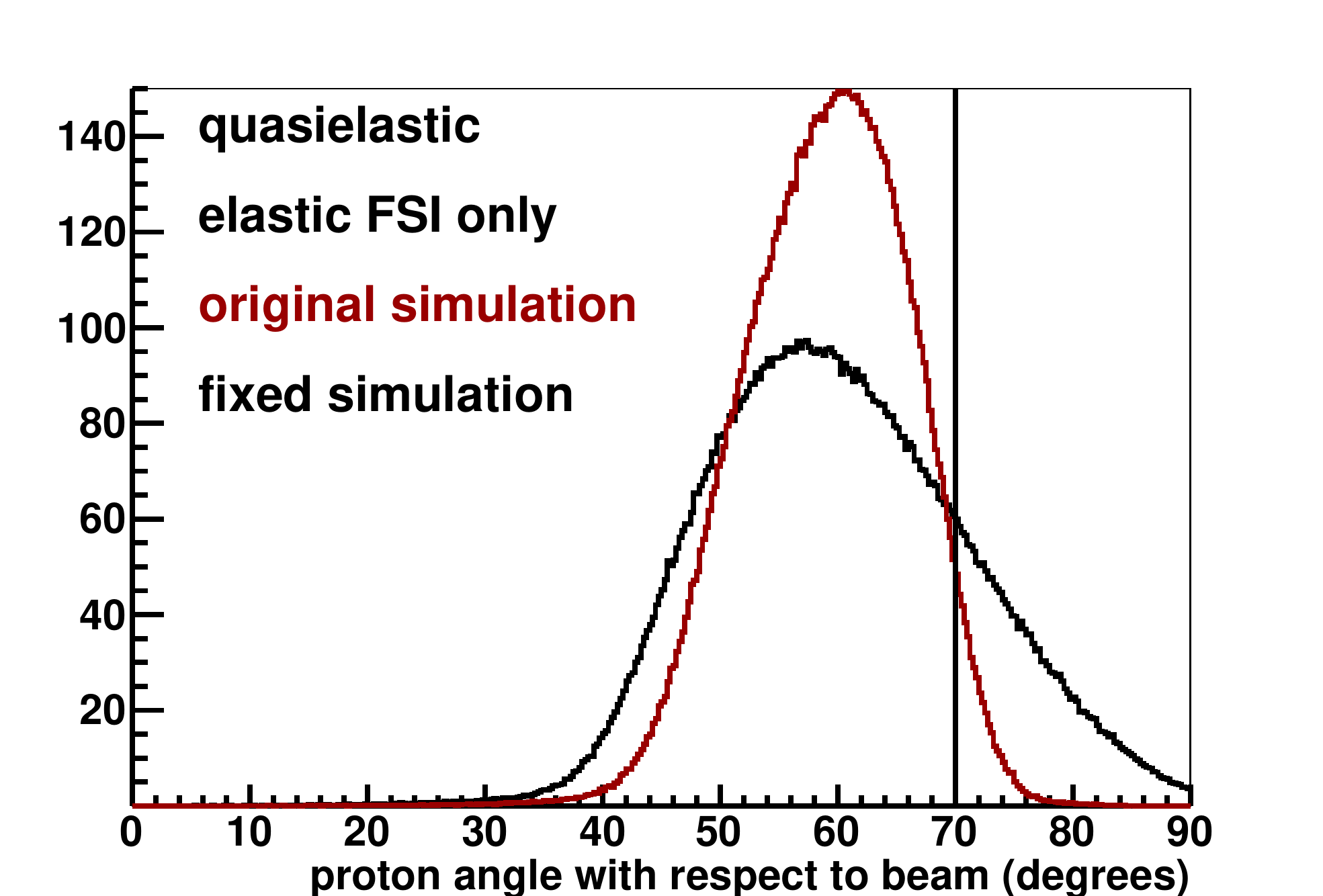}
\caption{These figures are for quasielastic events and elastic fate only:  the small energy distortion (left), the momentum distribution after muon cuts only (center), the angle distribution after muon and proton momentum selection (right).   Red is the original elastic code, black is the fixed version, both generated with {\small GENIE}v2.12.10+FSIfix.  The anomalous 2 MeV energy distortion has negligible effect on the proton momentum selection.  The angle distortion dramatically exaggerates protons passing the angle $< 70$ degree selection.
\label{fig:cuts}}
\end{center}
\end{figure*}

The original simulation produced an energy distortion that was less than 2 MeV, giving negligible distortion of proton momentum.   However, there is a significant angle distortion.   The latter caused significantly fewer events at high angles and more to be accepted into the sample with $\theta < 70$ degrees.  These are illustrated in Fig.~\ref{fig:cuts}.   For elastic only this is 11.7/9.5 = 25\% increase; for the whole sample it becomes only 6\% too many.    Reproducing the same for an alternate detector with 50 MeV kinetic energy thresholds and 4$\pi$ angle coverage does not change the results and interpretations that follow, except in this case it is always perfectly efficient to pass the angle selection.

\subsection{basic comparison of old vs. new}

\begin{figure*}[tbh!]
\begin{center}
\includegraphics[width=7cm]{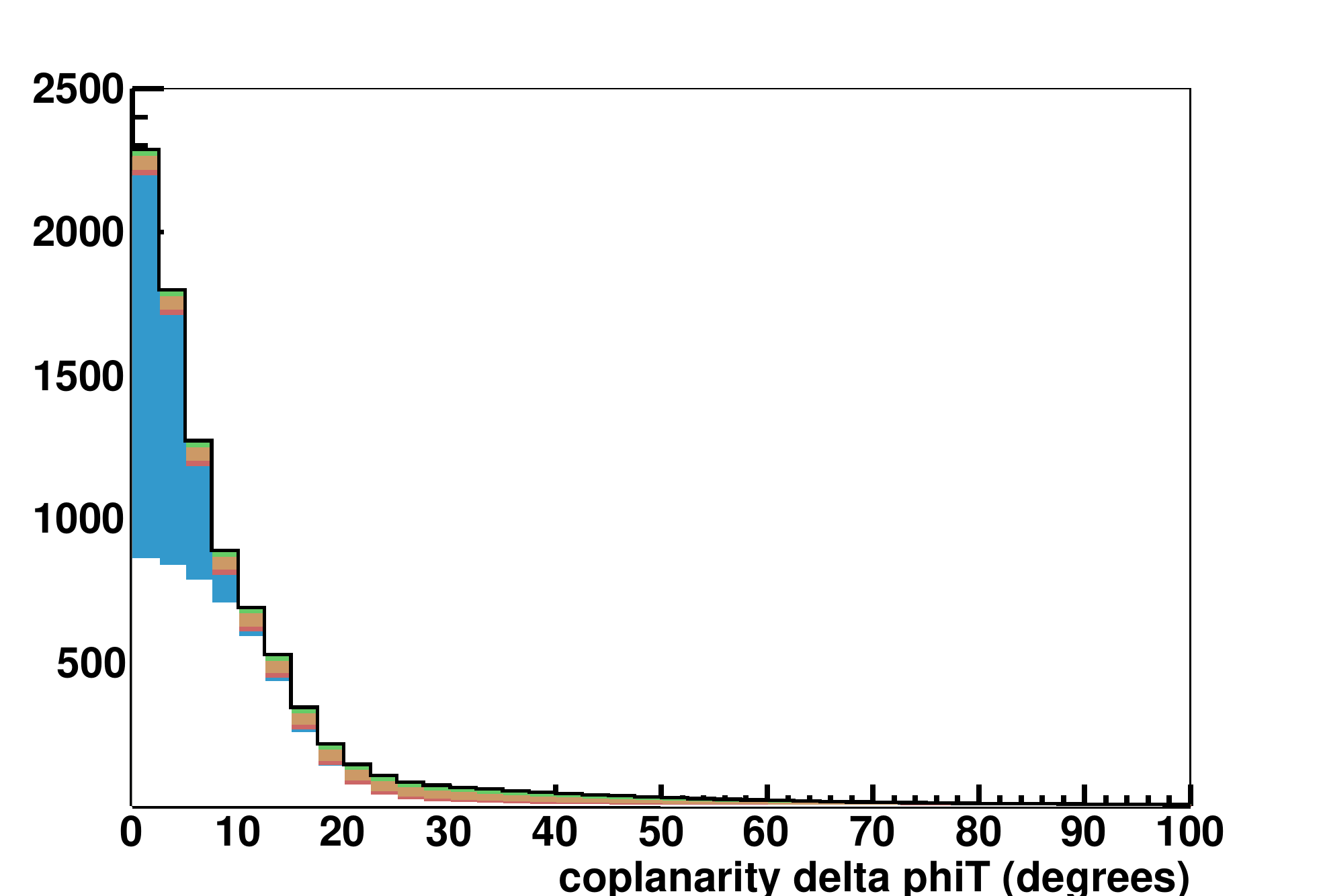}
\includegraphics[width=7cm]{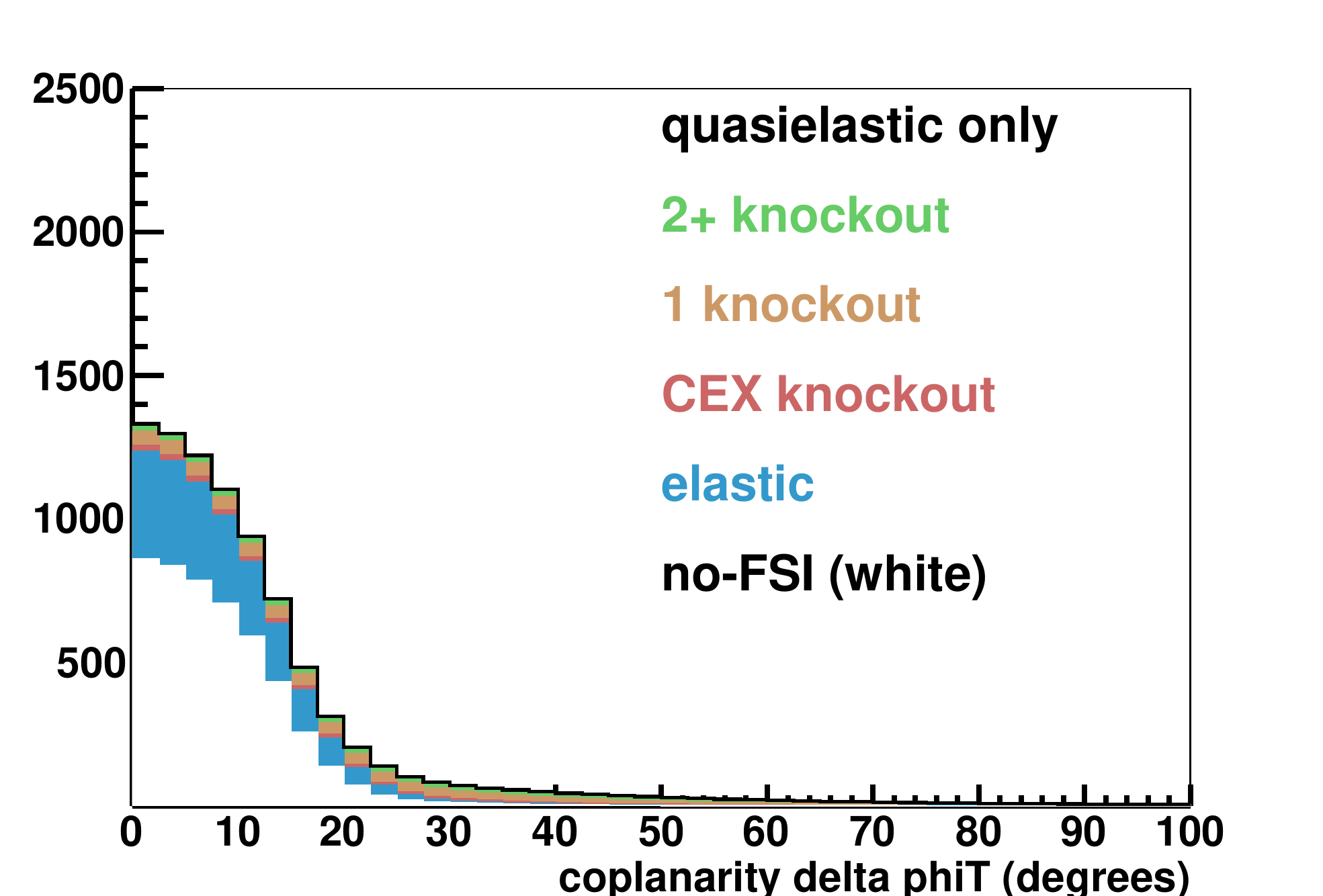}
\includegraphics[width=7cm]{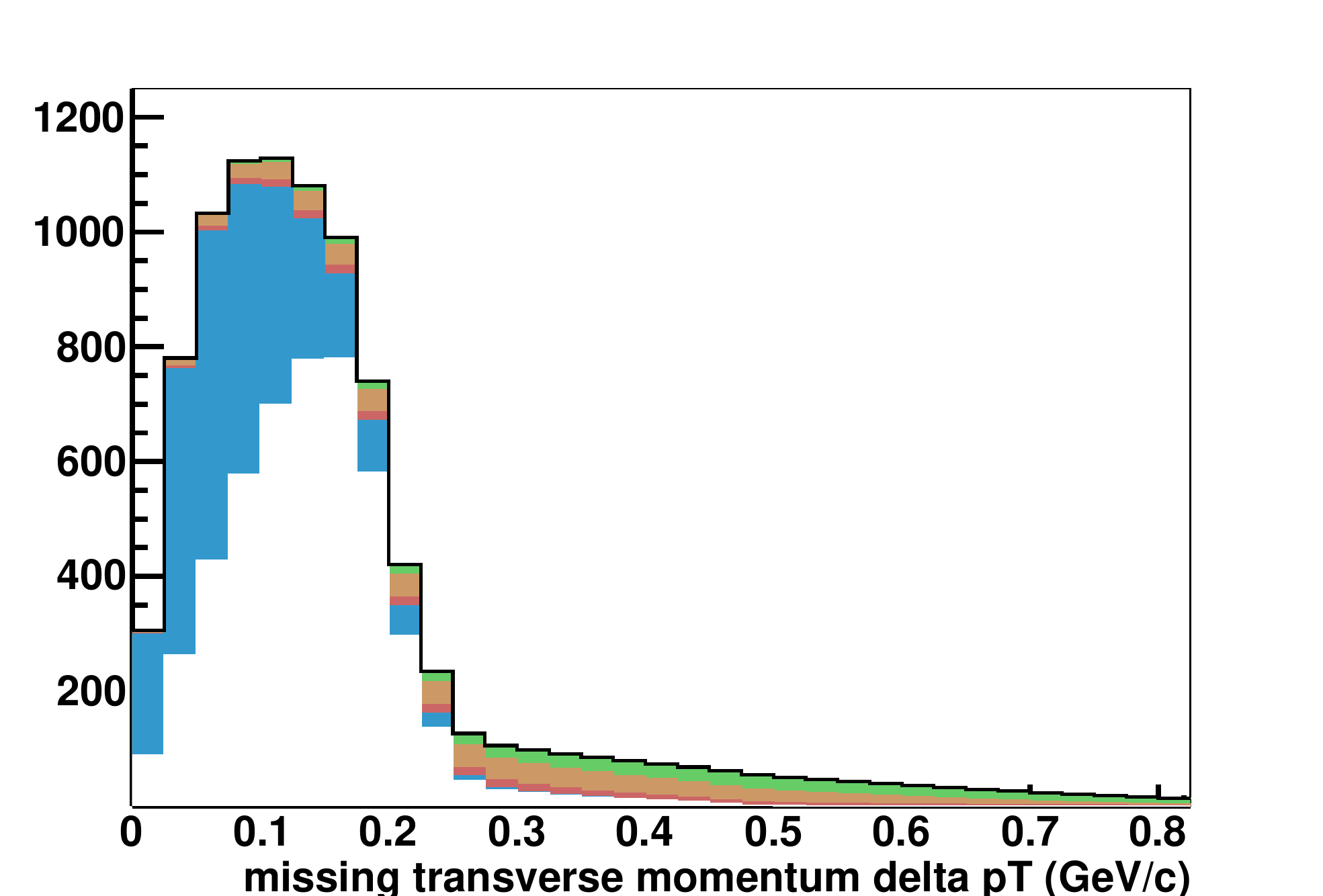}
\includegraphics[width=7cm]{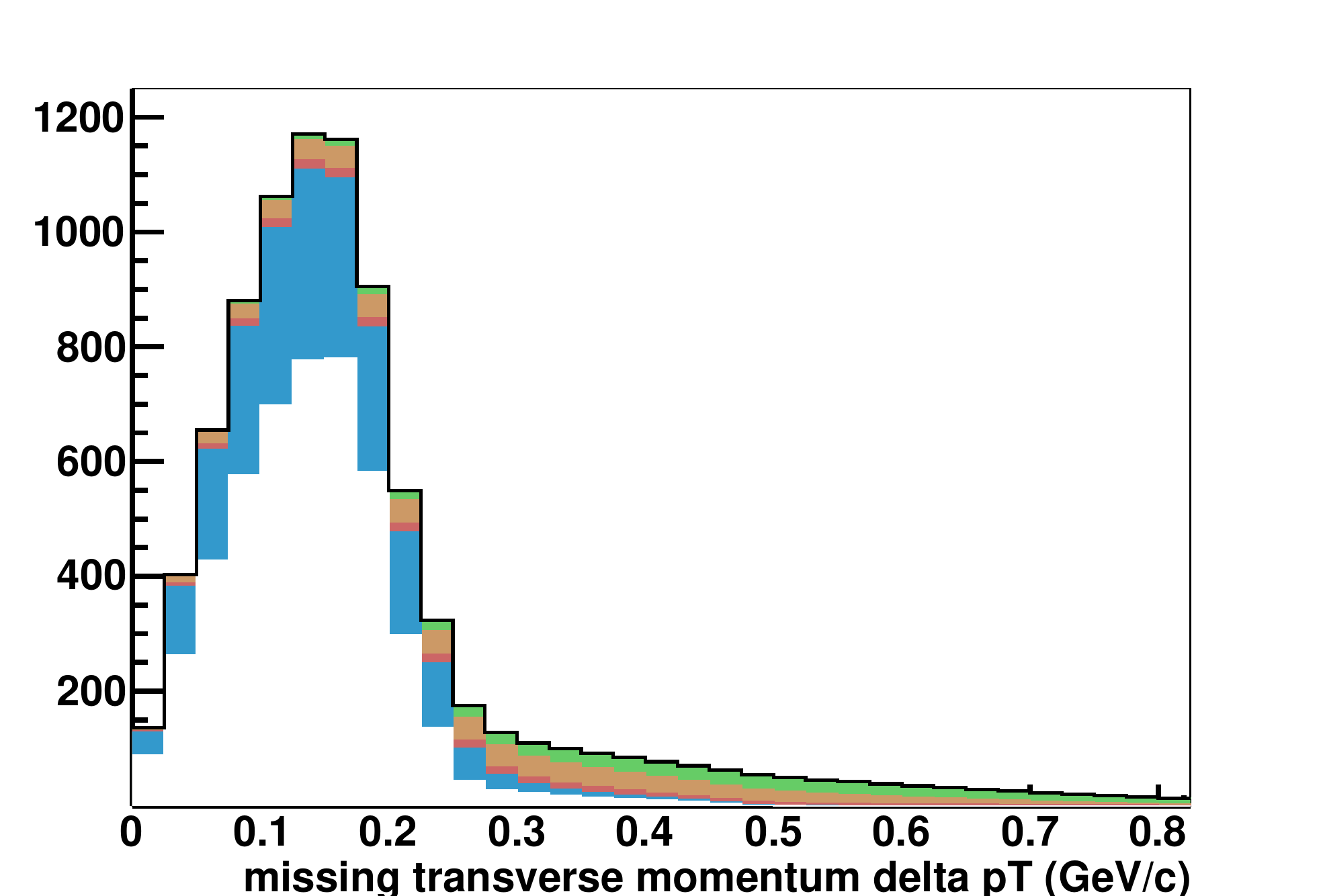}
\includegraphics[width=7cm]{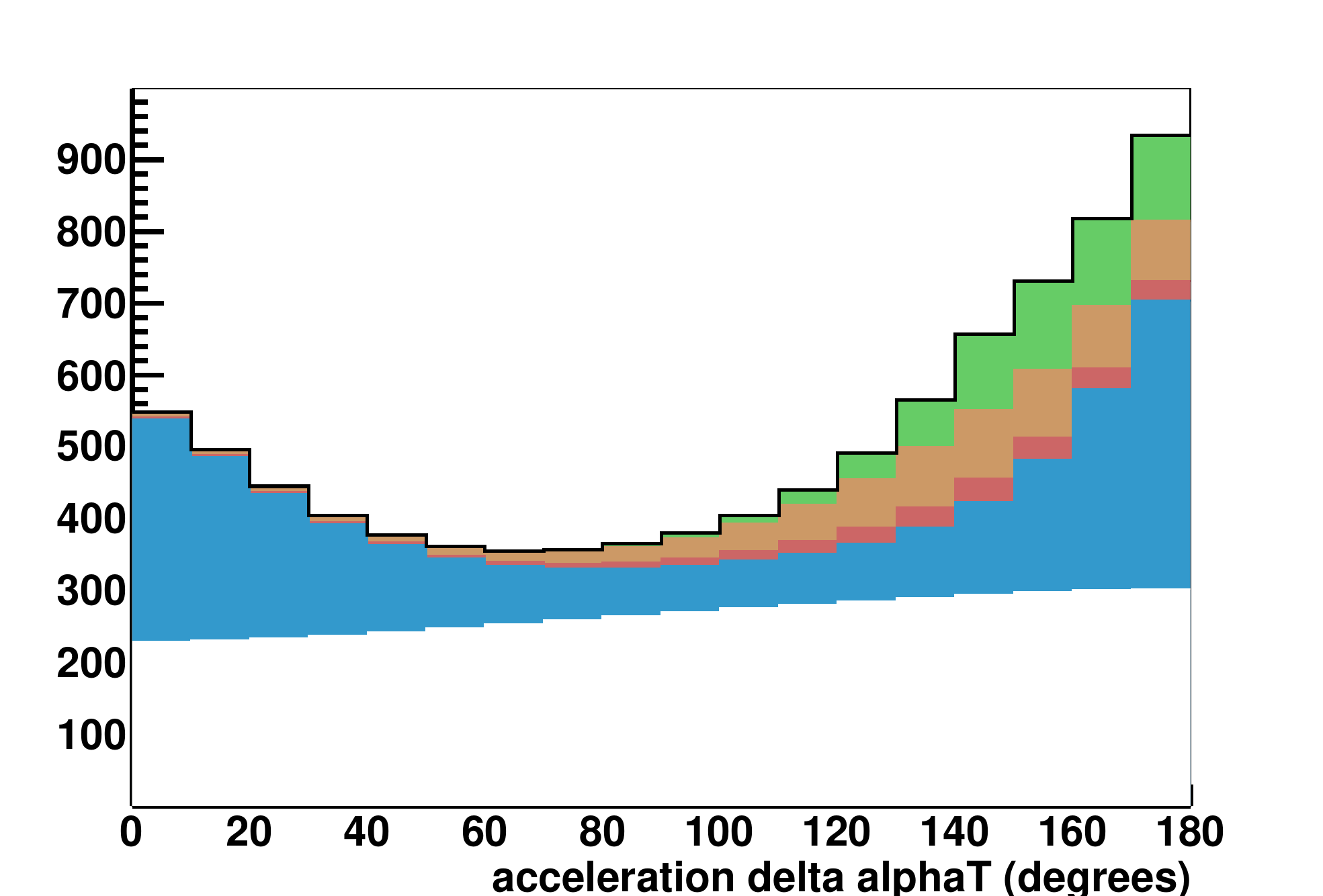}
\includegraphics[width=7cm]{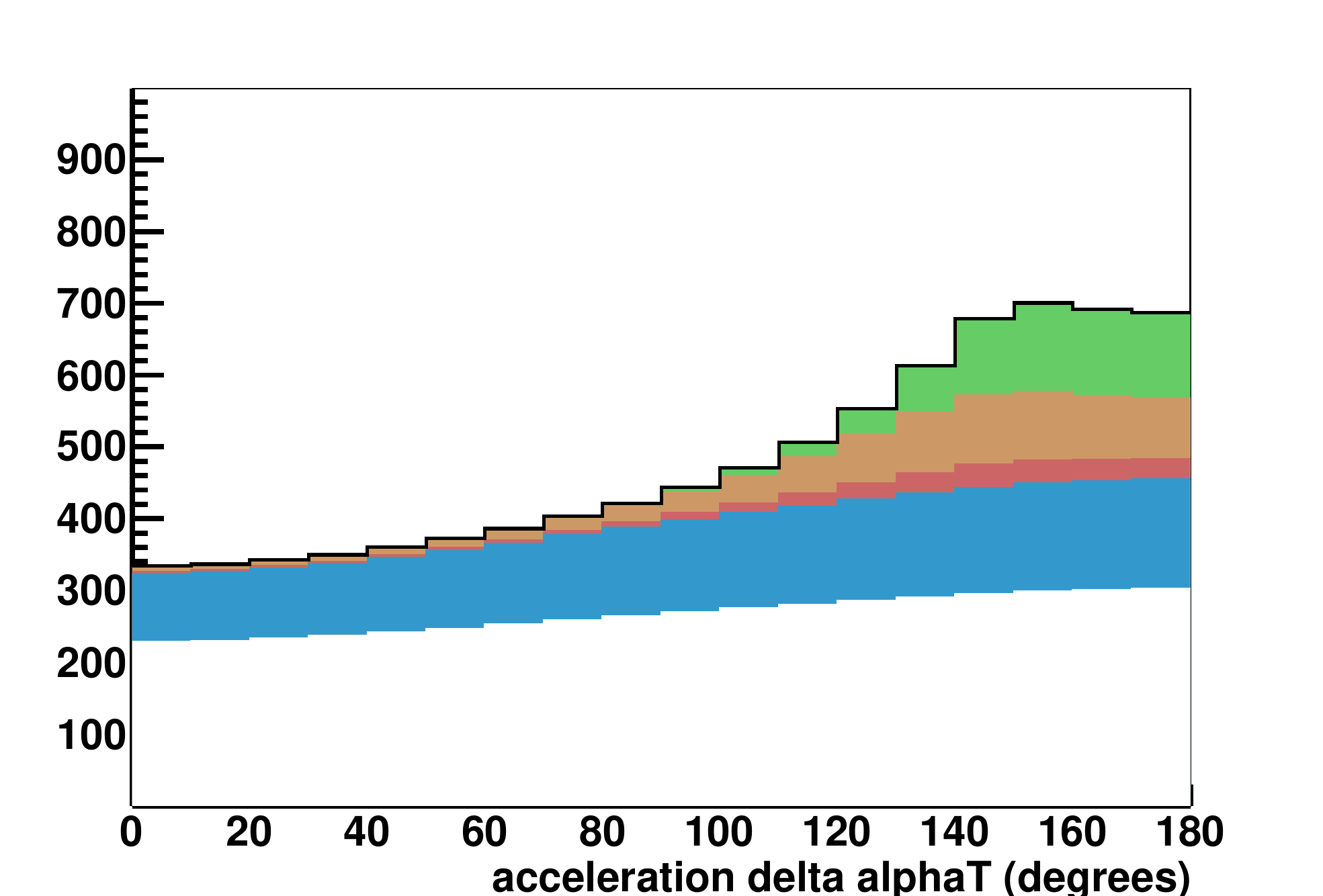}
\includegraphics[width=7cm]{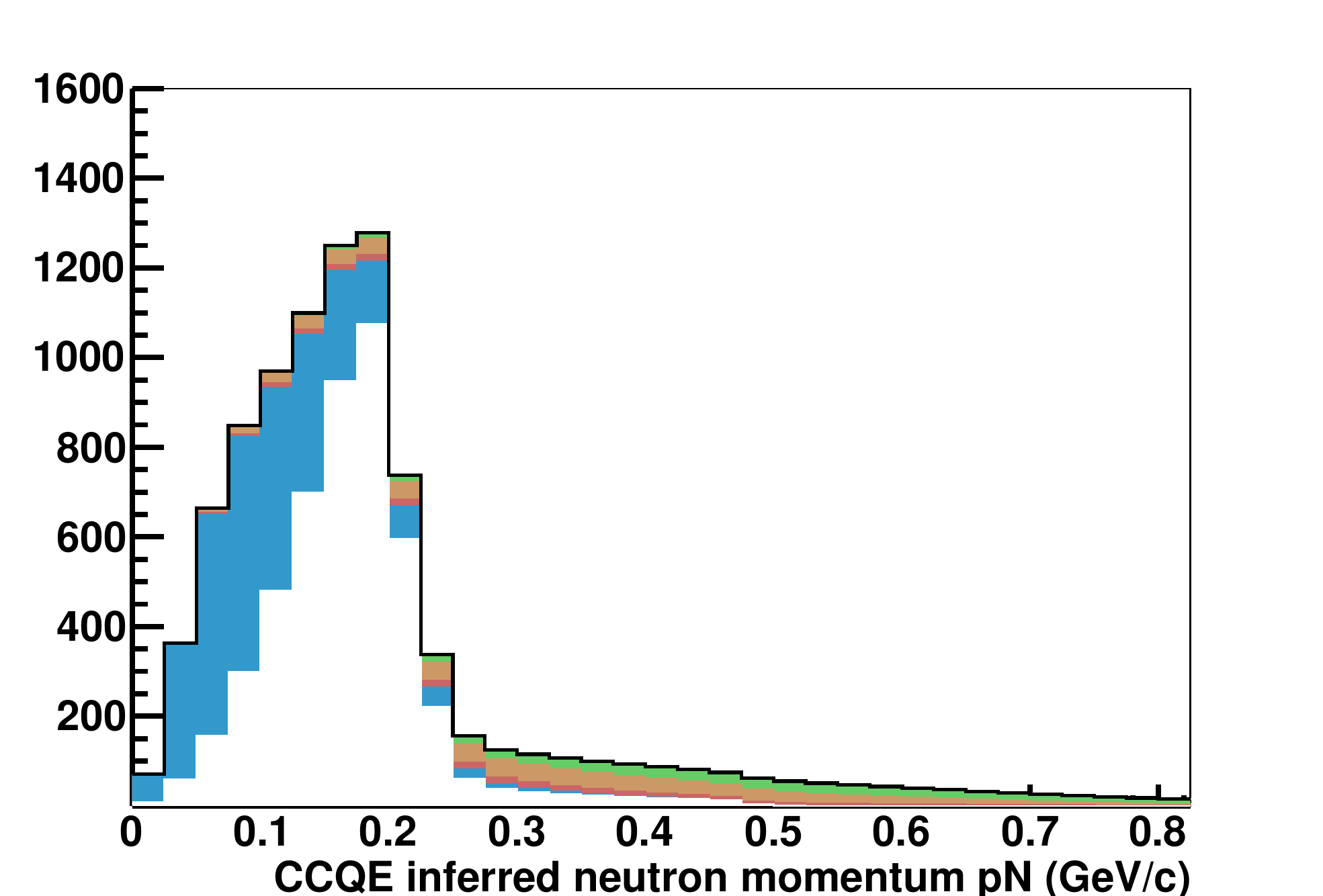}
\includegraphics[width=7cm]{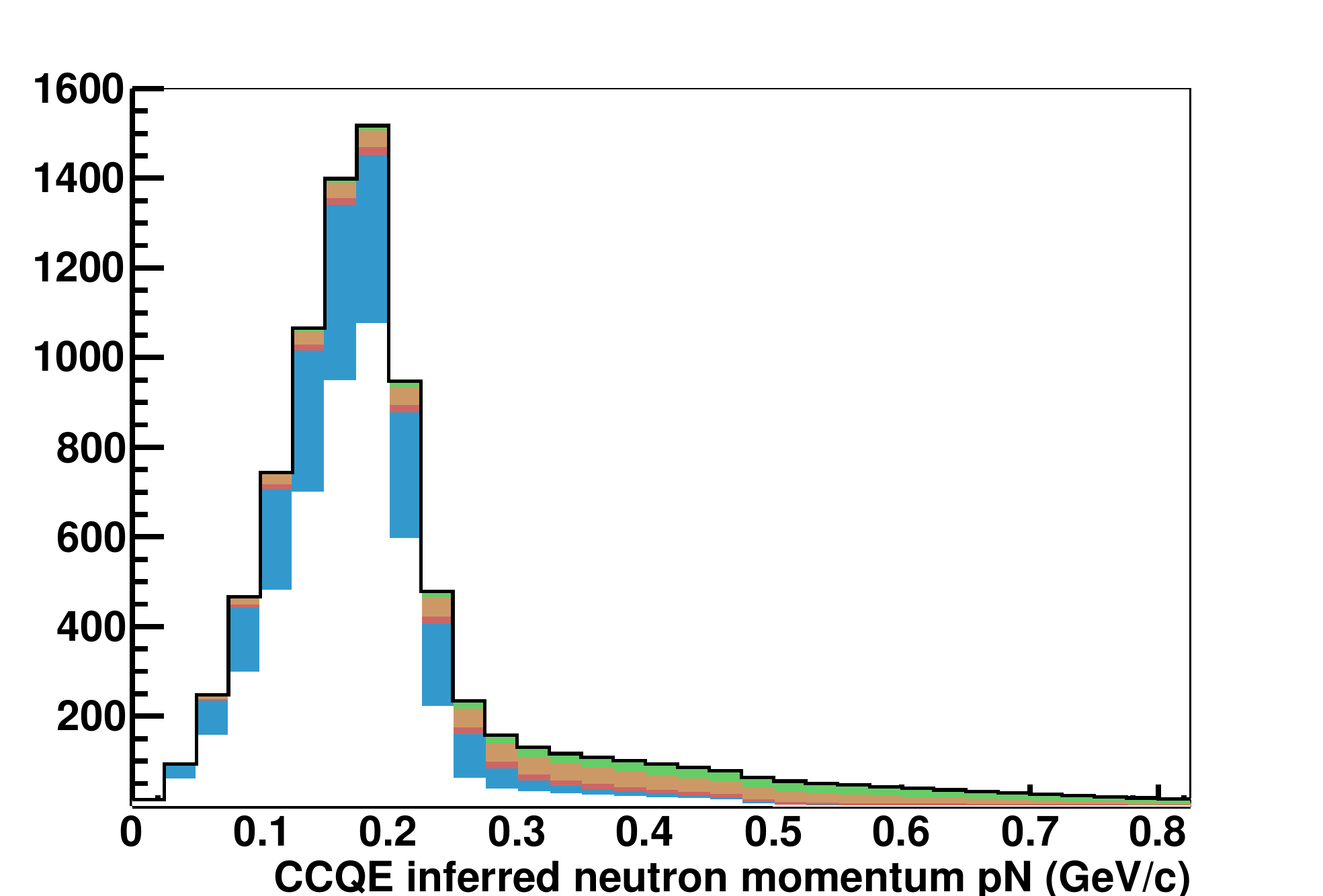}
\caption{{\small GENIE}v2.12.10+FSIfix original (left) and fixed (right) transverse kinematics distributions.  The elastic portion (blue) is shifts significantly and is overall somewhat reduced.  Distortions to the single nucleon knockout (red and brown) are not visible without using the ratios coming next.
\label{fig:oldnewsidebyside}}
\end{center}
\end{figure*}

The most prominent effect of fixing the TwoBodyKinematics function stands out immediately in the side by side comparison in Fig.~\ref{fig:oldnewsidebyside}.  The old simulation is on the left, the new simulation is on the right.   The elastic component is shown as the blue color stacked on the no-FSI component in white.   Single nucleon knockout with charge exchange (red) and without (brown) are also changed between the two simulations, though their effect will be more evident in the ratio plots later.   The multi-nucleon knockout (green) is unchanged between the two simulations and the pion absorption on two nucleons does not apply.

\clearpage

The top row left plot shows that the elastic fate piles up a few degrees around perfect coplanarity in the old simulation.  The no-FSI component (white in the stacked histogram) is smeared around coplanarity because of Fermi motion.   The elastic hadron+nucleus scattering component should also do this, but with small additional angle smearing on top, making its distribution slightly wider.   The fixed component in the right column of the plots shows exactly that character.   When anomalous coplanarity was recognized as unphysical, it led to the identification of the routine that was causing a problem.   

The fourth figure shows the most complex observable, $p_N$, the estimated momentum \cite{Furmanski:2016wqo} of the struck nucleon.   For protons that experience no FSI, it really is the estimator for the original nucleon momentum, and crucial for exploring the differences between a global Fermi-gas, a local-Fermi-gas, and models with spectral-function character.  The left plot shows an anomalous population to the left of the peak.   It is not obvious with only the stacked plots, but on the right Fermi-motion plus significant additional angle smearing via FSI migrate events from the peak into the tail.   

In the next subsections, ratios are used for comparisons of different approximations to the intended model.   Most 10\% distortions are only evident by taking ratios of the different components and their total.  At the present time, distortions smaller than this 10\% are probably not directly measurable.  However, they may play a role in the correct description of systematic uncertainties in other observables.
   
The structure of the four plots in each next subsection is the same.    The top portion of each plot shows the inputs to a ratio.  The model being tested in the numerator is shown as a stacked histogram.  The same fully fixed model (ElasticConfiguration = 3 in the UserPhysicsOption.xml file) is shown as the thicker black line and is always in the denominator.  Each fate is shown as a stacked histogram plot with the same color scheme and ordering from bottom to top:  white = no-FSI, blue = elastic + nucleus FSI, red = single nucleon knockout with charge exchange, brown = single nucleon knockout, green = multi-nucleon knockout, and later purple = protons from pion absorption.  Pion production is specifically excluded by the CC0$\pi$ signal definition, and is not shown in these initial comparisons.

In the lower sub-panels, the ratio of the alternate model to the fixed model is shown for the total (thick black line) and each subcomponent.  The easiest to spot details are wiggles in the ratio that indicate the peak above is unnaturally narrow because of the anomalous code OR the peak is flattened through additionally randomized hadron direction putting more strength at the edge of the peak or in the tails.   Some of the distortions need a very wide range of ratios, others are better observed with a narrow range around 1.0, so different vertical scales are used.   An excess in the ratio means the numerator (e.g. the old model, or a proposed temporary fix) overestimates the fully corrected model prediction.

\begin{figure*}[tbh!]
\begin{center}
\includegraphics[width=8cm]{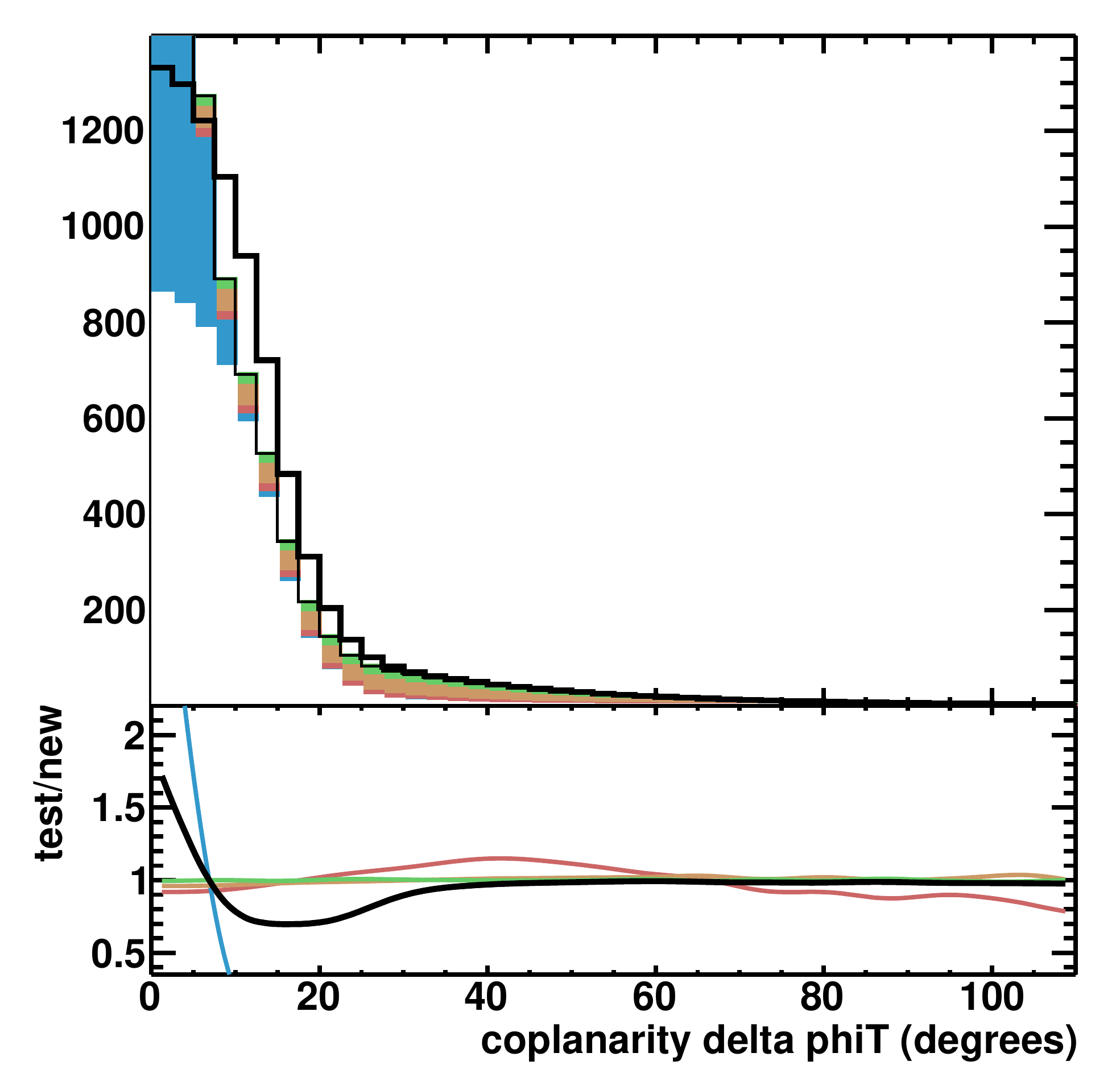}
\includegraphics[width=8cm]{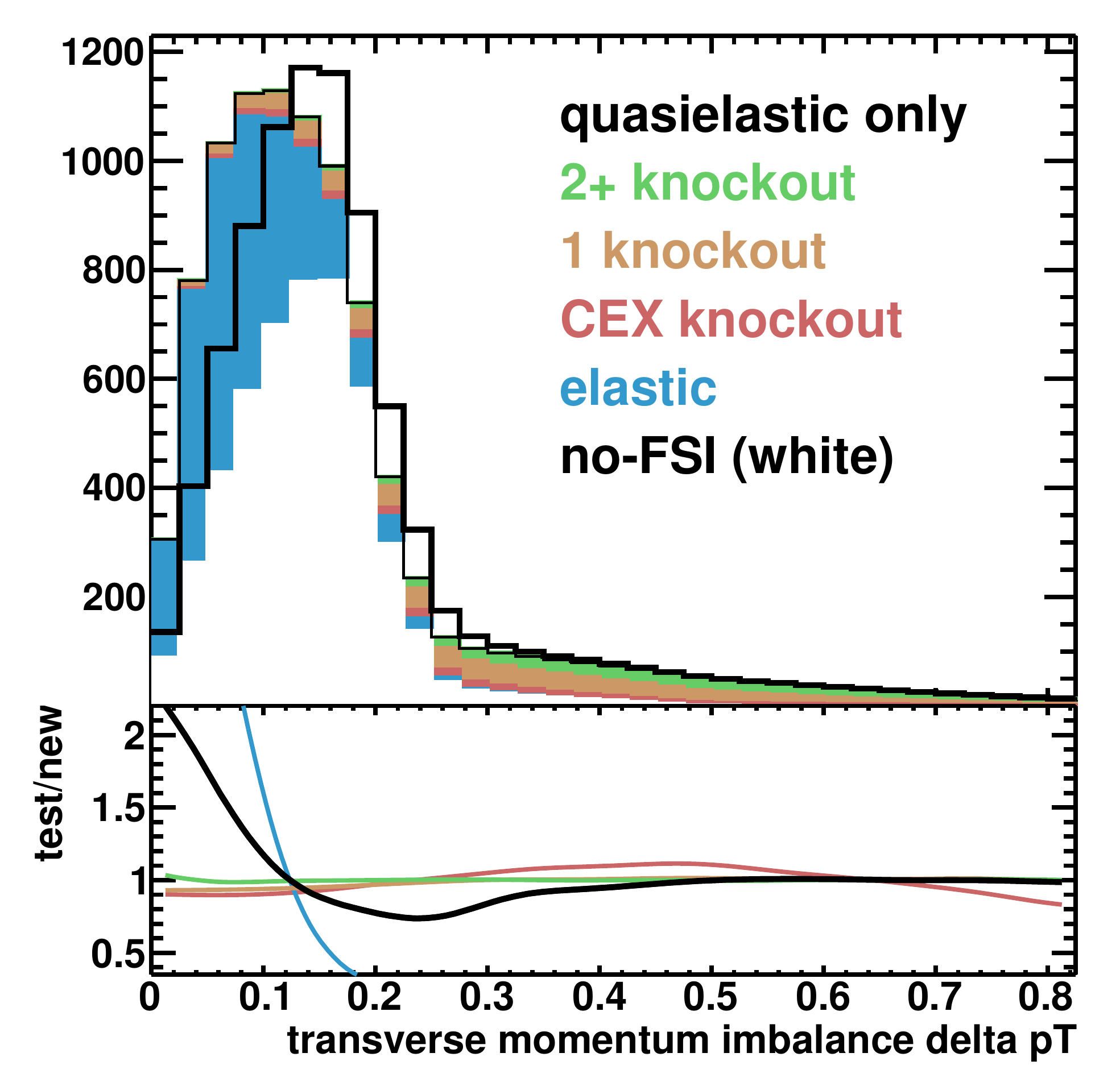}
\includegraphics[width=8cm]{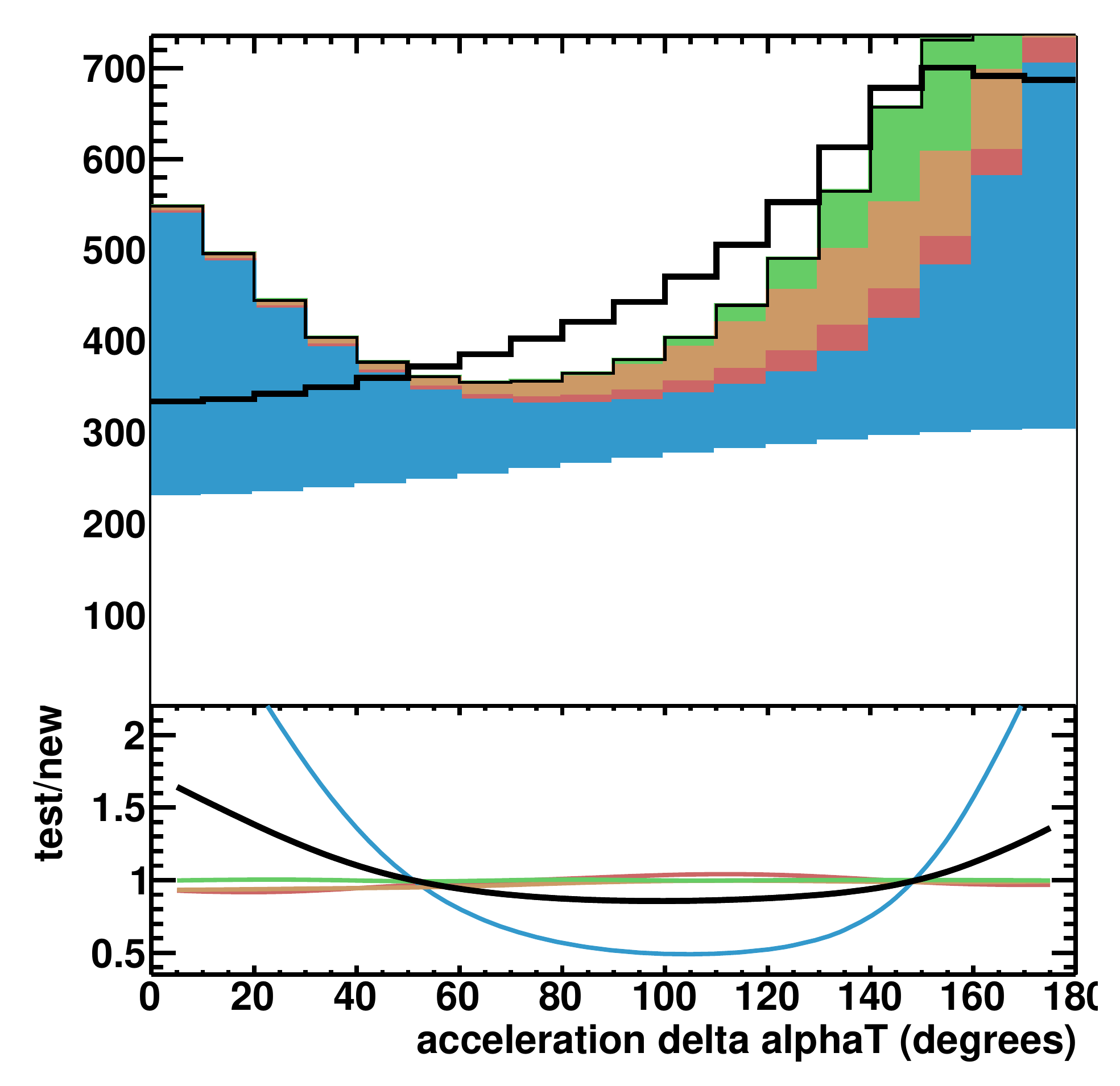}
\includegraphics[width=8cm]{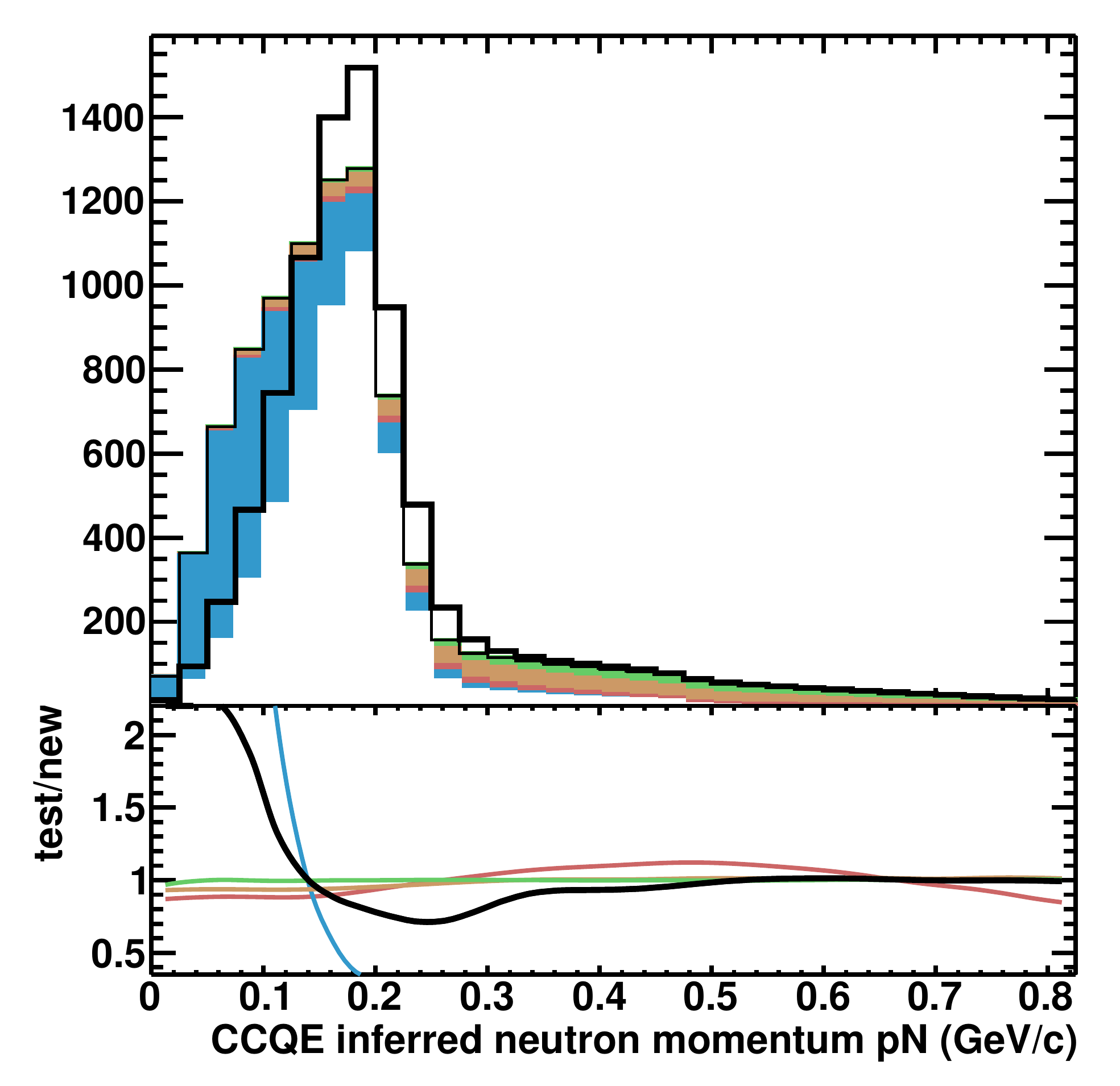}
\caption{{\small GENIE}v2.12.10+FSIfix original {\small GENIE} hA compared to the version with all fixes (thick black line).
The net distortion is primarily from the elastic component, shown with the blue line in the ratio.
It is so severe, lines exceed  the ratio vertical axis.  Same distributions as Fig.~\ref{fig:oldnewsidebyside}.
\label{fig:oldovernew}}
\end{center}
\end{figure*}

\begin{figure*}[tbh!]
\begin{center}
\includegraphics[width=8cm]{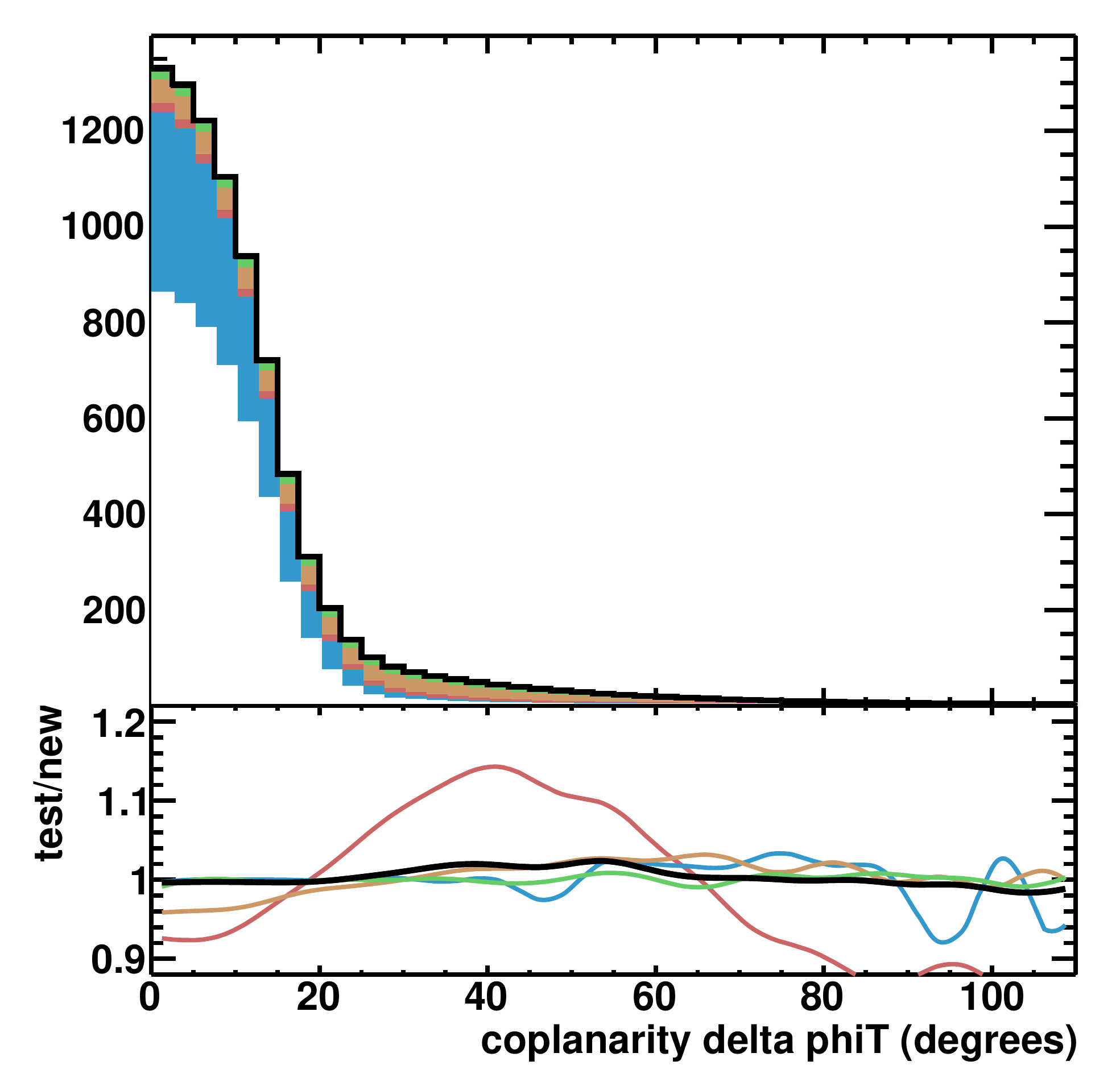}
\includegraphics[width=8cm]{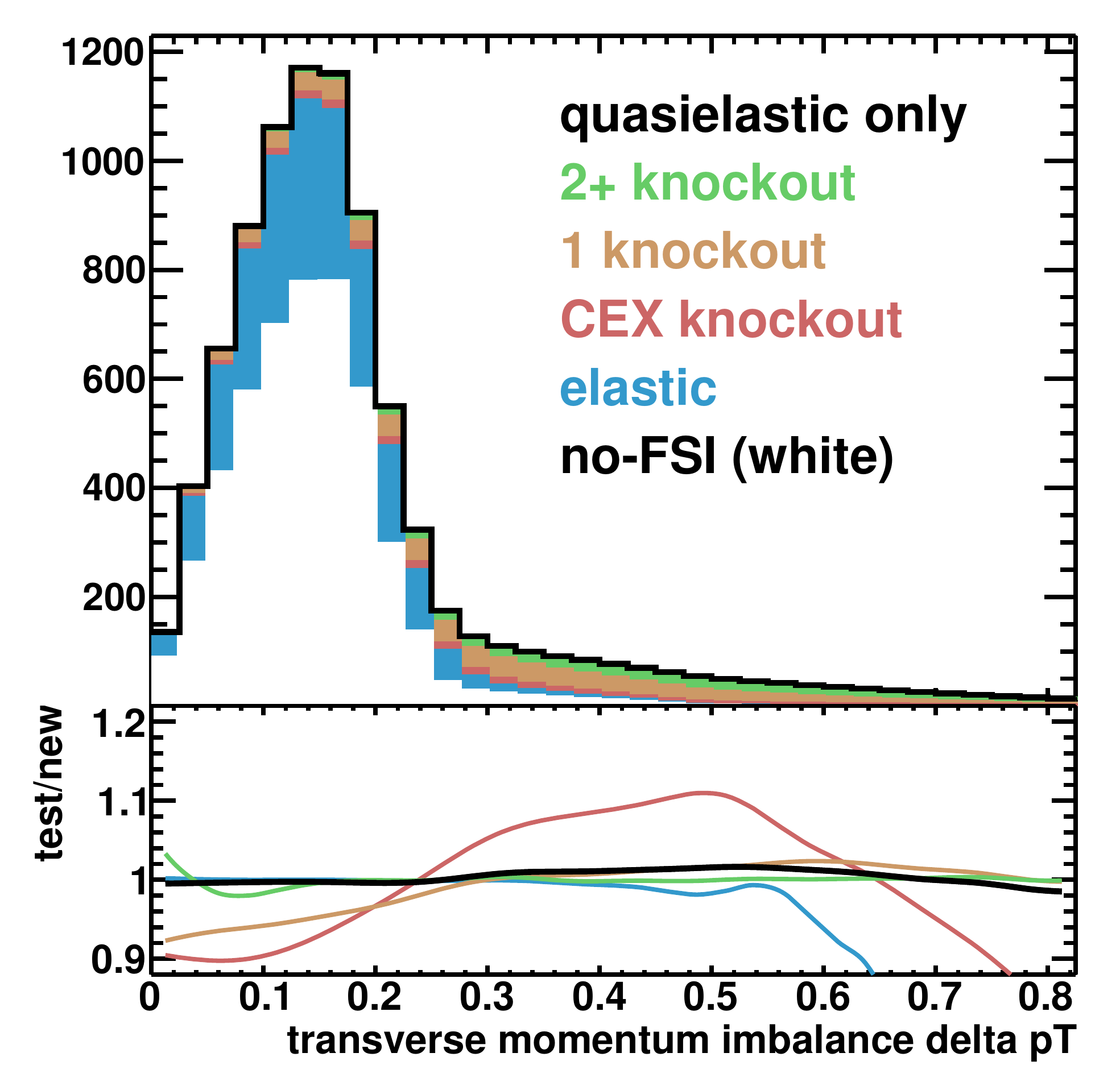}
\includegraphics[width=8cm]{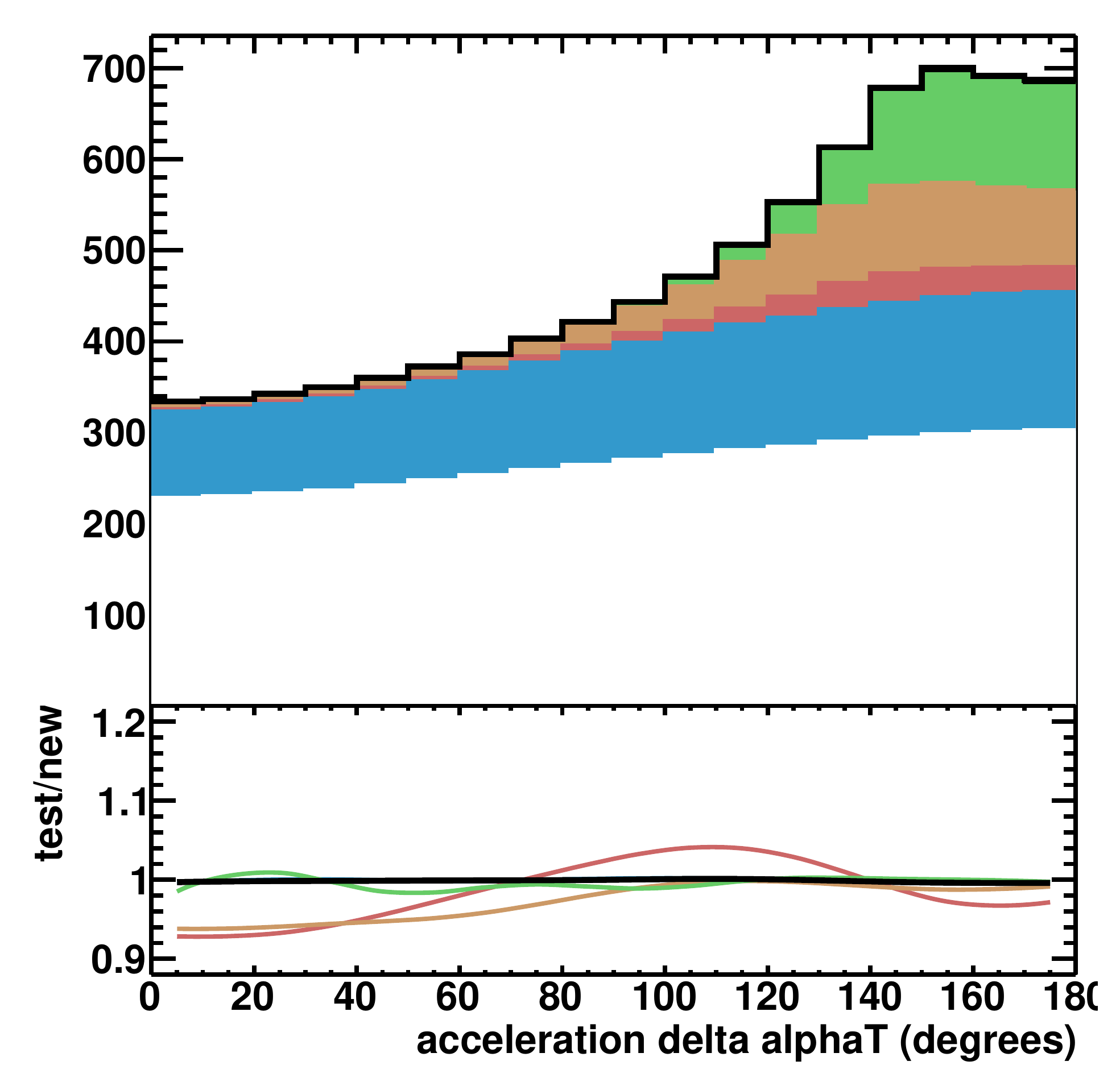}
\includegraphics[width=8cm]{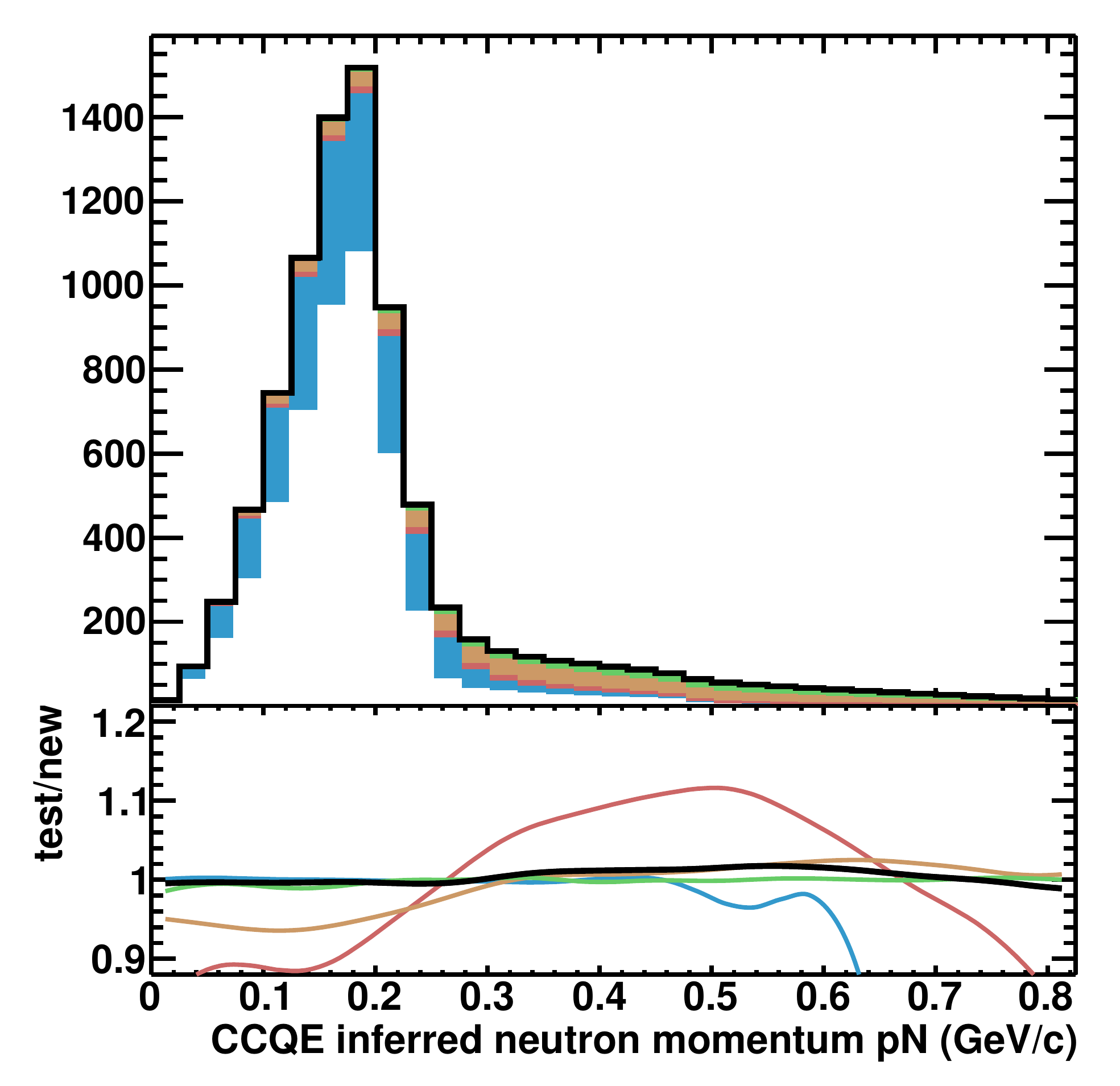}
\caption{{\small GENIE}v2.12.10+FSIfix hA with fix to the elastic only compared to the version with all fixes (thick black line).
This version highlights the differences in the code for the red and brown single-nucleon knockout components.
The differences are less than 10\% of these subcomponents and less than 2\% of the total, suggesting that fixing the inelastic FSI need not be a priority given current systematic uncertainties.
\label{fig:fixELovernew}}
\end{center}
\end{figure*}

\begin{itemize}
\item Sec.~IIC and Fig.~\ref{fig:oldovernew}: original vs. new  \\ dominated by the elastic distortion \\(ElasticConfig 0 vs. 3), same as comparison in Fig.~\ref{fig:oldnewsidebyside}
\item Sec.~IID and Fig.~\ref{fig:fixELovernew}: only fixed elastic vs. new \\shows milder distortion from single-nucleon knockout \\(ElasticConfig 1 vs. 3)
\item Sec.~IIE and Fig.~\ref{fig:nofsiovernew2}: elastic as no-FSI (inelastics not fixed) vs. new  \\ turning elastic to no-FSI as if reweighting existing MC \\(ElasticConfig 2 vs. 3)
\item Sec.~IIF and Fig.~\ref{fig:nofsiovernew}: elastic as no-FSI  (inelastics fixed) vs. new  \\ shows turning elastic to no-FSI is a pretty good approximation \\(ElasticConfig 4 vs. 3)
\item Sec.~IIG and Fig.~\ref{fig:2015overnew}: hA2015 vs. new \\ shows different distortion from turning elastic into inelastic or no-FSI\\({\small GENIE}'s hA2015 vs. ElasticConfig 3)
\end{itemize}

\pagebreak
\subsection{Elastic original vs. new for quasielastic reactions}

\clearpage
\pagebreak
\subsection{fixed elastic only vs. new for quasielastic reactions}

\pagebreak

\subsection{Elastic as no-FSI vs. new for quasielastic reactions}

\begin{figure*}[tbh!]
\begin{center}
\includegraphics[width=8cm]{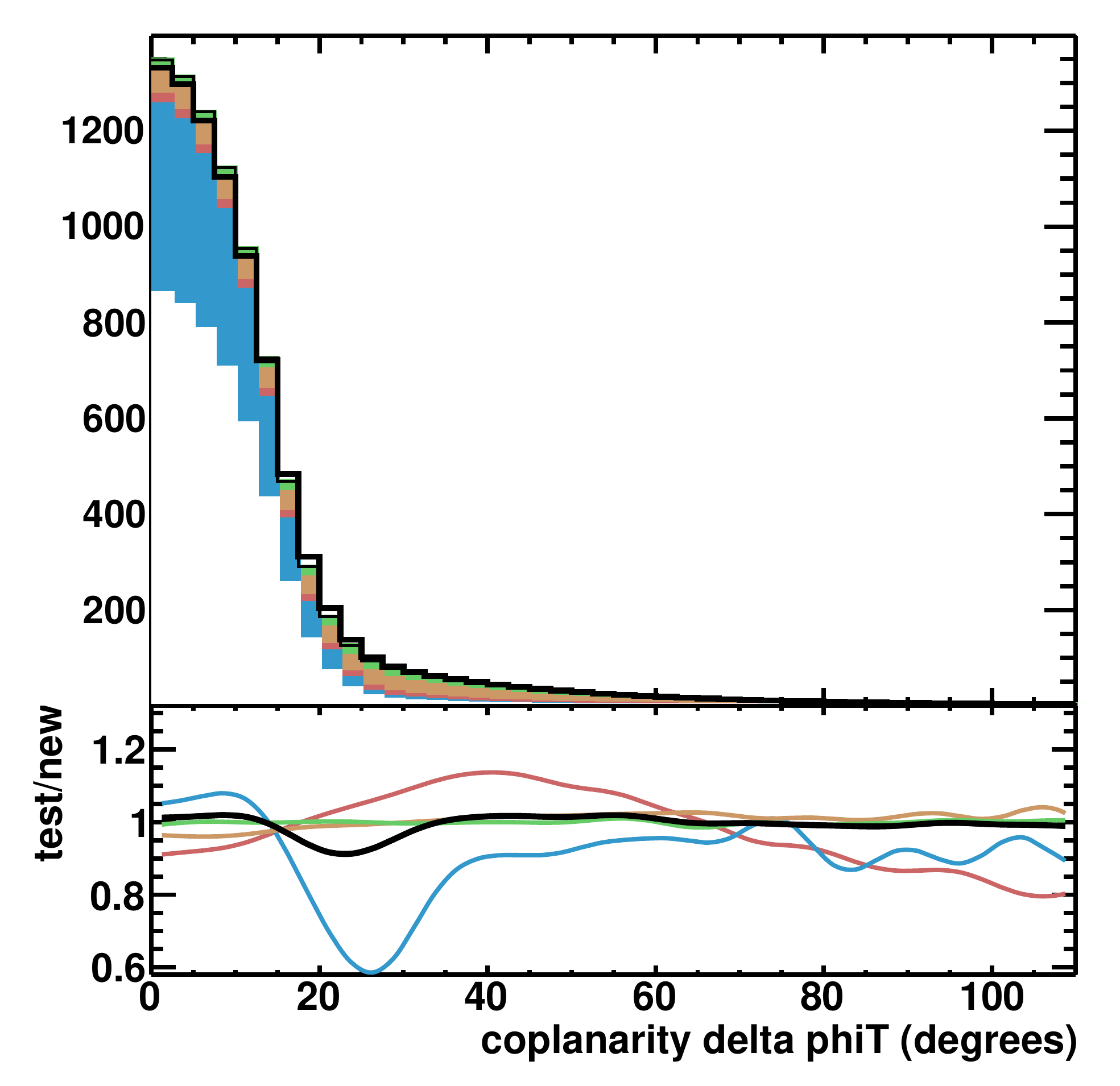}
\includegraphics[width=8cm]{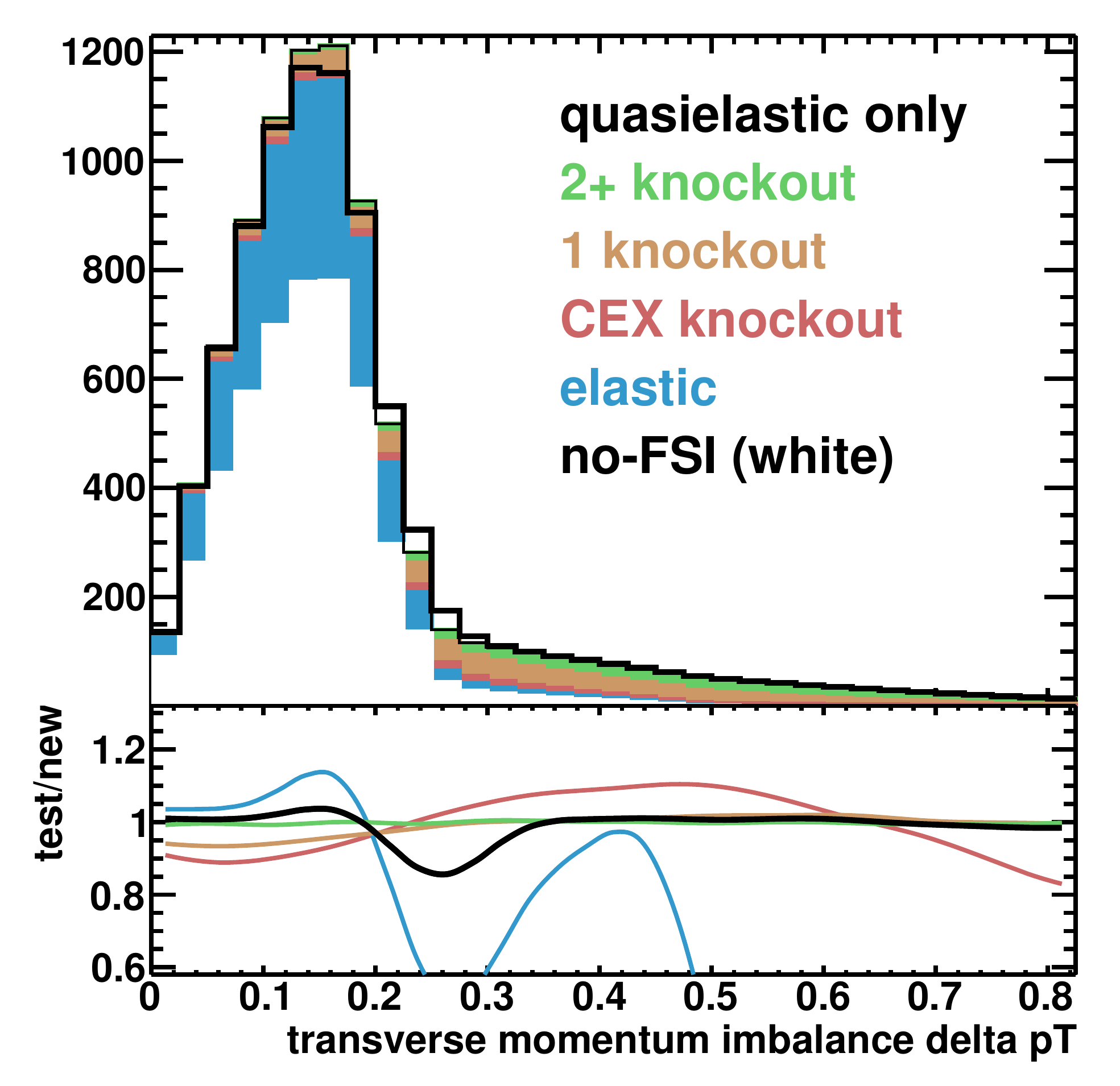}
\includegraphics[width=8cm]{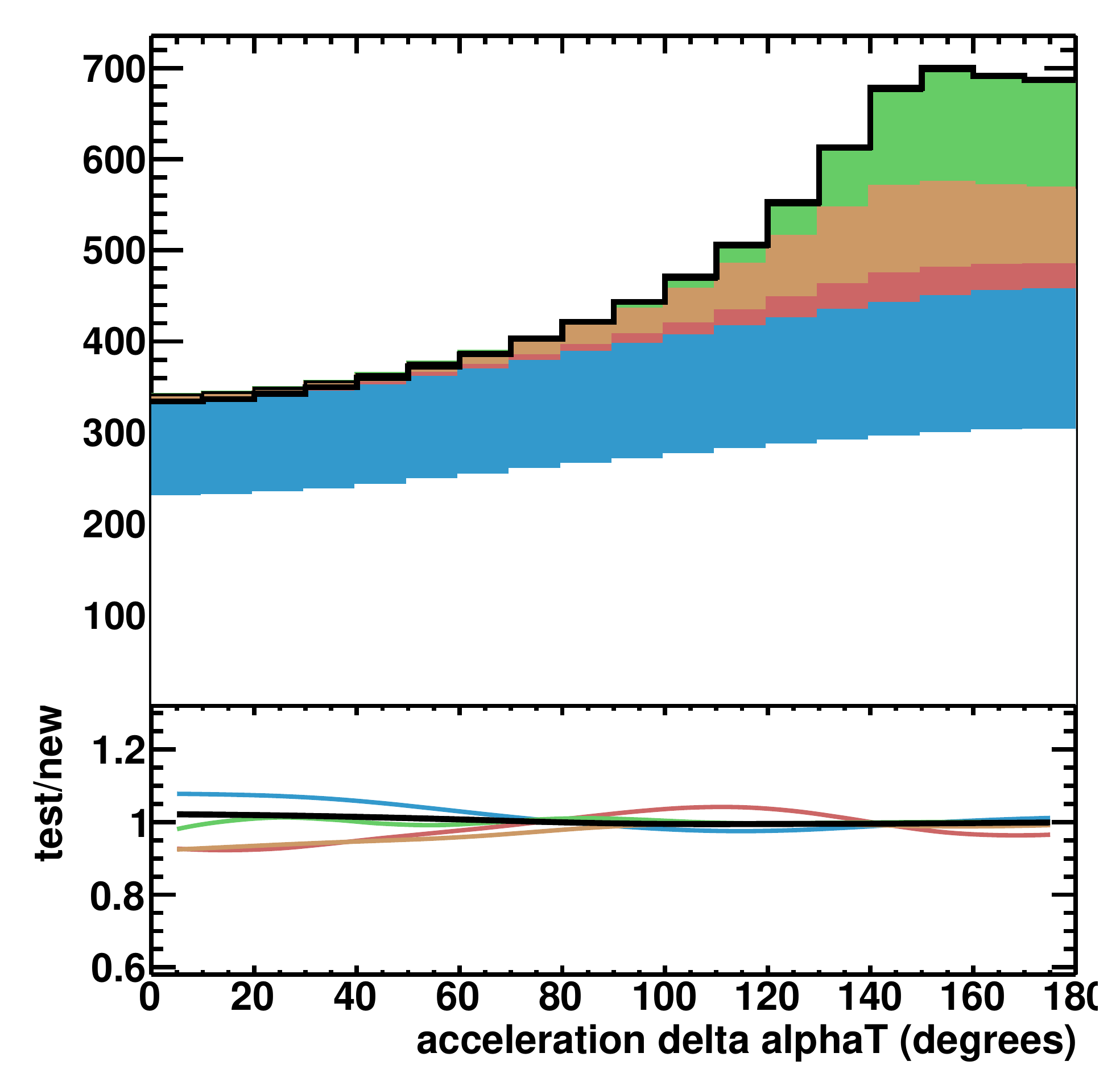}
\includegraphics[width=8cm]{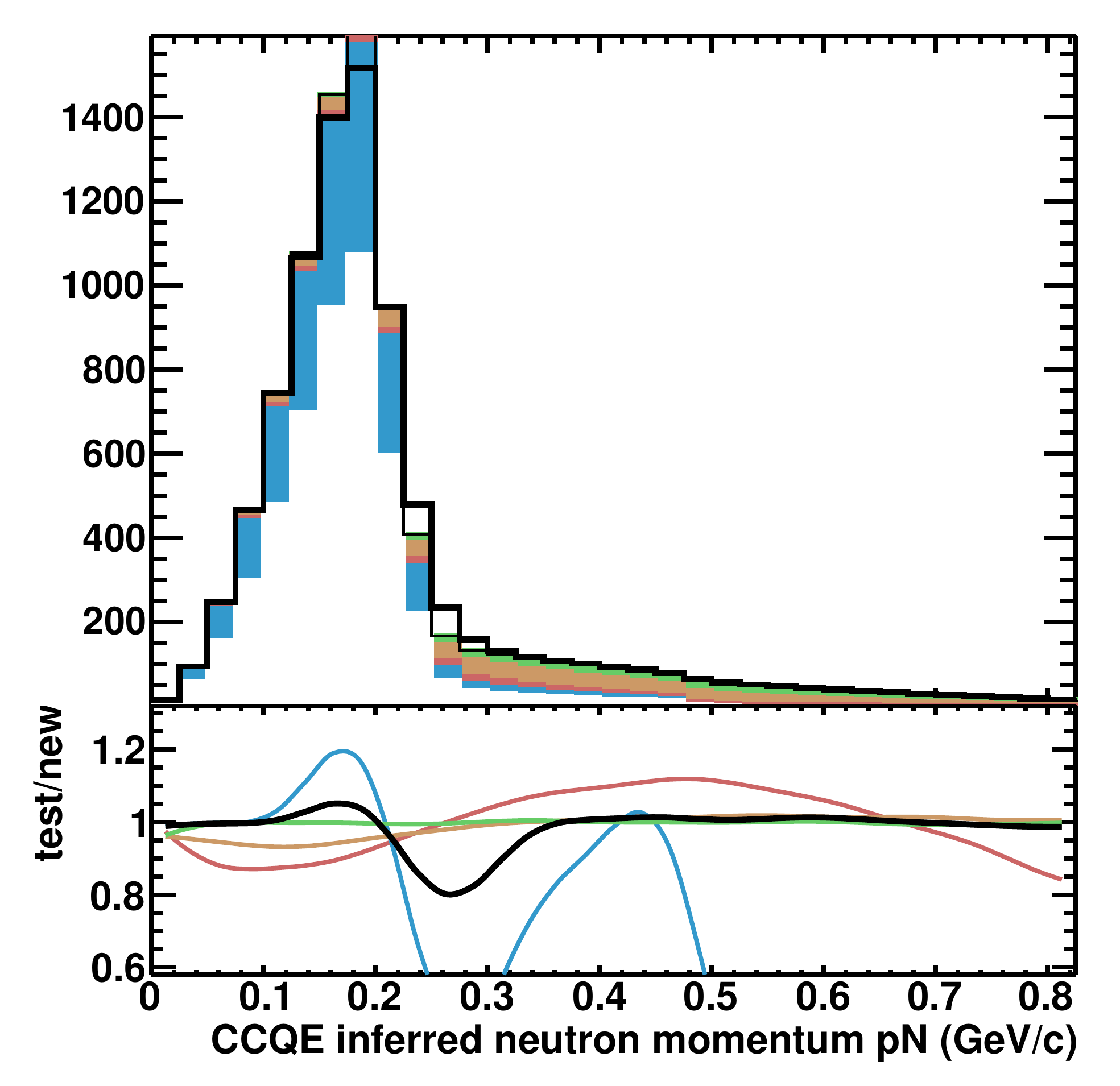}
\caption{{\small GENIE}v2.12.10+FSIfix hA with elastic scatters' $\theta_{CM}$ set to zero compared to the version with all fixes (thick black line).
The inelastic component is NOT fixed.
The black line in the ratio shows the net distortion if existing {\small GENIE} 2.12.10 Monte Carlo weights elastic FSI to no-FSI without regenerating samples.   Having no angle scattering dominates the net effect, the inelastic component has a tiny effect.   The predicted distortion from scattering is at most 20\% of the total CCQE sample at high-side base of the peak.  
\label{fig:nofsiovernew2}}
\end{center}
\end{figure*}

\pagebreak
\subsection{Elastic as no-FSI vs. new and inelastic fixed for quasielastic reactions}

\begin{figure*}[tbh!]
\begin{center}
\includegraphics[width=8cm]{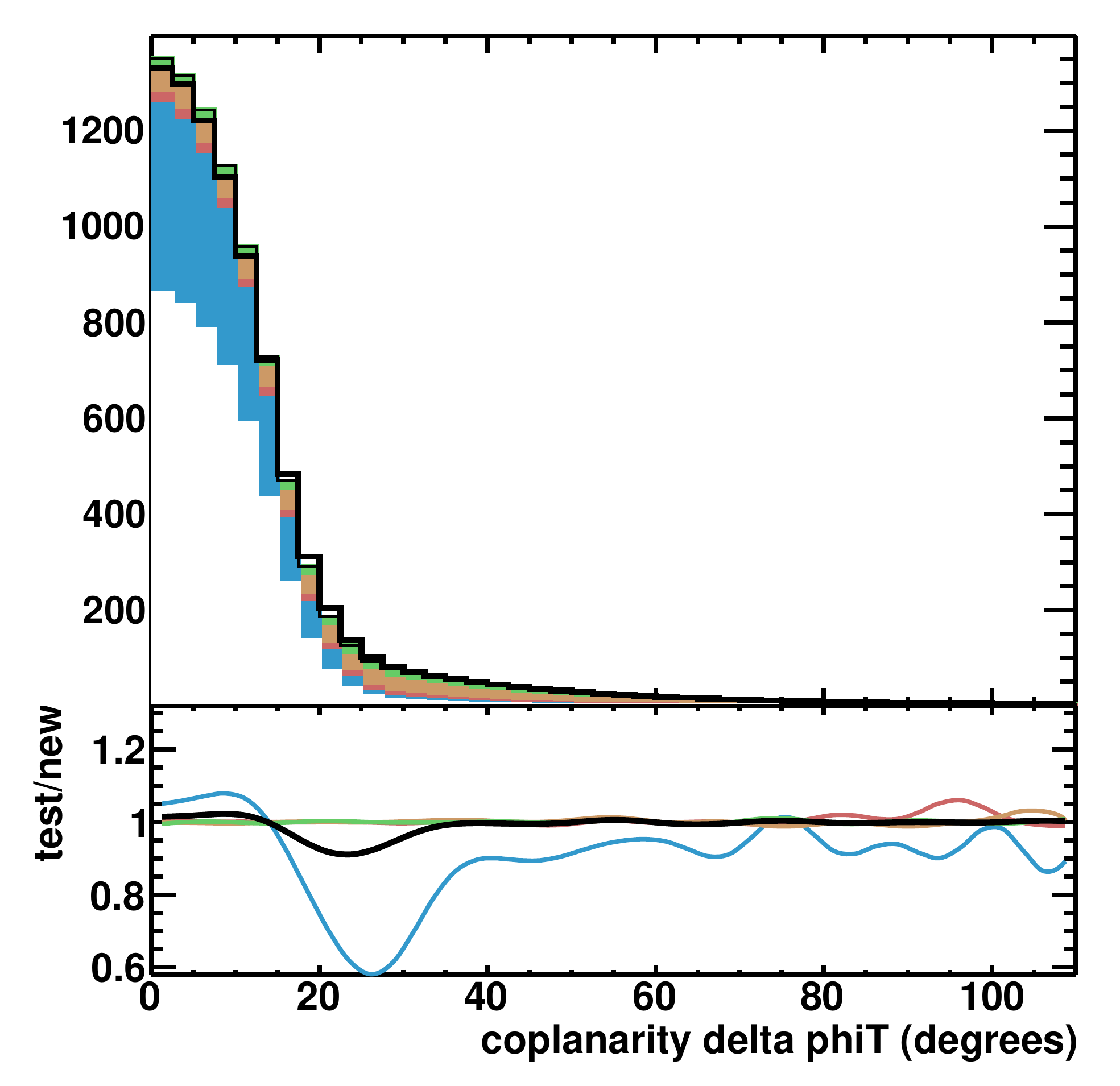}
\includegraphics[width=8cm]{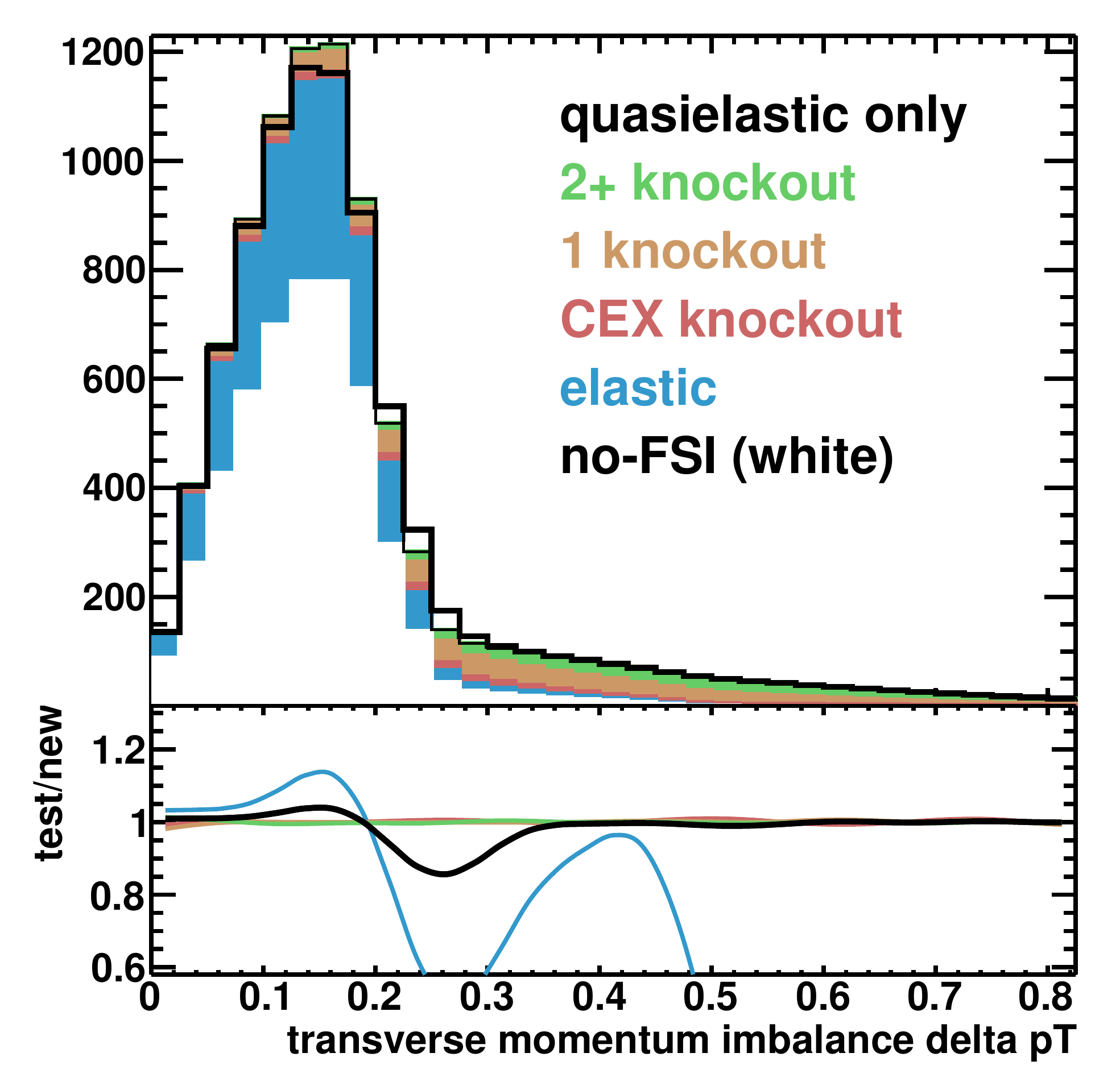}
\includegraphics[width=8cm]{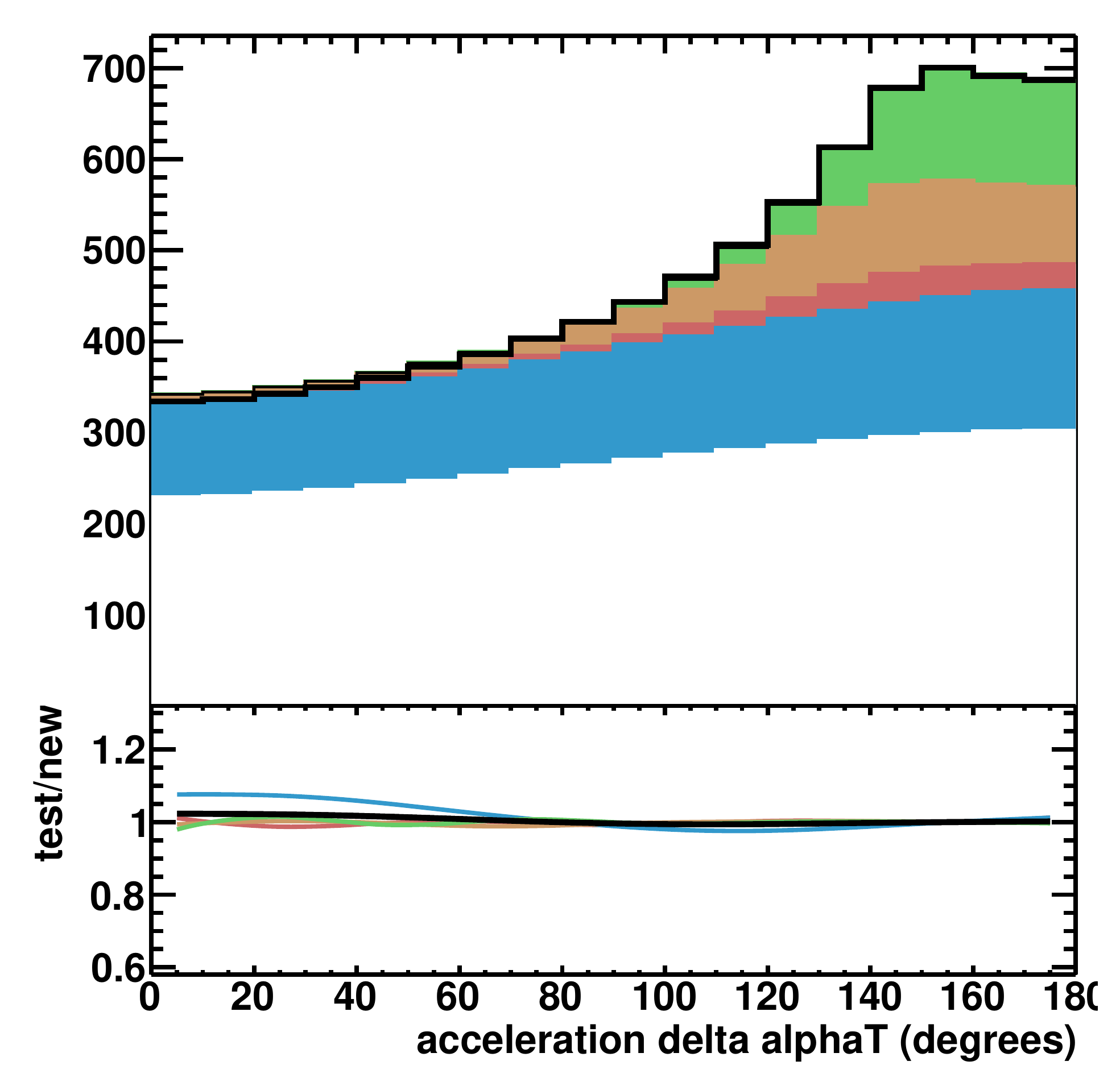}
\includegraphics[width=8cm]{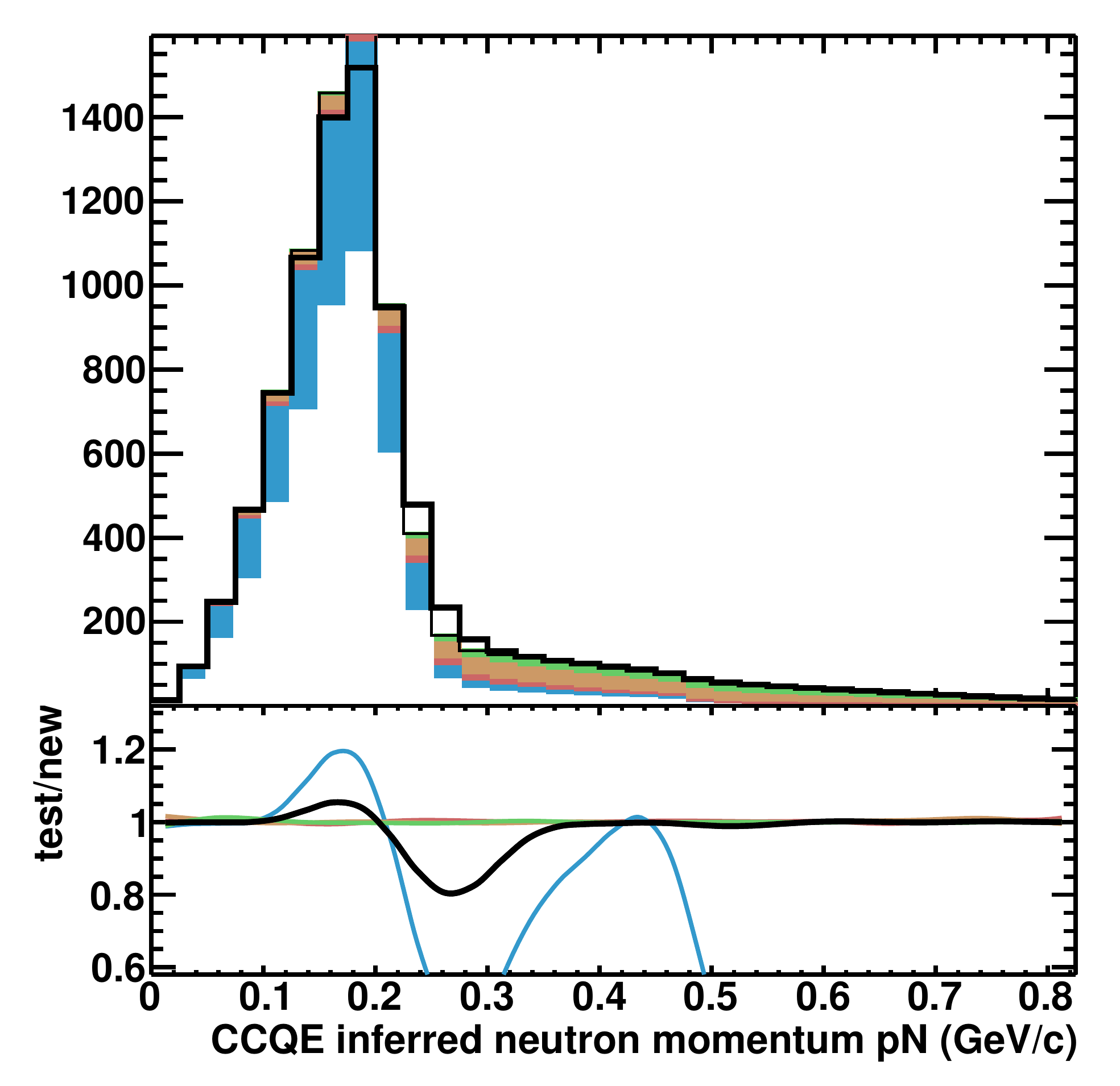}
\caption{{\small GENIE}v2.12.10+FSIfix hA with elastic scatters' $\theta_{CM}$ set to zero compared to the version with all fixes (thick black line).  
The inelastic component is the same fixed version for both.
This shows the effect of the scattering angle clearly broadens the peak of the elastic distributions, making an enhancement at the edge of the peak in the denominator histogram.  The predicted distortion from scattering is at most 20\% of the total CCQE sample at high-side base of the peak.  This may be the size of the effect missing from generators that do not implement elastic scattering at all.
\label{fig:nofsiovernew}}
\end{center}
\end{figure*}

\pagebreak
\subsection{Elastic hA2015 vs. new for quasielastic reactions}

\begin{figure*}[tbh!]
\begin{center}
\includegraphics[width=8cm]{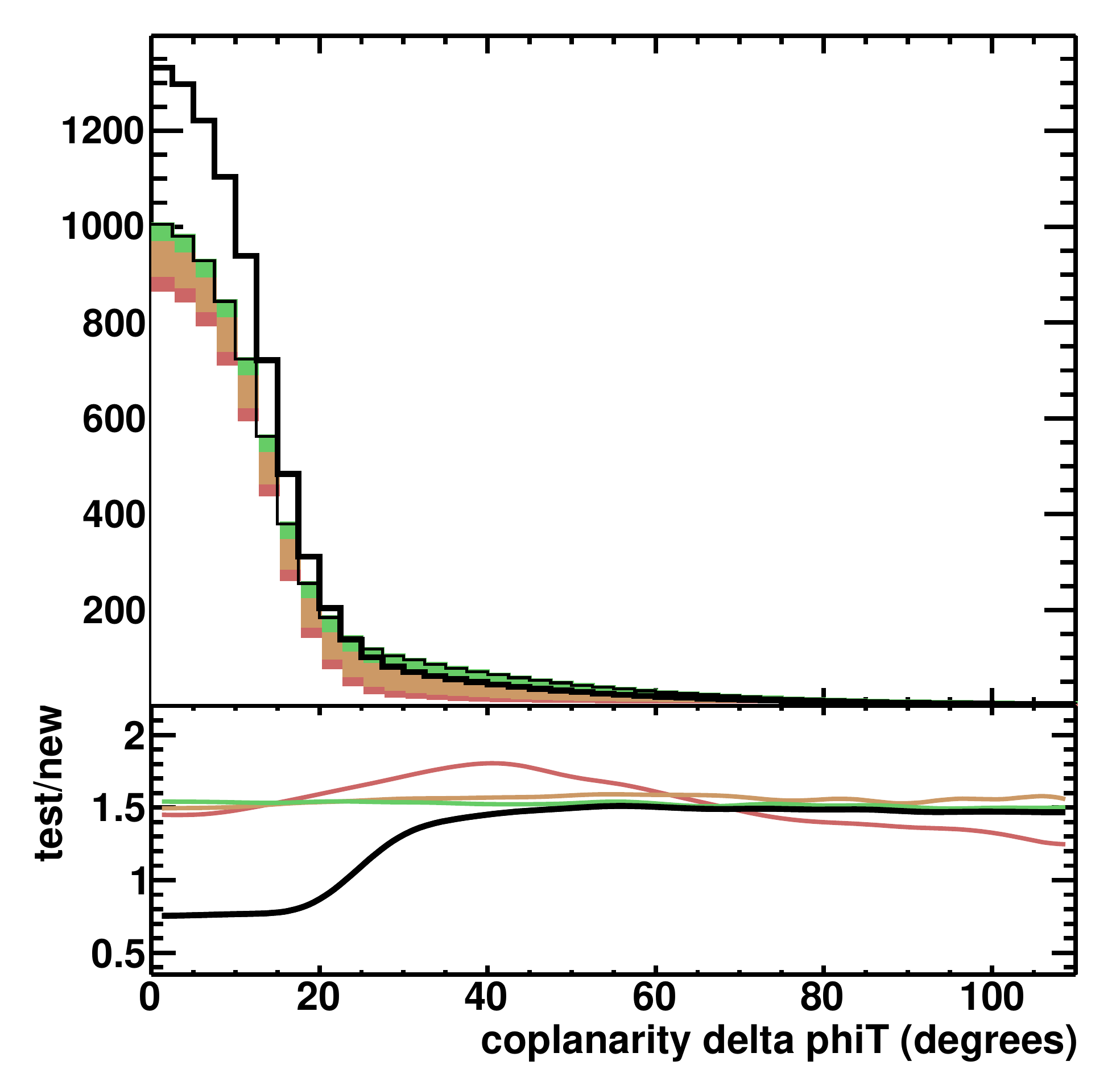}
\includegraphics[width=8cm]{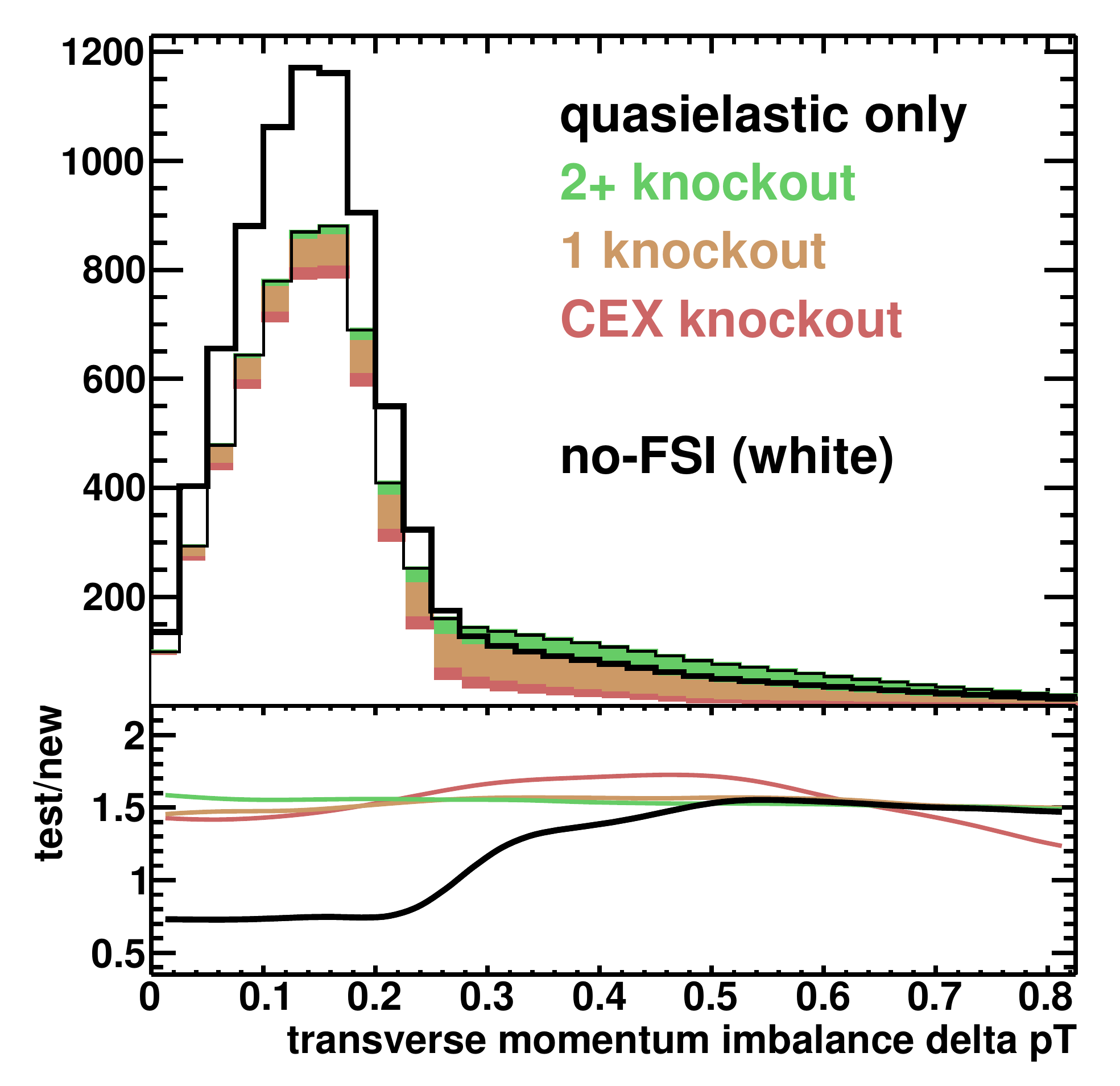}
\includegraphics[width=8cm]{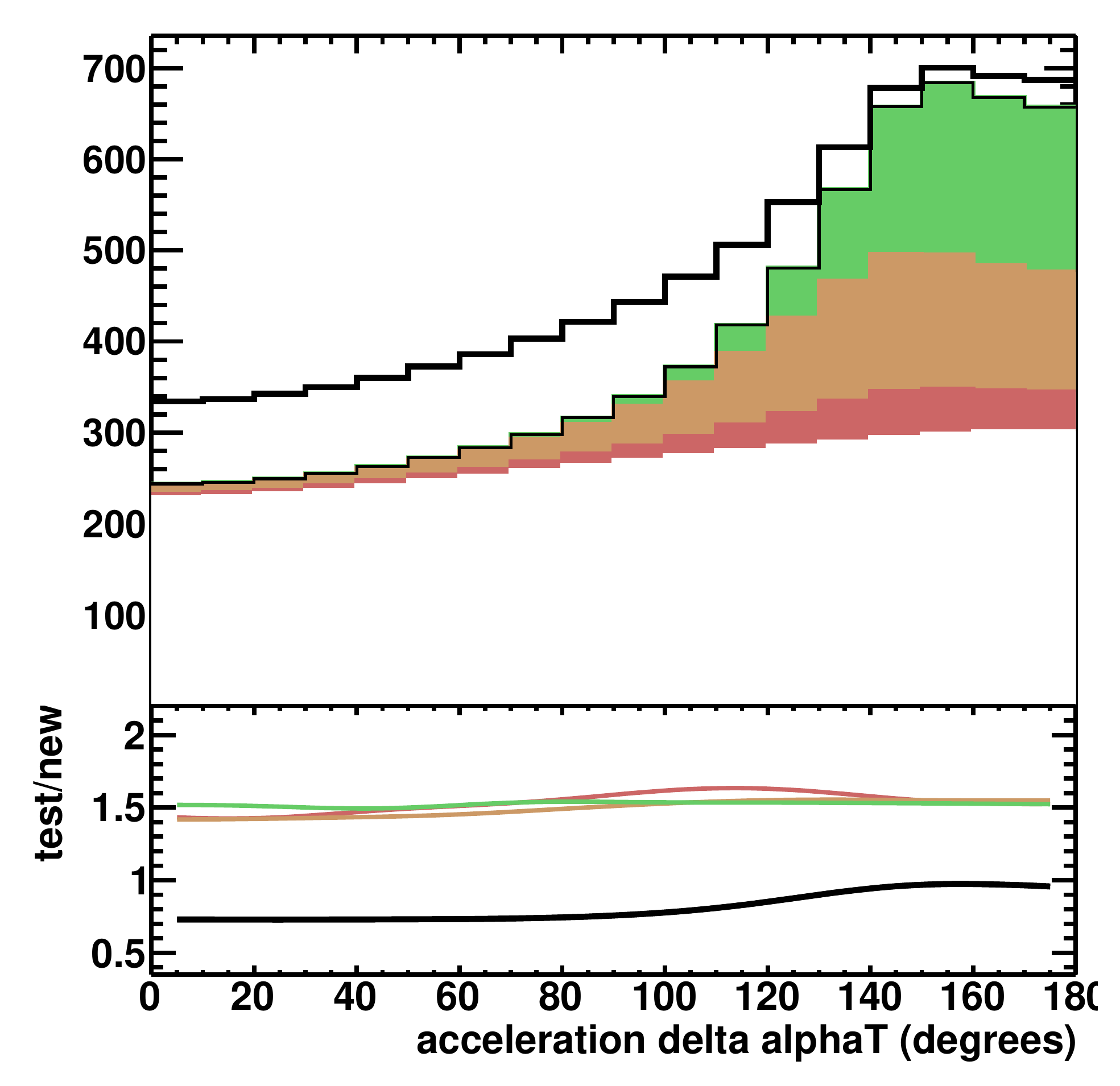}
\includegraphics[width=8cm]{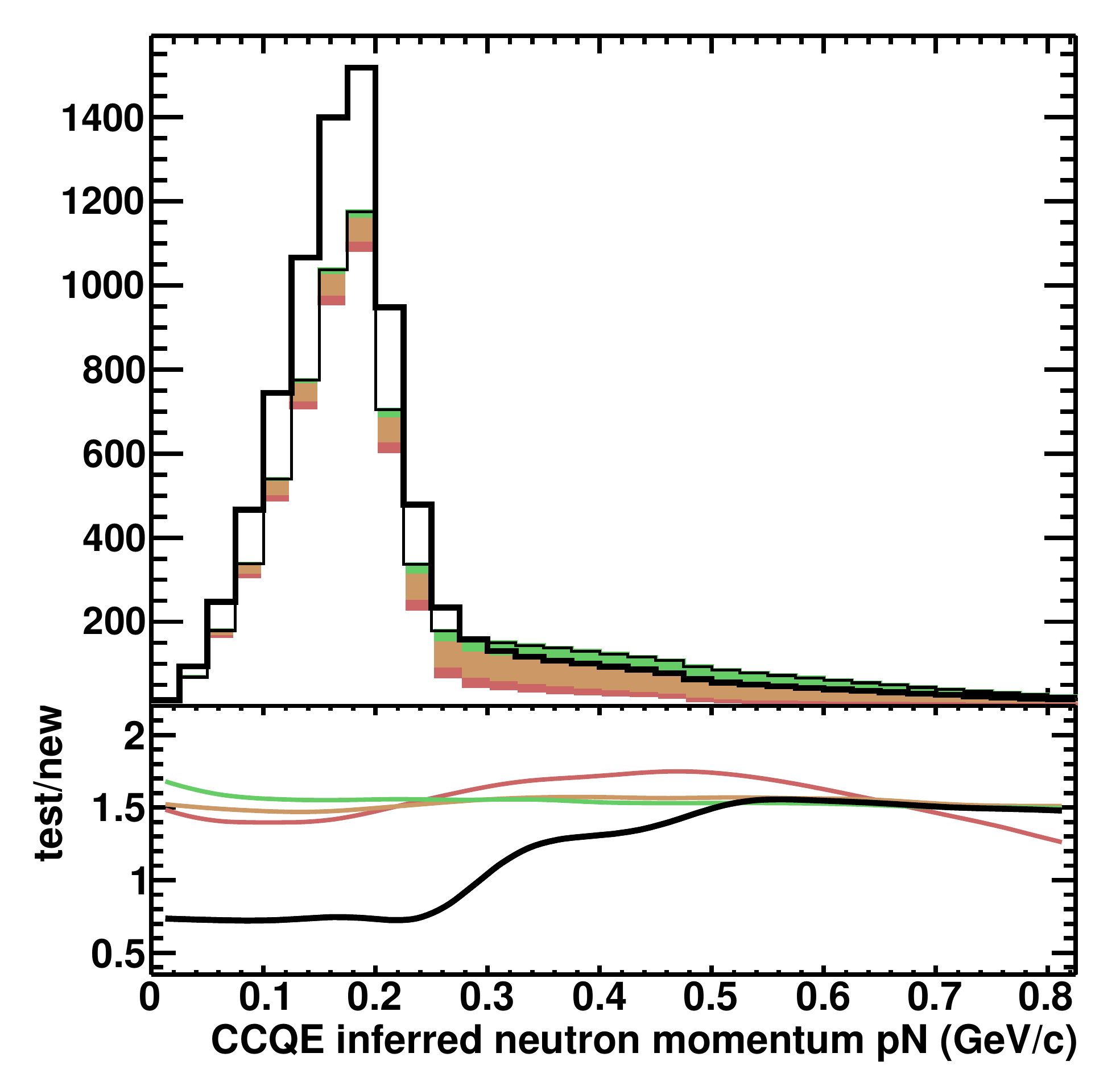}
\caption{{\small GENIE}v2.12.10+FSIfix hA2015 (elastic becomes other fates) compared to the version with all fixes (thick black line).
Because this implementation retains the hadron+nucleon mean free path from SAID, the former elastic events are replaced proportionally with the other inelastic fates.  The factor of 1.5 evident in the ratio and a corresponding reduction of strength where the elastic hadron+nucleus component once was.    This is a very different prediction than either the fixed elastic version in the denominator or the set $\theta_{CM}=0$ version (mode 4) in the previous Fig.~\ref{fig:nofsiovernew}.  
\label{fig:2015overnew}}
\end{center}
\end{figure*}

\pagebreak

\subsection{Discussion on eliminating elastic hadron nucleus scattering}

Supporting data analysis, all current experiments have fully generated Monte Carlo samples which are expensive to produce and maintain.  Directly turning elastic nucleus scattering into no FSI is close to the fixed model, so these samples can be approximately fixed using a reweighting scheme.   For the samples above, weighting the FSI=3 process to zero and weighting up the no-FSI process by 1.5 is a good approximation.  It should be adequate to only apply the weighting to the CCQE process, where there is always exactly one nucleon experiencing FSI.   For samples with lower energy nucleons, a weight that depends on nucleon energy should be constructed.   The result will be like Fig.~\ref{fig:nofsiovernew2} and similar to Fig.~\ref{fig:nofsiovernew}.

It is not clear what to make of the large difference in the previous Fig.~\ref{fig:2015overnew} (hA2015) compared to the modest difference in Fig.~\ref{fig:nofsiovernew} (ElasticConfig=4, $\theta_{CM}=0$).  Which one is closer to reproducing the physics implied by external hadron nucleus scattering data?  Or neutrino and (e,e'p) data?   This requires additional studies beyond the scope of this paper.  However, it is easy to further articulate the two paths.

The original combination of hA inputs (hadron nucleon cross sections from SAID, fate table from Mashnik), combined with a stepper scheme that starts from outside the nucleus, was able to reproduce external hadron data.   In this case, the ElasticConfig=4 reweighting the old elastic component to no-FSI will both be good approximations and the fixed version ElasticConfig=3 will be better, up to some uncertainty.   The hA2015 provides a very different prediction, enhancing knockout reactions by 50\%, and would seem incorrect.

Another possibility is that the use of nucleon level cross sections from SAID imply that the hadron+nucleus scattering process in Mashnik should not be included in these reactions.   In this case hA2015 is closer to what was originally intended, and the elastic component (if desired) should be carved out of the no-FSI outcomes instead or added randomly to all outcomes.   In this case, the weighting scheme for already generated Monte Carlo samples should weight up the other inelastic components and not the no-FSI component.

Agreement to MINERvA data could be sensitive to the above options.  Achieving a constraint requires understanding the interplay with the also uncertain predictions for the strength of the $\Delta$ resonance and 2p2h processes and also the spectral function tail of the nucleon's initial motion.  Some argue that elastic scattering does not make sense in the first place.  The classic beam experiment data was described by quantum-mechanical diffraction from a perfectly absorbing ``black disk'', which does not immediately seem to apply to the {\small GENIE} situation.  We come back to this point in Sec.~\ref{whyelastic} because Fig.~\ref{fig:nofsiovernew} suggests these highly peaked distributions are sensitive to additional small-angle scattering or lack thereof.

\begin{table}[h]
\begin{tabular}{c|ccccccc}
selection stage & total & no-FSI & CEX & elastic & single & multi & pi \\ \hline 
Before any cuts & 100 & 36.0 & 6.8 & 22.8 & 14.3 & 17.1 & 3.1 \\
pre-FSI energy & 58.8 & 24.5 & 3.7 & 12.5 & 7.6 & 10.2 & 0.5 \\ \hline
After all cuts new sim & 36.2 & 18.1 & 1.3 & 9.5 & 3.5 & 3.9 & 0 \\
After all cuts new sim $\theta_{CM}=0$* & 36.2 & 27.6 & 1.3 & 0 & 3.5 & 3.9 & 0 \\ \hline
After all cuts hA2015 & 32.1 & 18.1 & 2.2 & 0 & 5.7 & 6.1 & 0 \\
After all cuts hA2015* + elastic & 32.1 & 9.1 & 2.2 & 9.0 & 5.7 & 6.1 & 0
\end{tabular}
\caption{Same as Table~\ref{tab:neutroncounts} but with two versions of hA2015 without and with the elastic process being taken out of the no-FSI category.  The scenarios with * simply moves 9.5 or 9.0\% to or from the no-FSI column for illustration purposes, and are not from an actual configuration of the simulation.   
\label{tab:fraction}}
\end{table}

The {\small GENIE} 3.0 series has two recommended versions, a treatment hA2018 which is like hA2015, and a multi-step intranuclear cascade model hN2018.  Neither include hadron nucleus elastic scattering, and both are affected by the bug in the milder way only through the inelastic processes.  There may have been additional retuning for the {\small GENIE} 3.0 versions.  Revisiting the benchmark comparisons and direct comparisons to other intranuclear cascade codes may be instructive.  In all cases, the uncertainty on absolute and relative strength of FSI components should be taken seriously.  

\subsection{Energy change and calorimetry}

The new elastic+nucleus scatters result in less than 1 MeV energy change compared to the prescattered hadron, shown earlier in the left plot of Fig.~\ref{fig:cuts}.   This fluctuation is practically symmetric around zero and is conceptually similar to how ideal gas scattering produces both acceleration and deceleration such that the Maxwell velocity distribution is obtained.  Before the fix, the code produced a distribution that had between 1 and 2 MeV of acceleration (the scattered hadron was more energetic).   On the scale of contemporary measurements, this anomalous 2 MeV is smaller than other uncertainties such as the nucleon removal energy or experimental track-based or calorimetric hadronic energy scale uncertainties.   Only observables which use the hadron angle that are affected, not calorimetric quantities such as in \cite{Rodrigues:2015hik}.  

 In contrast, the inelastic components of FSI lead to significant energy loss by the initial hadron, leading to the interpretation of $90 < \delta\alpha_{T} < 180$ as deceleration. 

\subsection{Background reactions}

In a real experiment's data sample, the selected transverse kinematics distributions are accompanied by a background of non-quasielastic processes, most significantly 2p2h and resonance interactions.  They have much broader distributions reflecting the three-body nature of a process where the third particle was unseen, below reconstruction threshold, or otherwise not included in the analysis.   Because they are broad, additional smearing has only slight effect.  On the other hand, they are more significant in the tail of the inferred neutron momentum $p_N$ distribution as generated CCQE events are.

The fates represented in the figure may be distorted by the fact that two hadrons are experiencing FSI in these samples.   In the selection the most energetic proton in the momentum and angle range is used in this distribution.  A proton above the cut and another below the cut leads to the second being used to make these quantities.   Two protons in the momentum range lead to the most energetic being used.   In this way, the kinematic entry in the histogram may be affected by the fate of an unseen nucleon, so the distortion by fates may not have a simple dependence on the old or new routine that generated those fates.  The color scheme now refers only to the FSI fate that led to the selected proton.  The total combinatorics are too challenging to represent in a single figure.

\begin{figure*}[tbh!]
\begin{center}
\includegraphics[width=8cm]{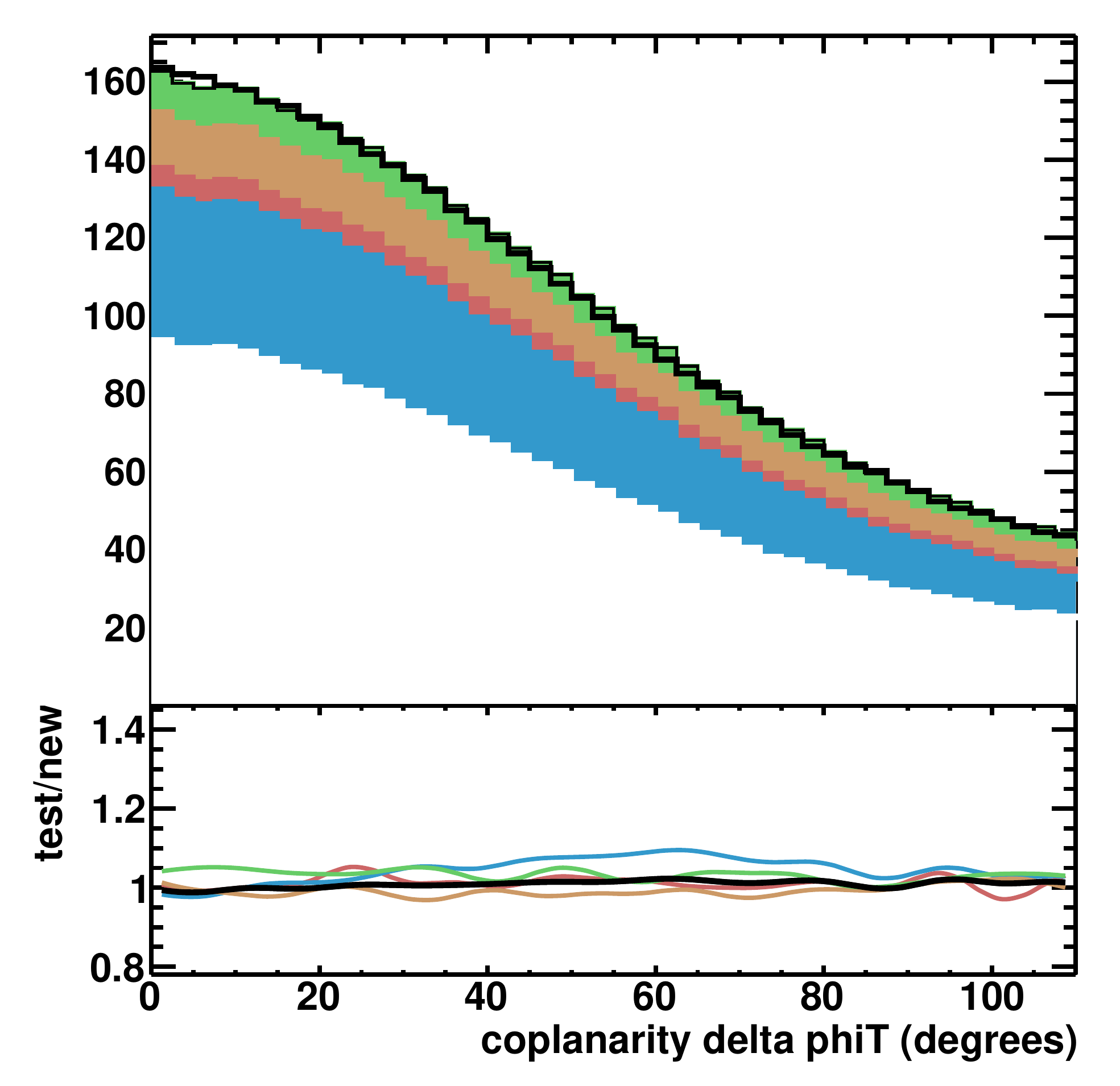}
\includegraphics[width=8cm]{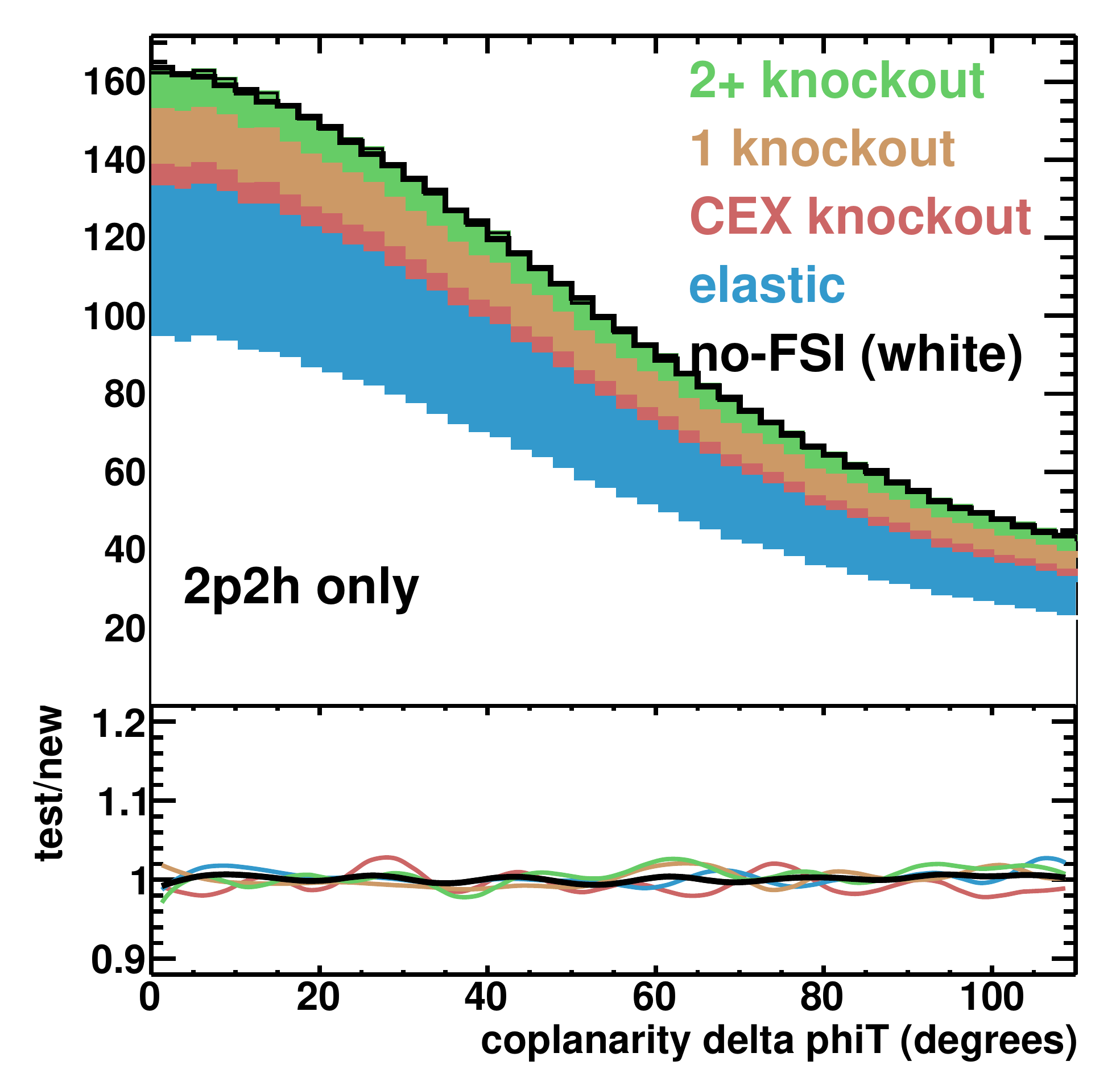}
\includegraphics[width=8cm]{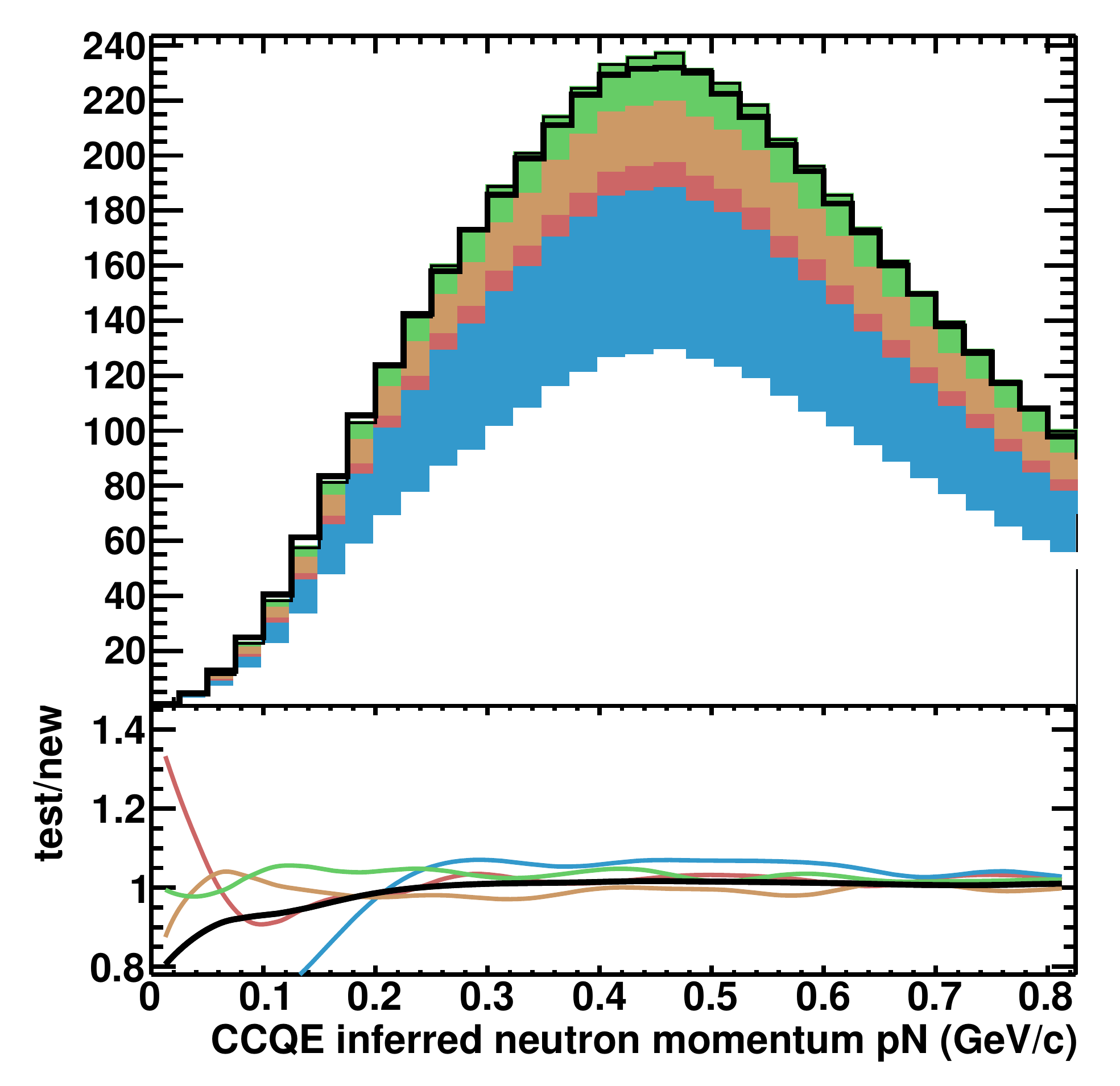}
\includegraphics[width=8cm]{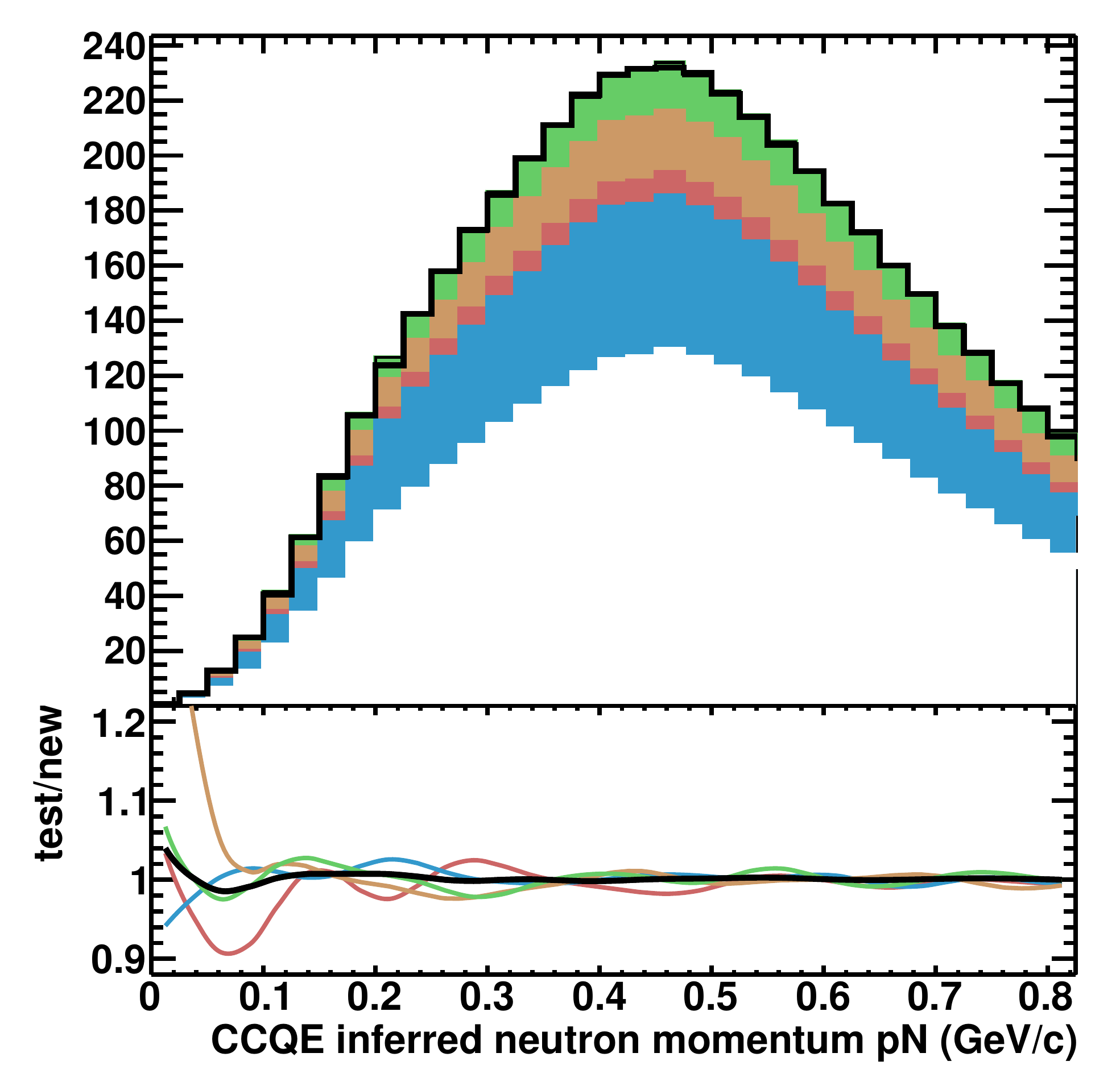}
\caption{{\small GENIE}v2.12.10+FSIfix coplanarity and inferred neutron momentum for 2p2h reactions with no pion and a proton that passes the transverse kinematic cuts.
The left column compares old code with the fixed code, the right comparison has  ElasticConfig=4: $\theta_{CM}=0$ for elastic but fixes for the inelastic processes, representing a shift of elastic scattering to no-FSI.  There is slight distortion toward a narrower peak in the lower left plot and a slight overall increase in acceptance with the old code.  No significant distortion is seen with zero scattering angle.
\label{fig:backgroundcoplanarity2p2h}}
\end{center}
\end{figure*}

The 2p2h events that pass this selection show relatively mild distortion in every case.   There is anomalous narrowing of the peak of the elastic component in the lower left plot of Fig.~\ref{fig:backgroundcoplanarity2p2h}, visible only in the ratio, but not as strong as in the CCQE case.  All other distortions, from the mistaken code, or a fix that would set  $\theta_{CM}=0$, are negligible for current experiments.   The absolute number on the vertical axis indicates this sample larger in the tail of $p_N$ than the quasielastic component is.

\begin{figure*}[tbh!]
\begin{center}
\includegraphics[width=8cm]{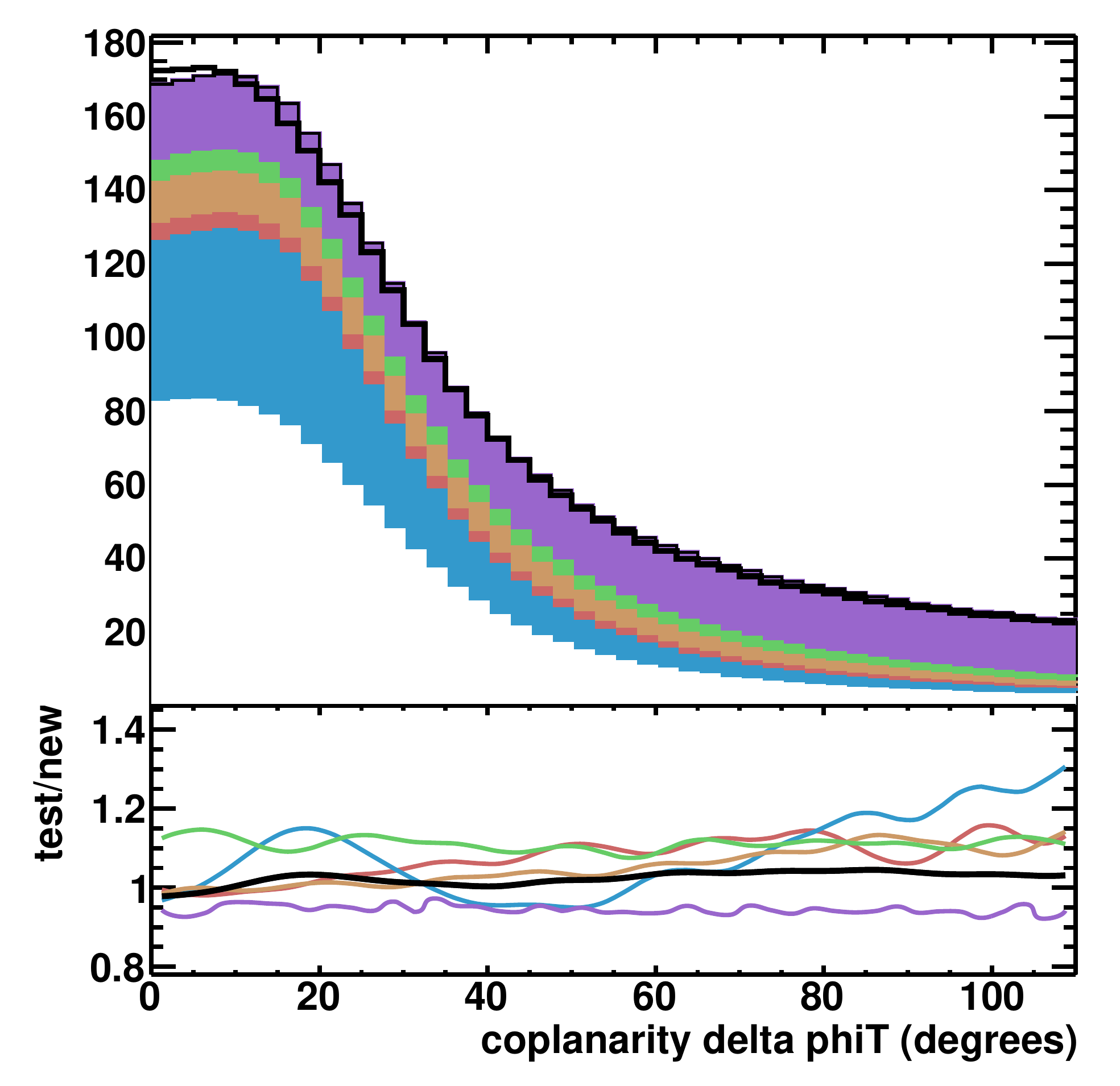}
\includegraphics[width=8cm]{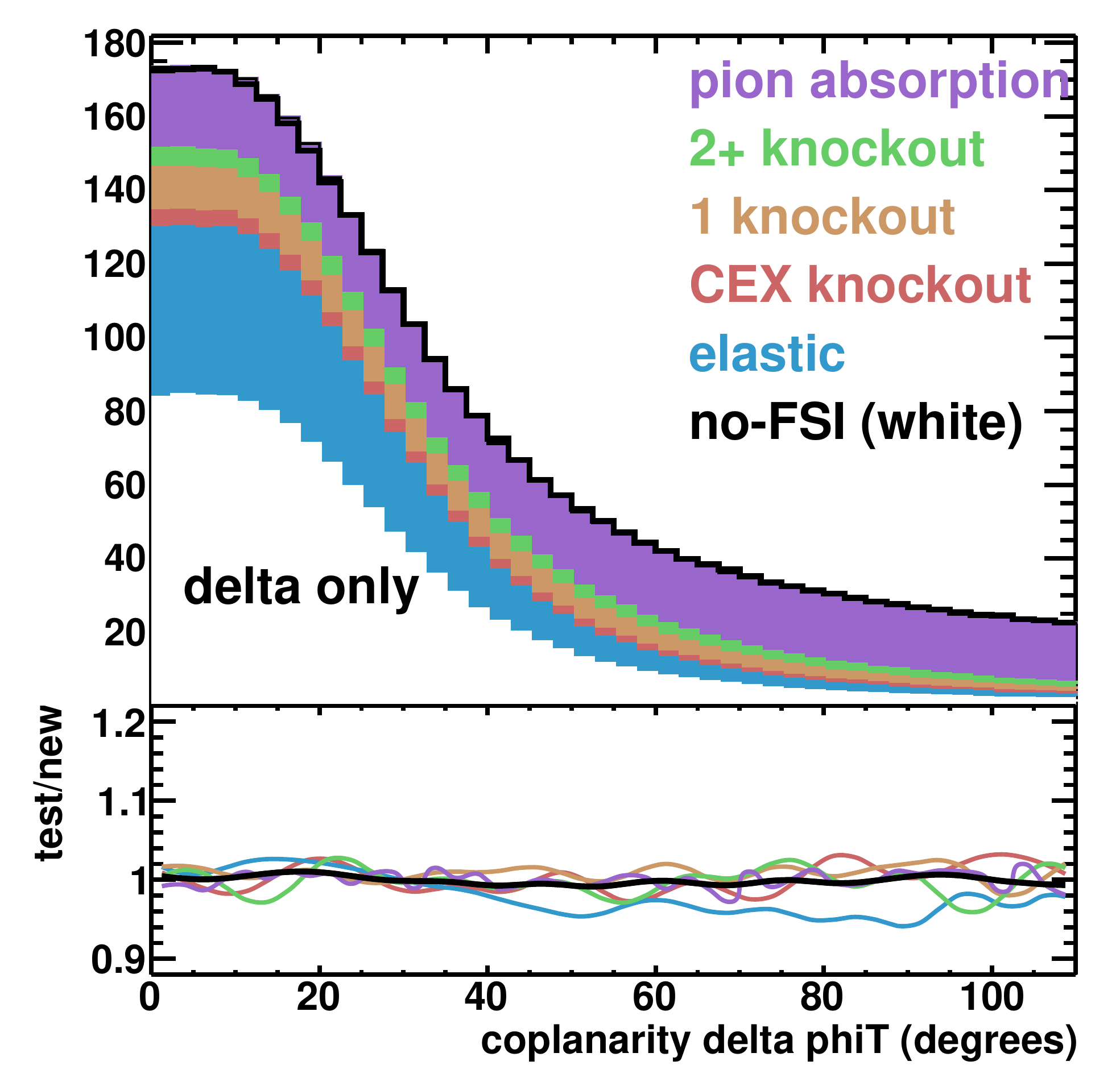}
\includegraphics[width=8cm]{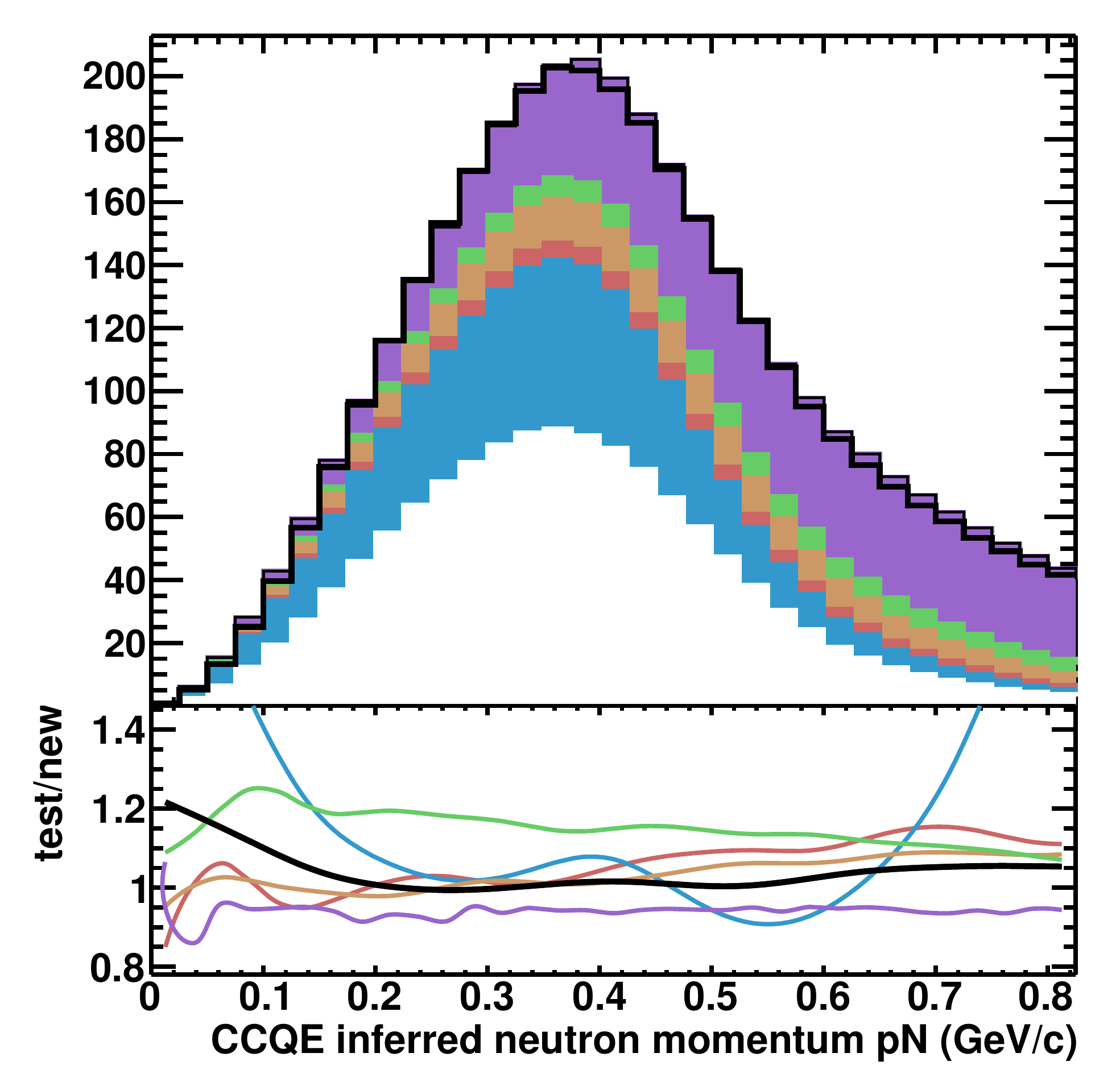}
\includegraphics[width=8cm]{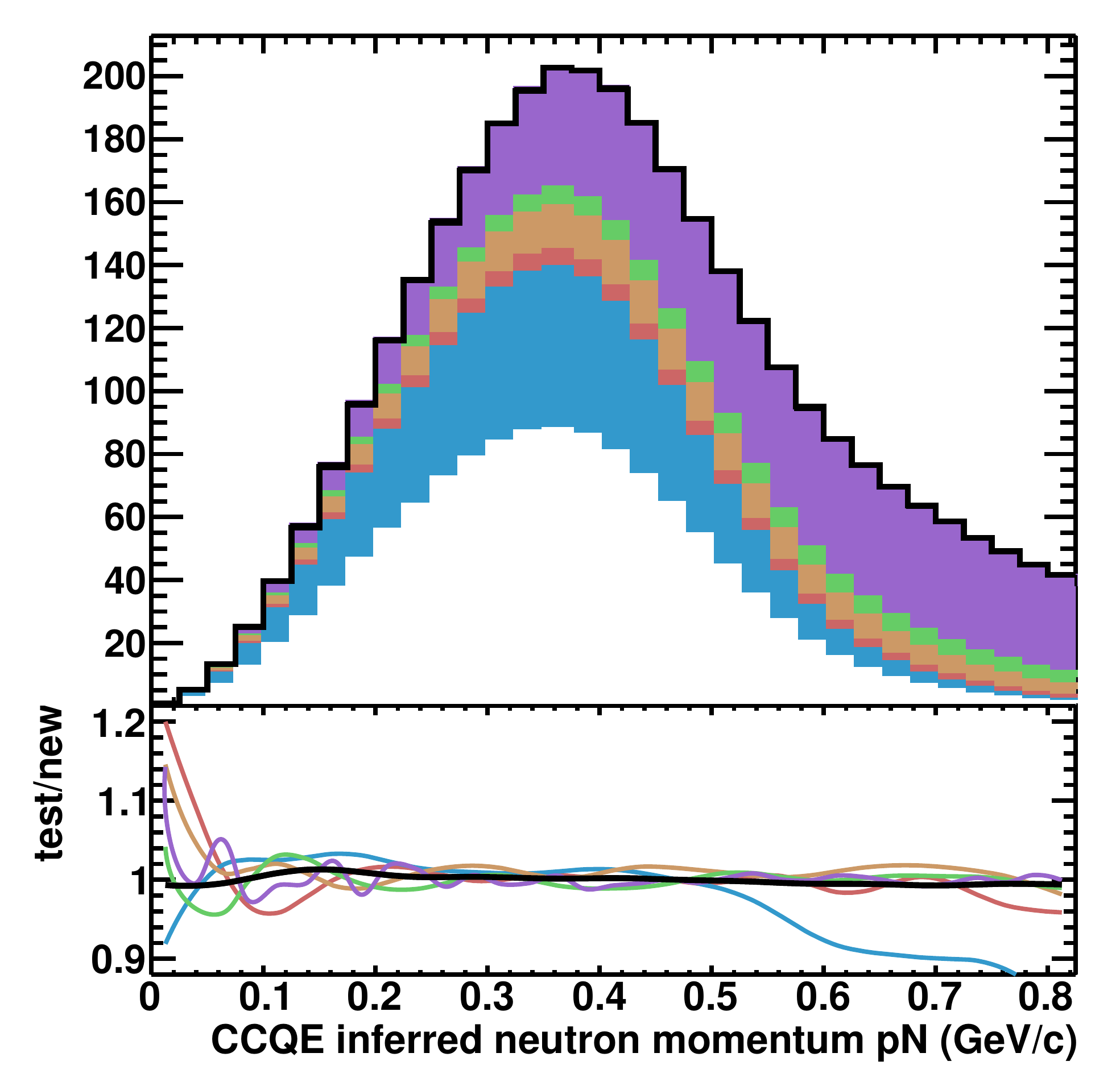}
\caption{{\small GENIE}v2.12.10+FSIfix coplanarity and inferred neutron momentum for $\Delta$ reactions with no pion and a proton that passes the transverse kinematic cuts.  In addition to the previous color scheme for the measured proton, a purple shade represents protons produced by the pion absorption process, which dominates the contribution in the high side tails.
The left column compares old code with the fixed code, the right comparison has elastic $\theta_{CM}=0$ (ElasticConfig=4).  
\label{fig:backgroundcoplanarityDelta}}
\end{center}
\end{figure*}

The top-left plot in Fig.~\ref{fig:backgroundcoplanarityDelta} figure compares the old and new simulations for $\Delta$ resonances where the pion was absorbed and a proton passed the selection, producing a CC0$\pi$ event.   The wiggle ratio indicates the blue elastic component in the old simulation is both narrower in the peak and enhanced in the tail compared to the fixed simulation, which is more complicated than for true CCQE.  A similar trend is shown in the lower-left plot of the inferred neutron momentum.  When the simulation sets $\theta_{CM}=0$ (right plots), the distribution is simply, but slightly narrower than the fully fixed version, as expected.    A new purple color stacked on top represents protons that were produced following pion absorption, which includes the before and after fix to the pion absorption on two nucleons process.

\section{Delta reactions}

The elastic hadron+nucleus and single-nucleon knockout process (with and without charge exchange) apply to simulated pions also.   For simple distributions of pion angle, the net effect is quite mild because those distributions are so broad and so many processes contribute in addition to the elastic hadron+nucleus process.  For a distribution that specifically constructs a transverse imbalance, the distortions are once again severe.  

\begin{figure*}[tbh!]
\begin{center}
\includegraphics[width=8cm]{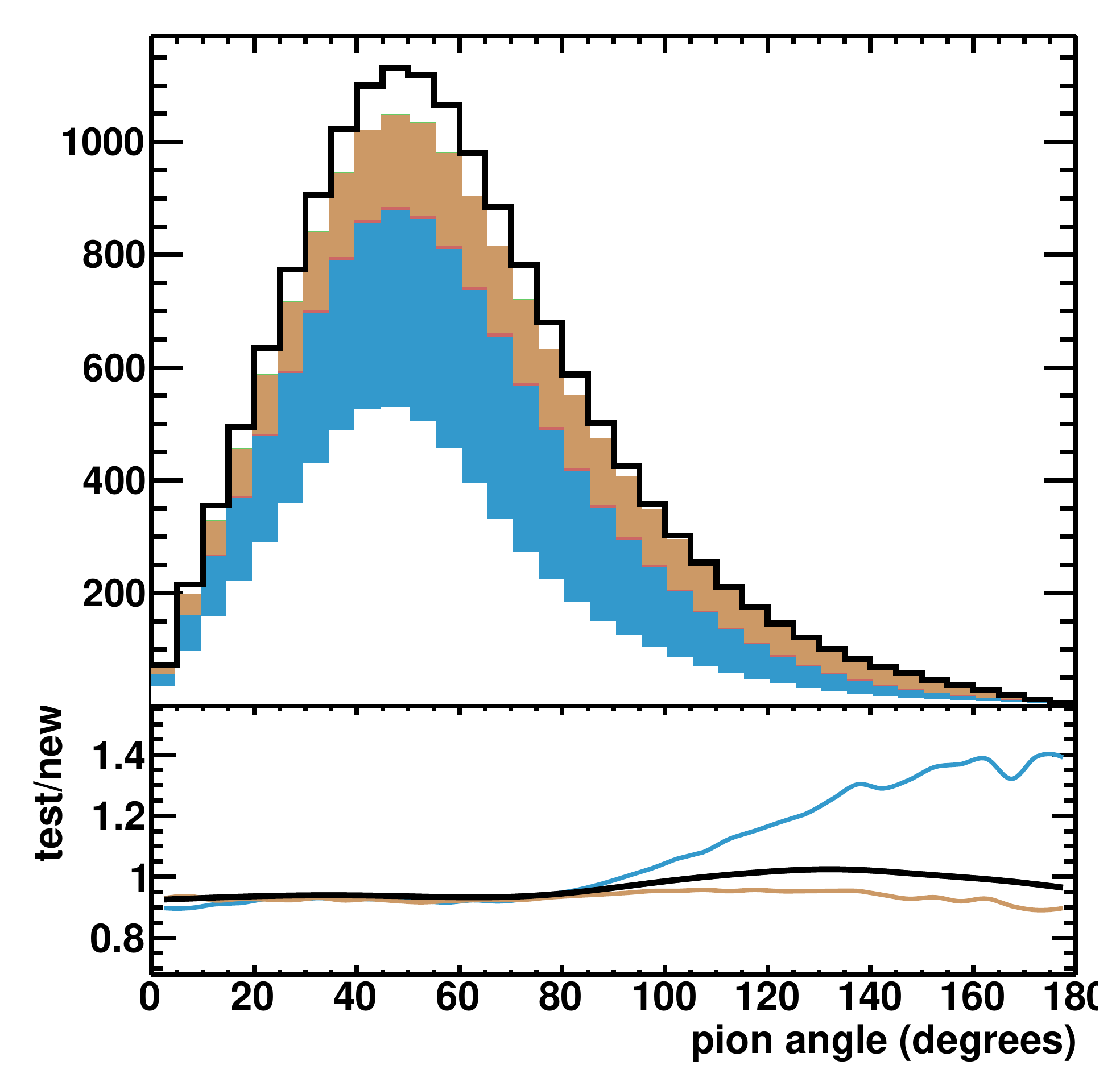}
\includegraphics[width=8cm]{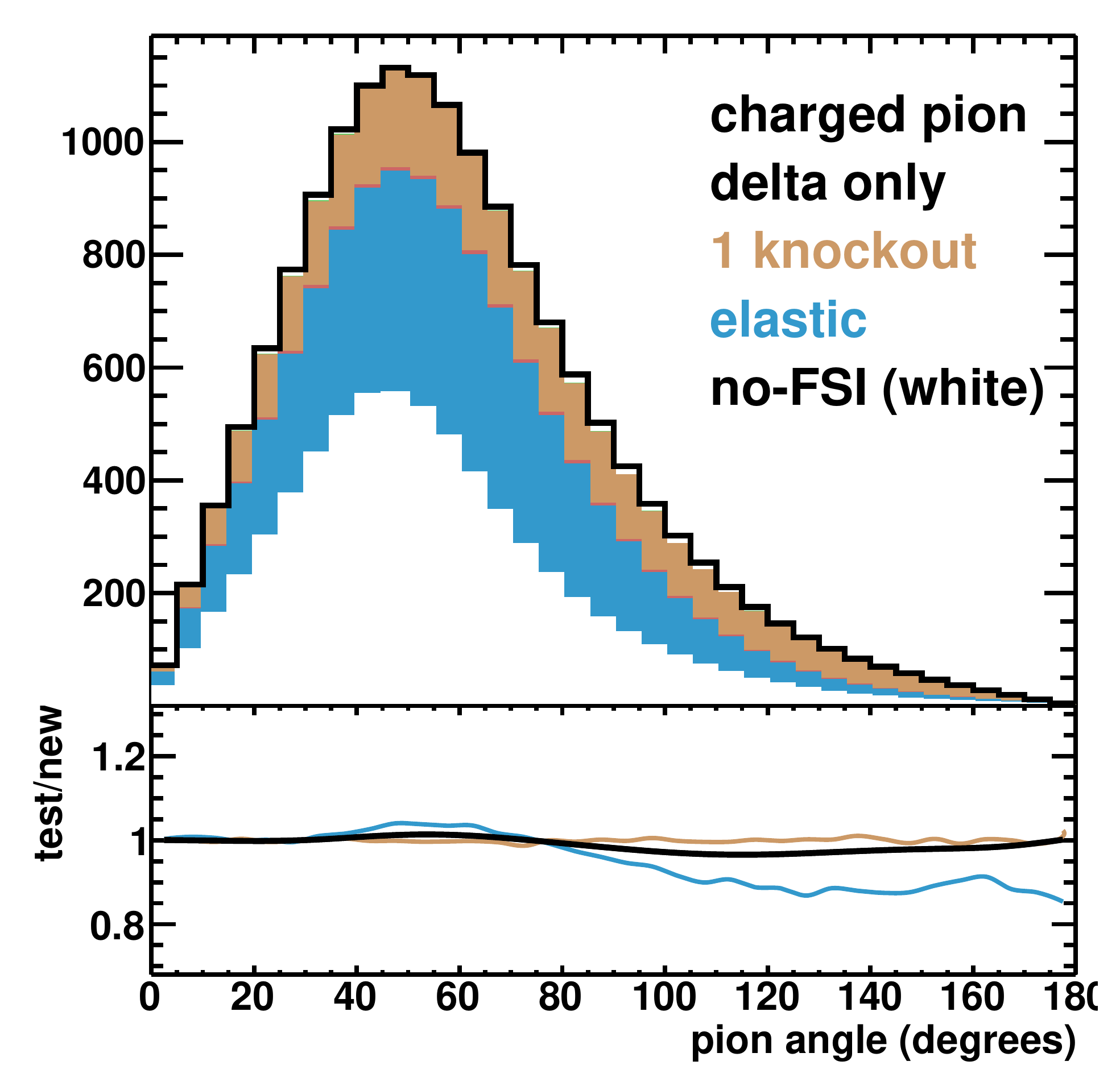}

\caption{{\small GENIE}v2.12.10+FSIfix pion opening angle with respect to the neutrino.  The old elastic scatter code (left) put anomalously too many pions in the backward direction, though in absolute terms they are not so numerous.   Overall, a few percent more events pass the new selection.   Making elastic scatters have $\theta_{CM}=0$ (right) again narrows the elastic peak and produces (different scale for the ratio) a milder 15\% distortion in the tail.   The right plot also shows the effect of fixing the single-nucleon knockout processes (brown) remove distortions that were less than 10\%. 
\label{fig:pionneutrino}} 
\end{center}
\end{figure*}

\begin{figure*}[tbh!]
\begin{center}
\includegraphics[width=8cm]{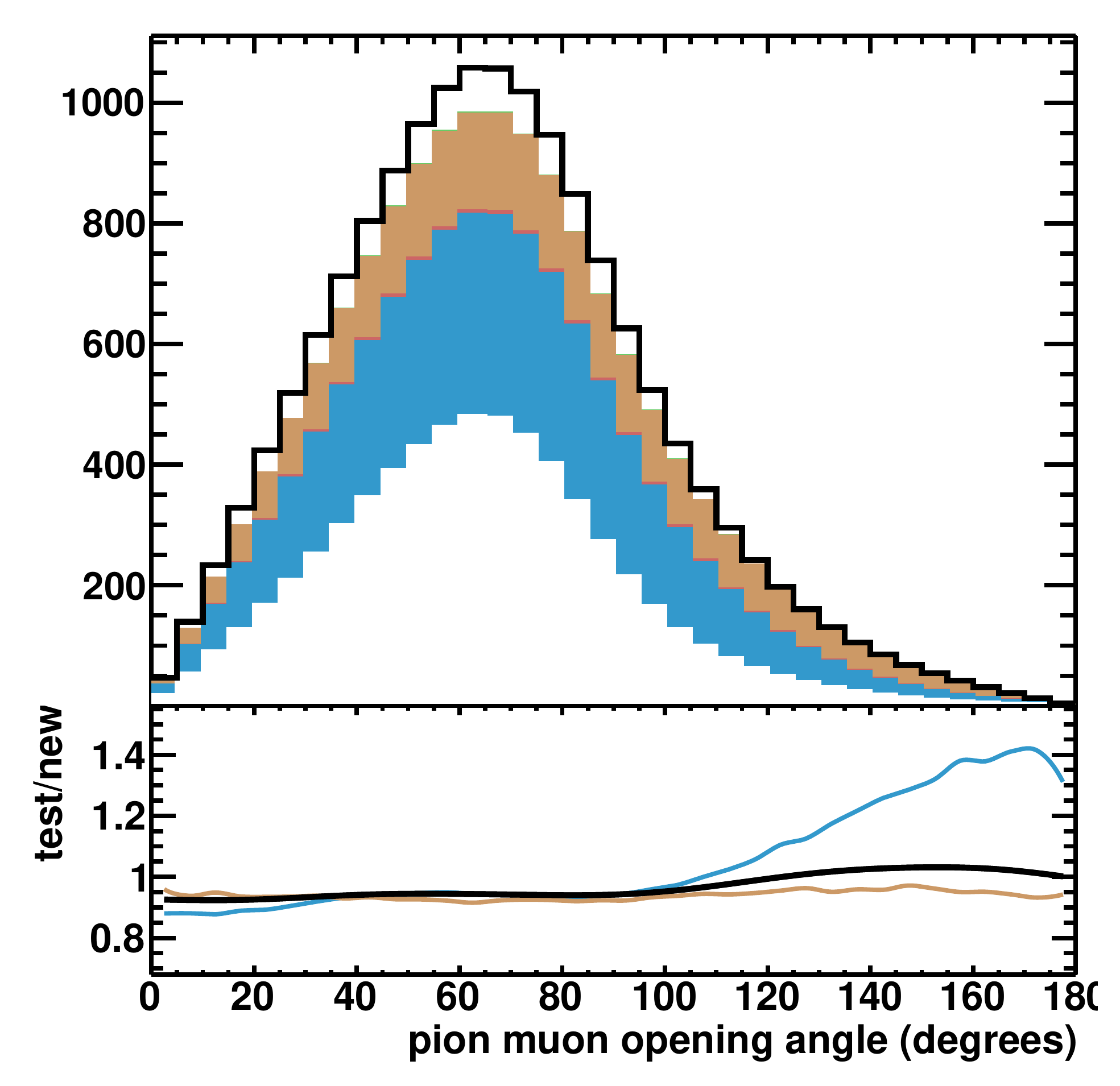}
\includegraphics[width=8cm]{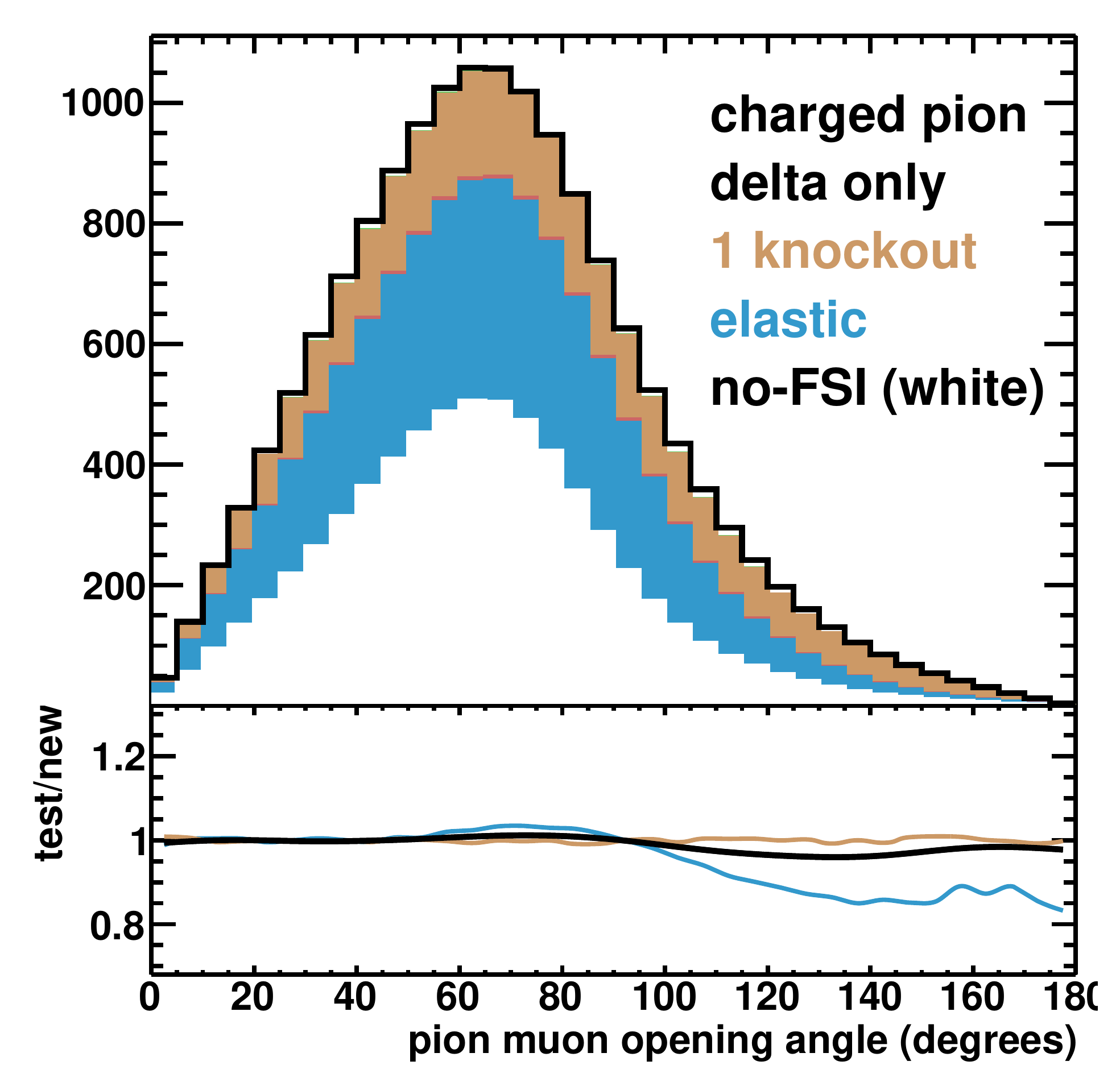}
\caption{{\small GENIE}v2.12.10+FSIfix pion angle with respect to the outgoing muon.  Original configuration 0 (left) and elastic as no-FSI config 4 (right) compared to the fully fixed configuration 3.  The conclusions are the same as the previous Fig.~\ref{fig:pionneutrino}.
\label{fig:pionmuon}}
\end{center}
\end{figure*}

\begin{figure*}[tbh!]
\begin{center}
\includegraphics[width=8cm]{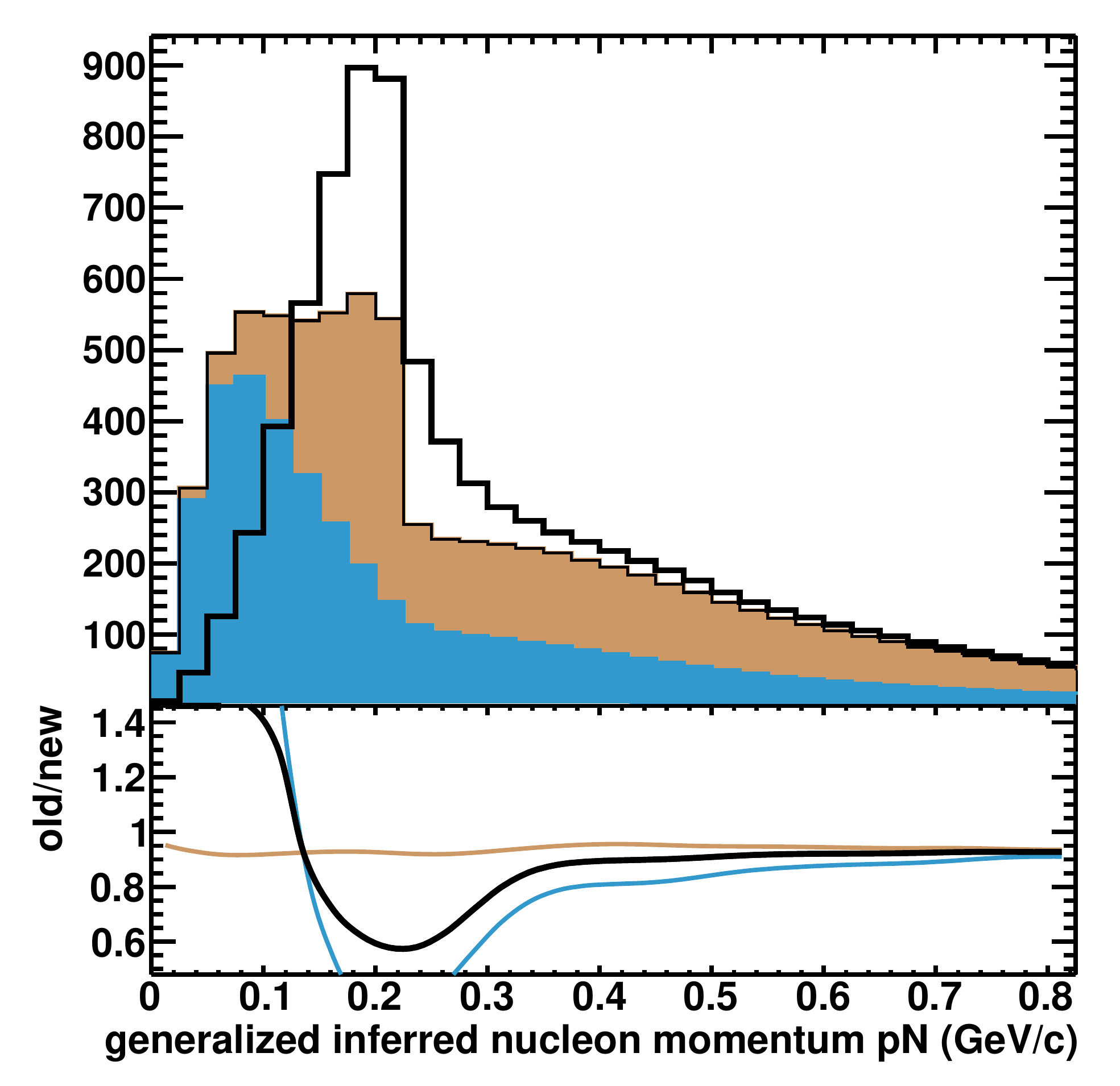}
\includegraphics[width=8cm]{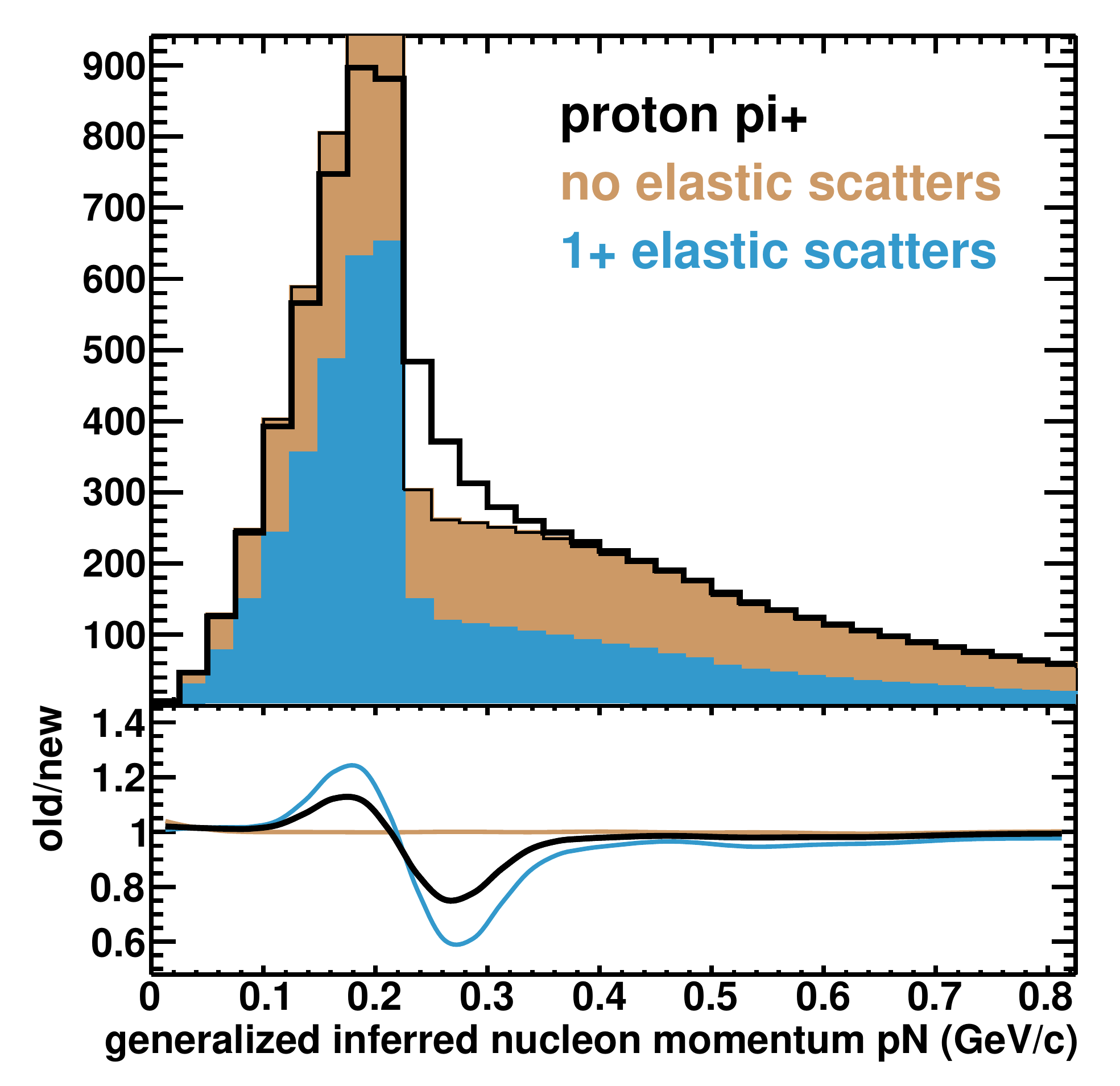}
\includegraphics[width=8cm]{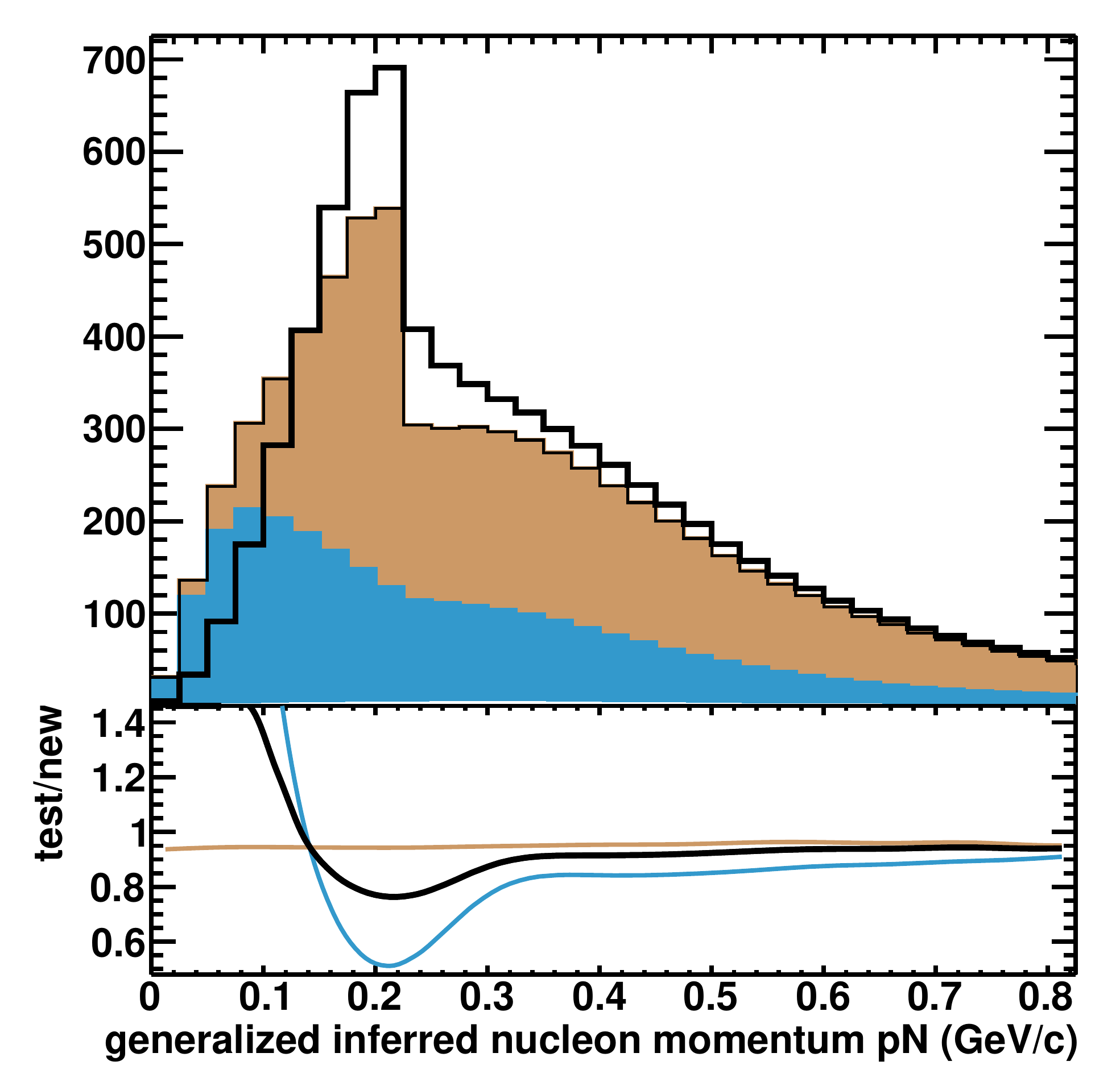}
\includegraphics[width=8cm]{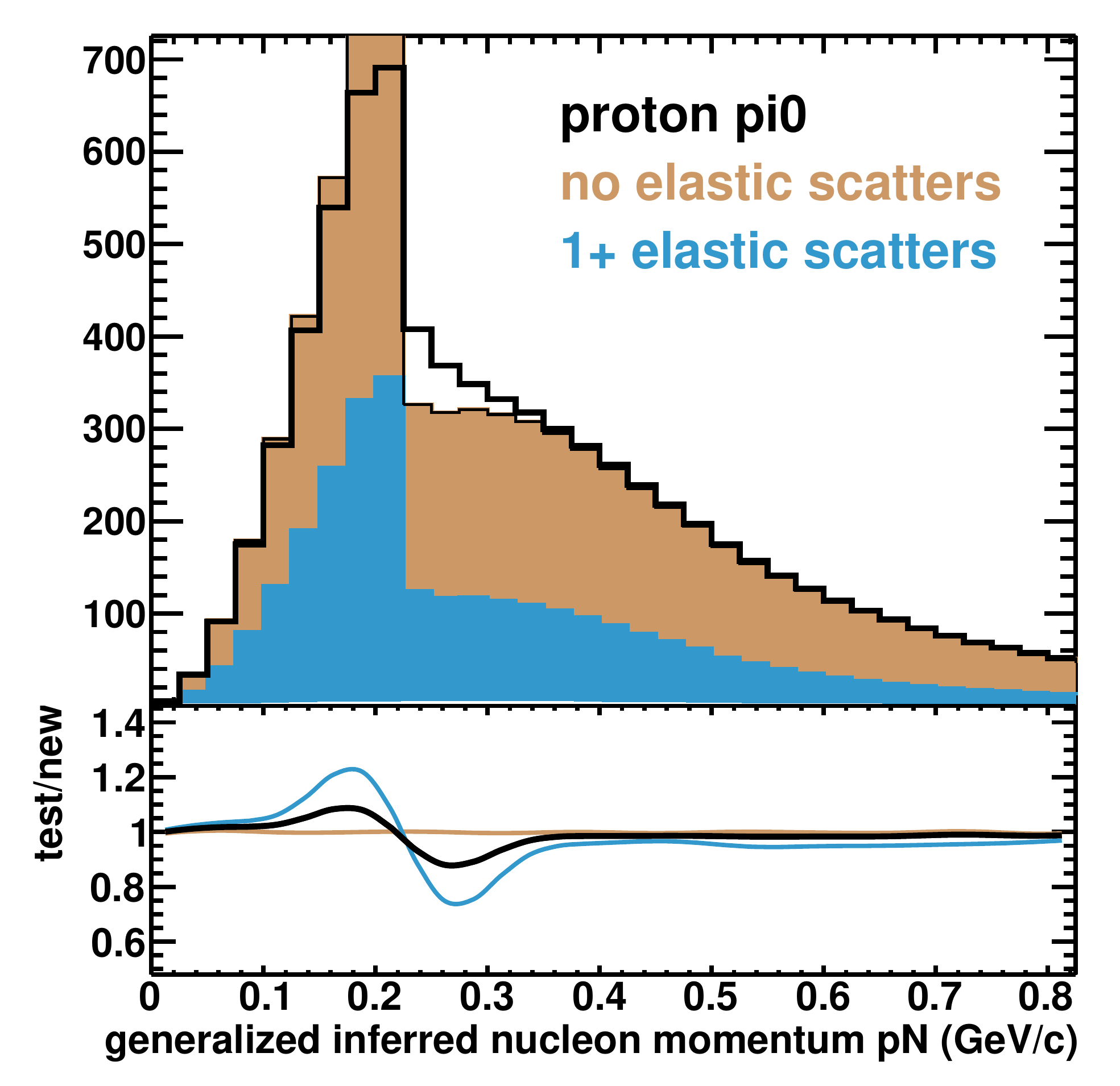}
\caption{The generalized inferred nucleon momentum from \cite{Lu:2019nmf} for events with one pion and one or more protons.  {\small GENIE}v2.12.10+FSIfix original (left) and elastic $\theta_{CM}=0$ config 4 (right) compared to fully fixed version.  All reactions are included, resonance and DIS dominate.  The component where any hadron experienced an elastic fate is shown in blue, and any other combination is shown in brown.   The severe distortion on the left follows the elastic component, while $\theta_{CM}=0$ on the right is also a significant distortion.
\label{fig:generalized}}
\end{center}
\end{figure*}

To focus on the most constrained kinematics in Fig.~\ref{fig:pionneutrino} and Fig.~\ref{fig:pionmuon}, we require a $\Delta$ resonance followed by a charged pion with at least 75 MeV kinetic energy in the final state, and any number of nucleons.   For CC neutrino reactions, this can occur directly via a $\Delta^{++}$ and $\Delta^{+}$.  The charge exchange process to or from a $\pi^0$ modifies the resulting spectra.   Two angle distributions are given:  the angle with respect to the neutrino direction and the angle with respect to the outgoing muon.  The fixed pion absorption process does not appear here. Its strength is not changed with the fix, and the kinematics of the outgoing protons are not used until Fig.~\ref{fig:generalized}.


Like the inelastic component for nucleons, these distortions are much smaller than the elastic case.   We think the nature of the bug in the code has a subdued effect as the total transverse momentum is divided among more particles.

Not surprisingly, when all three particles from a $\Delta^{++}$ are reconstructed and transverse kinematic imbalances can again be constructed and are meaningful.  The severe distortion returns, shown in the left plot of Fig.~\ref{fig:generalized}.   All processes are allowed to contribute directly or via FSI, though in practice resonance and DIS are the only significant ones.    All conclusions apply also to the $\Delta^+ \rightarrow p+\pi^0$ (lower plots) and  $\Delta^0 \rightarrow p+\pi^-$ channels (not shown).

When at least one proton and charged pion are reconstructed and other pions are below threshold or not present, the samples are enriched in $\Delta^{++}$ (top) or $\Delta^{+}$ lower  reactions with little missing energy or momentum.  The struck nucleon's momentum can again be inferred, described in \cite{Lu:2019nmf} as a generalized form of $p_N$ from \cite{Furmanski:2016wqo}.   This process is also considered a good candidate for a subsample with high resolution reconstructed $E_\nu$ desired by oscillation experiments.  Here again a MINERvA-like selection is used.  A pion with $0.075 < KE < 0.4$ GeV of kinetic energy is required, accompanied by no lower energy pion that would be tagged by the Michel positron from $\pi$ to $\mu$ to $e$ decay.  At least one proton in the range $0.450 < p_p < 1.200$ GeV is required, with any number of additional protons or neutrons allowed.

With two hadrons in the final state, either of them could experience any of the available fates.   In this figure, the blue color refers to situations when at least one hadron experiences an elastic hadron+nucleus scatter.  The brown is all other combination where the elastic fate did not occur.   The ratio emphasizes that the distortion clearly follows the elastic fate, as expected.   

Setting $\theta_{CM}=0$ (right plots), as many modern generators do, produces a significantly different prediction for this distribution, one that would probably affect interpretations of models for Fermi motion, nucleon removal energy, and short range correlation effects.  We return to this point in Sec.~\ref{whyelastic}.


\section{reactions with two-protons}

Another distribution of interest is the opening angle between two protons as observed by ArgoNeuT \cite{Acciarri:2014gev}.  This distribution is sensitive to scattering off correlated pairs of nucleons, such as a 2p2h process (simulated here with \cite{Nieves:2011pp}) or a short range correlated pair \cite{Subedi:2008zz}.  It is also to sensitive to regular CCQE and resonance events that experience FSI.   The distortions prior to fixing the {\small GENIE} TwoBodyElastic function may barely be significant for this observable, not as dramatic as the transverse kinematic imbalance distributions.    The overall broader angle distributions and lack of anchor to the lab frame for the two-body process probably explains the difference.   When the only modification is elastic $\theta_{CM}=0$ (right plot) no substantial distortion is observed.

The same 3 GeV neutrino CC reactions are used.   When there are no pions and exactly two protons the final state (minimum 50 MeV kinetic energy), the opening angle between the protons is calculated.   This threshold is lower than MINERvA can achieve, but consistent with liquid and gaseous argon time projection chambers.  This technology has both the lower thresholds and can better record highly transverse hadrons.  In addition to ArgoNeuT, this threshold is similar to what MicroBooNE and T2K achieve in their lower energy neutrino beams.   Also, highly transverse protons are not cut in this signal definition as they are for the main MINERvA transverse kinematic imbalance analysis.

\begin{figure*}[tbh!]
\begin{center}
\includegraphics[width=8cm]{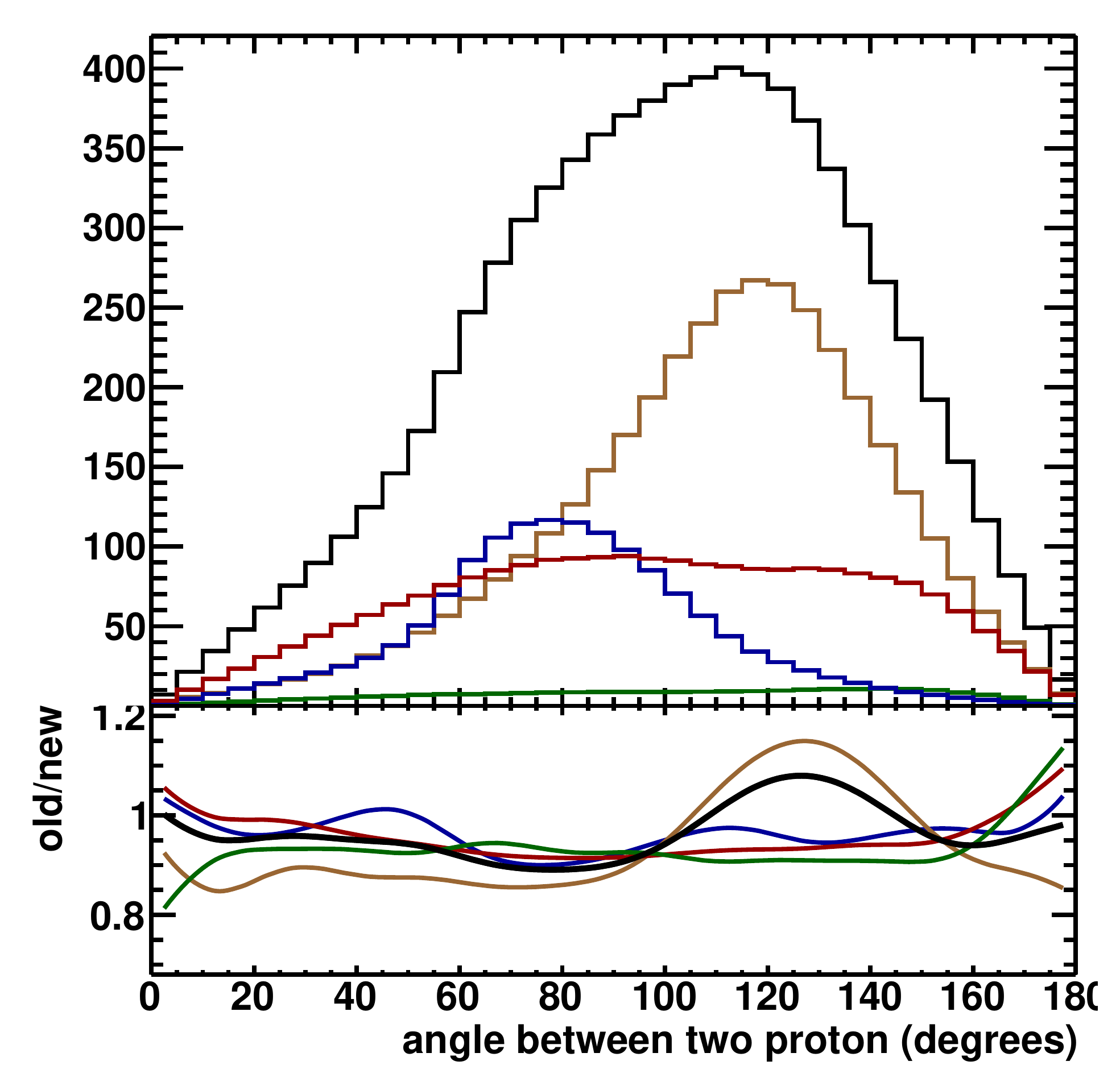}
\includegraphics[width=8cm]{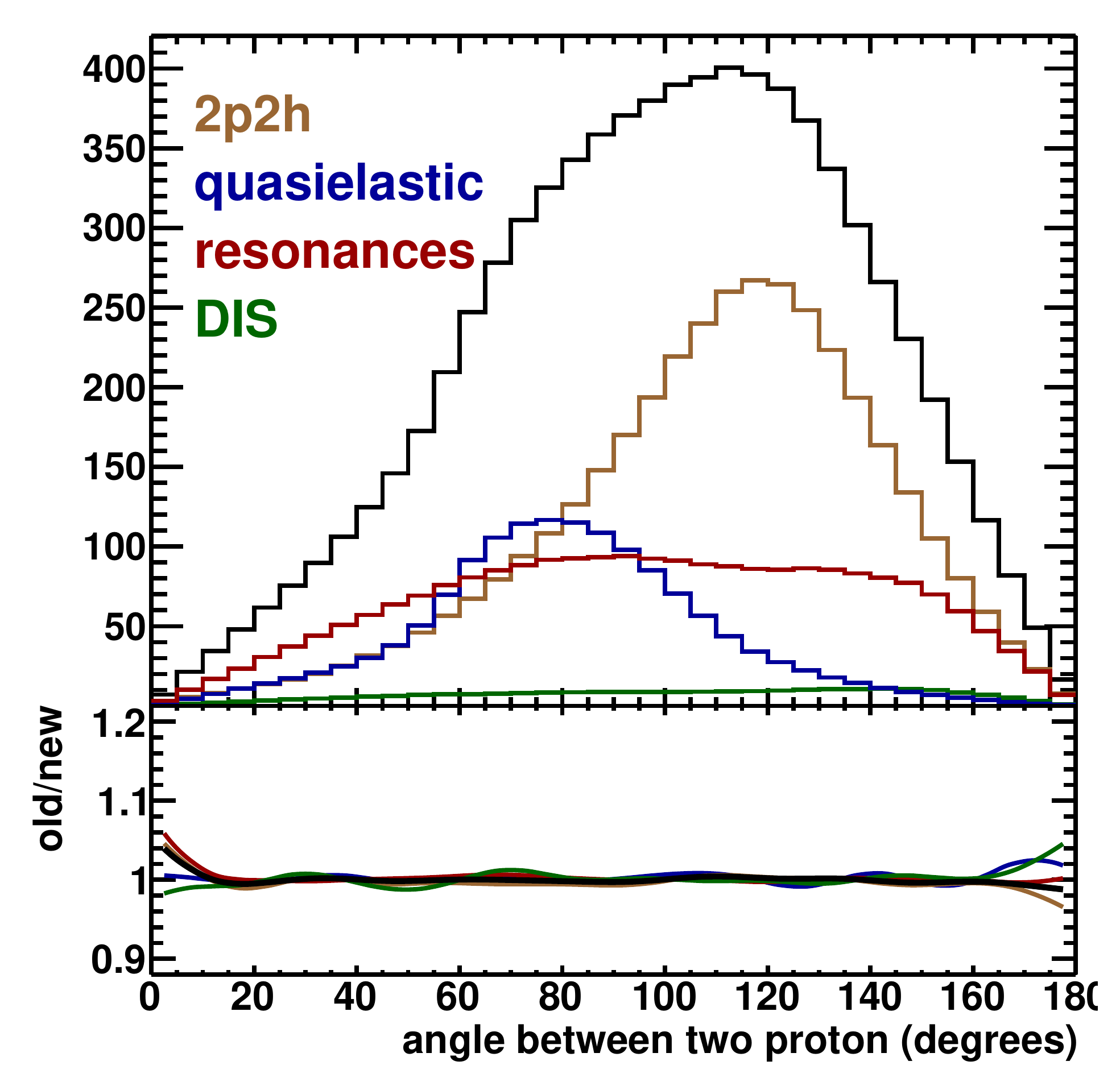}
\caption{Angle between two protons when exactly two protons are above threshold and no pions are in the event. {\small GENIE}v2.12.10-FSIfix original configuration 0 (left) and elastic $\theta_{CM}=0$ configuration 4 (right) compared to the fully fixed configuration 3.  Unstacked histograms break down the total by interaction process rather than FSI fate better show the detail.   blue = CCQE, red = resonance, brown = 2p2h, green = DIS.  This model for the 2p2h process produces especially back-to-back protons, competing with the broad distribution of resonances.   See text for discussion. 
\label{fig:pionoldovernew}}
\end{center}
\end{figure*}

The top part of the figure is constructed differently than prior figures.   Because each event experiences two or three fates, simple division of the sample by fates is not well defined like for CCQE.  Instead, the divisions are by {\small GENIE}'s definition of the original neutrino interaction process:  dark blue = CCQE, dark red = resonance, brown = Valencia 2p2h with MINERvA enhancement, and dark green = DIS.   Also, the histograms are not stacked, making it easier to see where each component resides in distribution from back to back (180 degrees) to co-linear (0 degrees).  The ratios are as before, the old model over the new fully fixed model, and are provided for each component.

Two protons with no pions from resonance and DIS interactions follow different codepaths in {\small GENIE} hA.  Pion absorption happens on two nucleons (which uses either an unfixed or fixed version of TwoBodyKinematics) or three or more nucleons, with relative probability obtained from \cite{Mckeown:1980wu,Mckeown:1981pw}.  Also one nucleon is produced from the decay of the resonance.   If it was a proton, it could knock out a second proton, or a neutron can knockout a proton.   If at least one of the two pion absorption processes happens the resulting three or more nucleons could meet the requirements for two protons above 5 MeV and no final-state pions.  The resonance and DIS distributions are broad because they have a component where both nucleons came from the absorption of the same pion and are by construction more back-to-back which produces a bump closer to 180 degrees.  The main part is when the two protons came from separate FSI processes producing a more uniform distribution.

The effect of pion absorption on two nuclei comes into play directly.   The ratio in the left plot shows the dark green (DIS) and dark red (resonance) components are enhanced 10\% at angle of 180 degrees where they combine to be slightly more than the 2p2h prediction of the rate.     The previous version of this code, when this special case passed an angle of zero, would produce an energetic nucleon and a less energetic nucleon exactly back to back in the lab frame, seemingly ignoring the Fermi-motion of the struck nucleons.  What was expected was the two nucleons would be back to back in the center of momentum frame and practically never so in the lab frame.   However, an additional random scattering angle is applied, similar to the one for hadron nucleon elastic scattering, thus giving more variation, and perfect back to back in the lab frame is never obtained.   The new code produces two nucleons back to back in the CM frame, oriented with that additional scattering angle, then boosts back to the lab frame.  

Overall, the CCQE portion of the sample peaks around 70 degrees; this quasi-elastic NN scattering typical of the kinematics of any CCQE process such as the familiar neutrino CCQE.  The CCQE distortion (blue in the ratio) between the two FSI models is consistent with extra randomness broadening of the peak in the old simulation.  To be in this sample, the CCQE events must have undergone either single or multi-nucleon knockout, not elastic scattering.

In contrast, the 2p2h process by construction prefers back to back protons \cite{Sobczyk:2012ms} that are sharing the energy and momentum in their two-proton CM frame, before rescattering occurs in two separate two-nucleon CM frames.   After scattering, the distortion instead has the flavor of being anomalously narrow in the old version instead of the width that no scattering would have produced.   Here, both elastic scattered and nucleon knockout processes contribute, similar to the distortion of the CCQE process in the previous subsections.   However, the distortion apparent here is milder, enhancing the peak by 10\% at the expense of decreasing the rate in the tails by 10\%.  The mildness may partly be because the peaked distribution is more gentle than the really sharp distributions in the CCQE transverse kinematics samples.

\begin{figure*}[tbh!]
\begin{center}
\includegraphics[width=8cm]{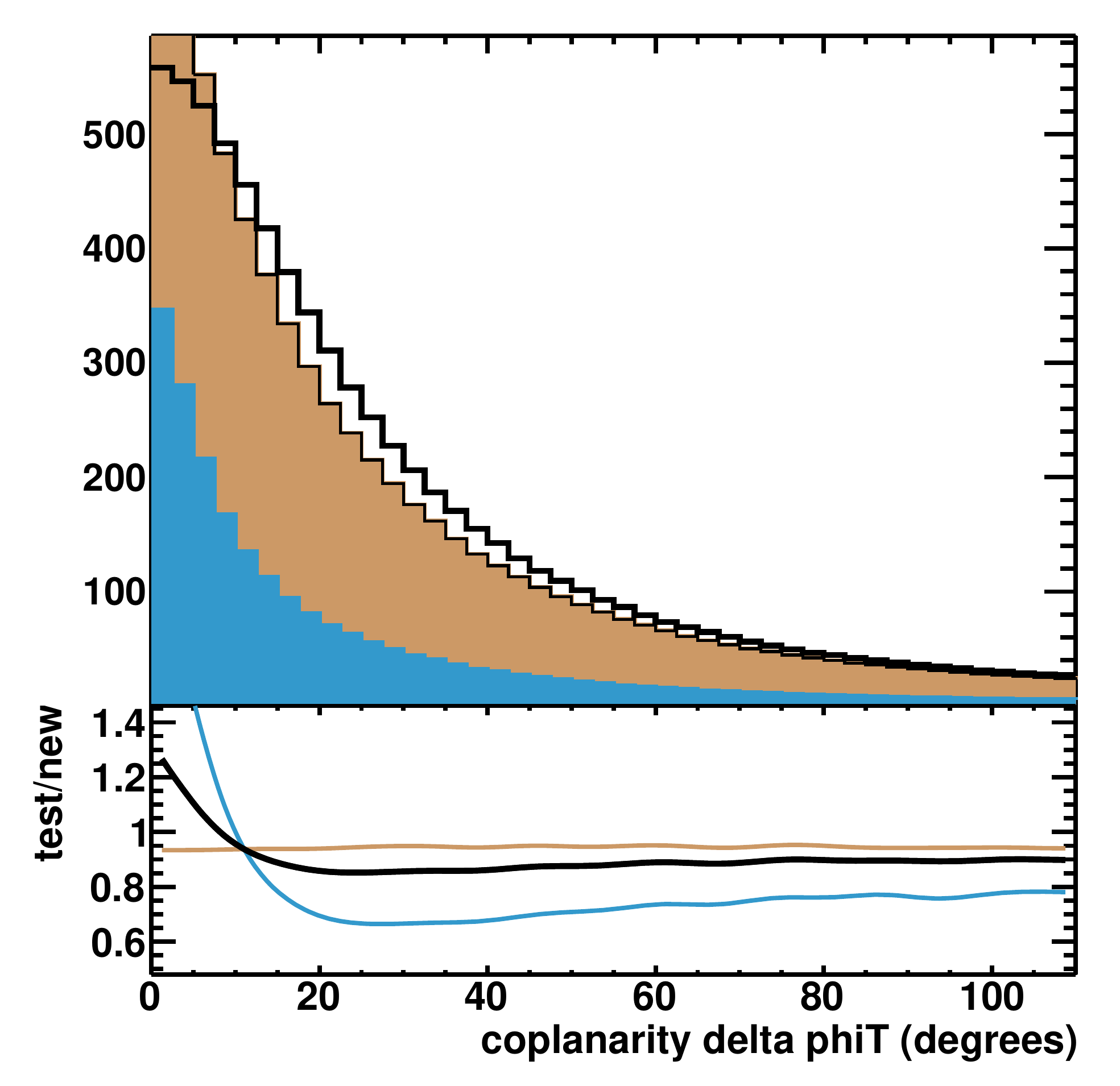}
\includegraphics[width=8cm]{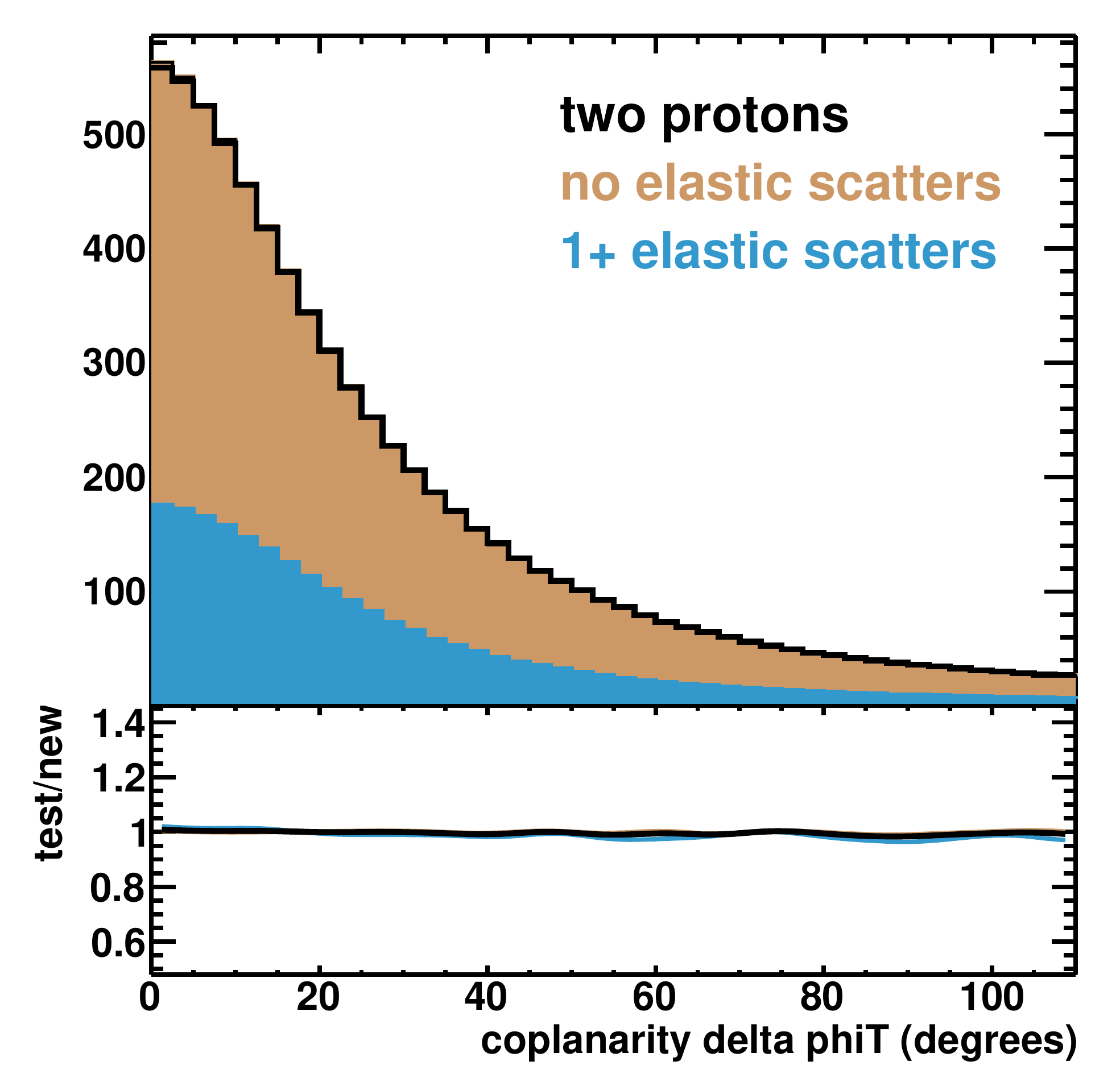}
\caption{Coplanarity angle between the combined proton momentum and the muon.  {\small GENIE}v2.12.10-FSIfix original configuration 0 (left) and elastic $\theta_{CM}=0$ configuration 4 (right) compared to the fully fixed configuration 3. Stacked histogram has events where at least one hadron undergoes an elastic scatter (blue) and all other combinations (brown).  The old elastic simulation shows the characteristic anomaly to be too coplanar. 
\label{fig:protoncoplanarity}}
\end{center}
\end{figure*}

Finally, if the two protons in the final state were the only two particles and were unscattered, we would expect their combined momentum to distribute around perfect coplanarity with the muon according to some Fermi-motion.  The distributions are shown in Fig.~\ref{fig:protoncoplanarity} with the same color scheme as Fig.~\ref{fig:generalized}, blue have an elastic scatter for at least one hadron anywhere in the event and brown is all other combinations.   The old simulation shows the anomalous excess near zero.  The excess is severe when the old elastic scattering model was being used, and shows no distortion otherwise.   Setting $\theta_{CM}=0$ in the right figure negligibly reduces the little extra scattering provided by the fixed elastic routine.

\section{Elastic hadron nucleus scattering revisited}
\label{whyelastic}

Many neutrino interaction generators, including the latest version 3.0 of {\small GENIE}, ignore elastic hadron+nucleus scattering completely.    We have shown that a correct implementation is not a leading FSI effect and if necessary can be approximated as no interaction at all.  Among the transverse kinematics distributions being measured today, the inferred nucleon momentum yields such peaked distributions that it may be sensitive to the presence or absence of the hadron+nucleus process, if other uncertainties are constrained.

The equivalent process studied decades ago is hadron plus a ground state nucleus goes to the same hadron plus nucleus still in its ground state.   The verification of the ground state was done by requiring small energy transfer, less than the minimum to get to the lowest excited shell model states.   Typically these experiments produced a measurement of the center of momentum scattering angle $\theta_{CM}$, which is empirically just what we need.  The process was well described by fitting parameters from an optical model to the data, invoking the ``black disk'' approximation for quantum diffraction.  This produces a circular diffraction-like pattern familiar from introductory optics.

The following figures show the data and model used in simulations since the precursor to the FSI code in {\small GENIE}, which was the {\small INTRANUKE} code within NEUGEN \cite{Gallagher:2002sf}.  The choice of a heavy nucleus for pion scattering reflects NEUGEN's origin for iron calorimeter experiments Soudan2 and MINOS.  Because spectrometer data does not go to angles arbitrarily close to the beam, that region was filled in using a rate slightly higher than the lowest angle data point.  For present FSI purposes, the smallest angle distortions are of no concern themselves, only the relative amount of the high angle scattering could be important.

\begin{figure*}[tbh!]
\begin{center}
\includegraphics[width=9cm]{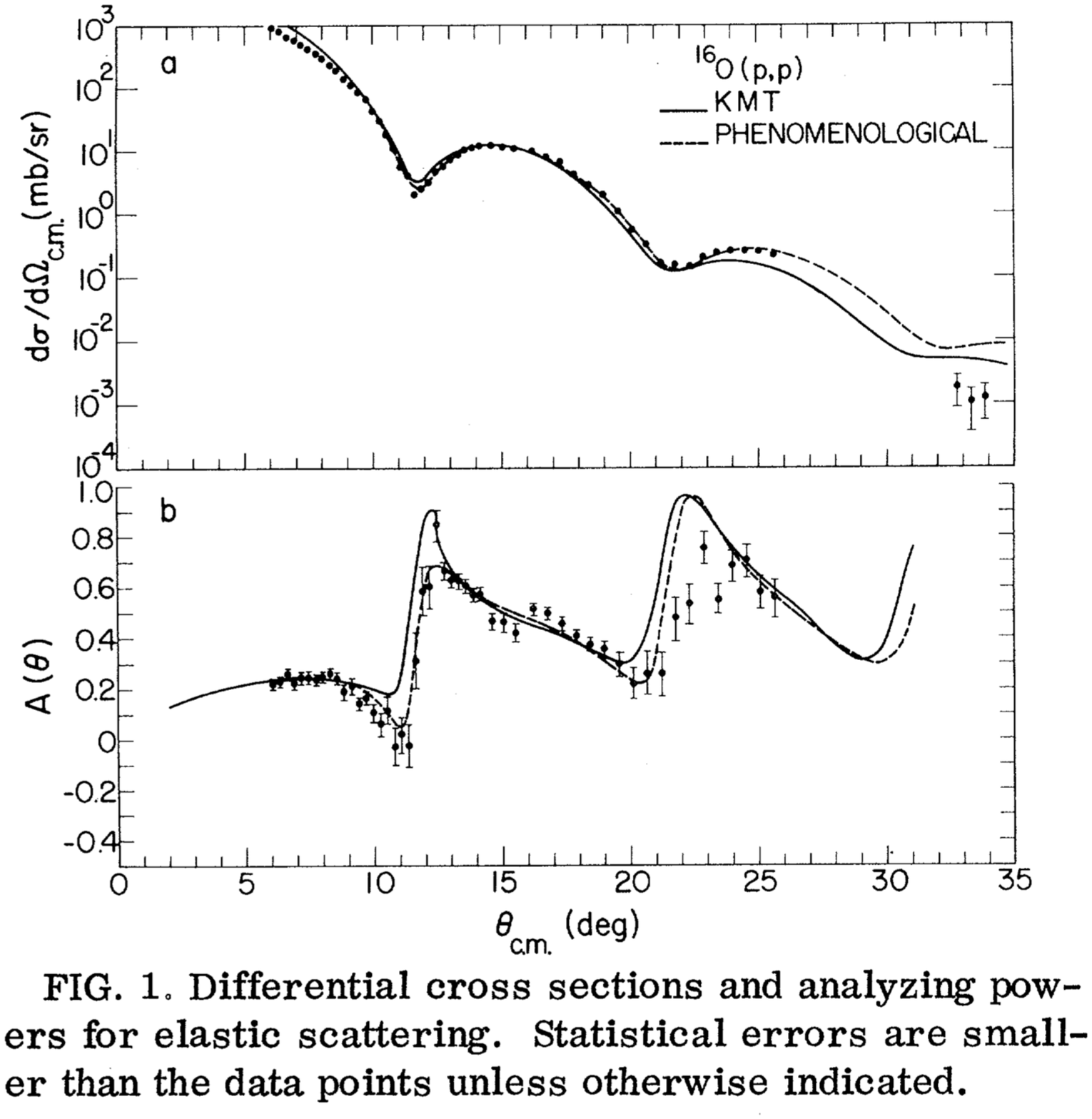}
\caption{Elastic proton-$^{16}$O scattering at 800 MeV.  LAMPF data, analysis, figure, and caption from \cite{Adams:1979iv} with a related publication \cite{Adams:1980by} of data on heavier nuclei.  The original pre-{\small GENIE} code is modified to use digitized data from www.nndc.bnl.gov/exfor , convert d$\sigma$/d$\Omega$ to d$\sigma$/d$\theta$, and draw randomly from the resulting cumulative distribution function.  The new version also does not have a problem that truncated the original distribution at 6.5 degrees.  The resulting angles are less than 5.5 (8.0) degrees 68\% (90\%) of the time.
\label{fig:dataxsn}}
\end{center}
\end{figure*}

\begin{figure*}[tbh!]
\begin{center}
\includegraphics[width=12cm]{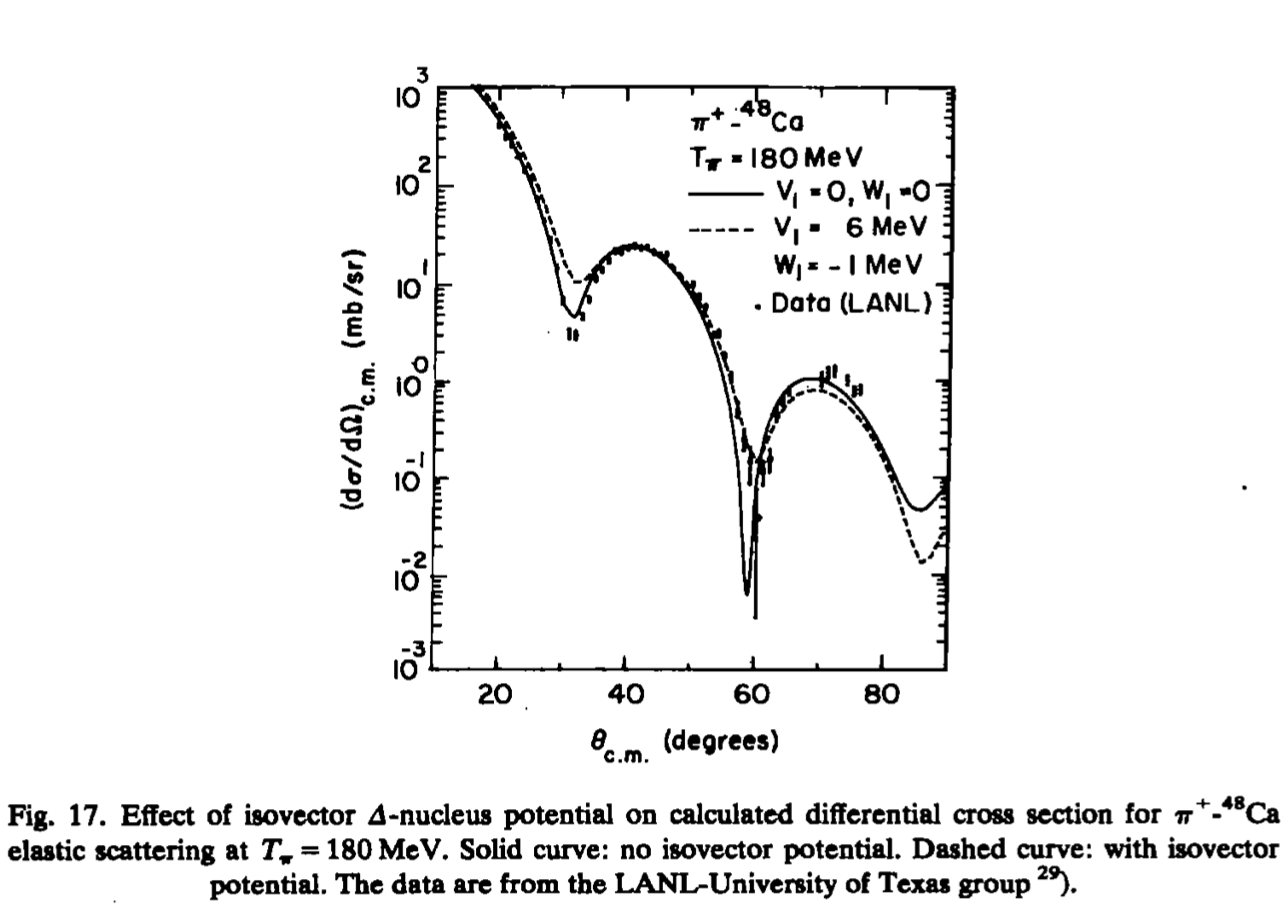}
\caption{Elastic $\pi^+-^{48}$Ca scattering at 180 MeV.   LAMPF data from \cite{Boyer:1981jr} optical model, figure, and caption from \cite{Freedman:1982yp}.  The original pre-{\small GENIE} code was modified to   convert d$\sigma$/d$\Omega$ to d$\sigma$/d$\theta$, and draw randomly from the resulting cumulative distribution function.   The resulting angles are below 13 (20) degrees 68\% (90\%) of the time.
\label{fig:dataxspi}}
\end{center}
\end{figure*}

For neutrino FSI, the hadron is born in the nucleus, so we expect some kind of modification of the resulting angle distribution width and diffraction pattern features.  Even if the ``black disk'' approximation is unsatisfactory, the underlying diffraction effects are probably present.    Without additional theory input, it is hard to guess whether the angle distribution should be narrower or wider, more prevalent or less.  The data consider only ground state nuclei, but we presumably chould include scattering that further excites the already excited nucleus, everything short of nucleon ejection.   Intended as a first approximation, {\small GENIE}'s angle distribution is taken from these two distributions to represent all relevant energies and nuclei, leading to additional uncertainty.

The on/off effect is large in the transverse kinematic imbalance studies in the first section plus Fig.~\ref{fig:dataxspi} and Fig.~\ref{fig:protoncoplanarity}.  It seems likely to be visible in analysis of (e,e'p) data from the CLAS collaboration.  Some colleagues have an interest (separately) in both hadron nucleus scattering and optical models and also neutrino interactions.  They include Carlotta Giusti \cite{Vorabbi:2018bav} and Jerry Miller who coauthored the pion scattering model work in Fig.~\ref{fig:dataxspi}.   It is likely a number of theoretical papers on different scattering processes already included an optical potential to describe this effect on the outgoing hadron wavefunction, possibly without special mention of its effects.



\section{Conclusion}

We describe a version of {\small GENIE} where the TwoBodyKinematics function is fixed.   This document offers a roadmap for other users of {\small GENIE} 2.12.10 to identify where their analysis and interpretation may be affected.   The hA elastic hadron + nucleus scattering process is most affected; the fixed version looks a lot like no FSI (zero $\theta_{CM}$), but not exactly.  The hA scattering using the ``2015'' configuration eliminates the elastic fate in a different way that might not be compatible with the intended tuning with the underlying hadron nucleus scattering data.  Regardless, significant uncertainties on these processes still apply.   

The effect of the mistaken code for hadron+nucleus elastic scattering is modest for the (possibly back-to-back) angle between two protons when no pions are present, and negligible simpler distributions such as pion angle or the nonCCQE components of the CC0$\pi$ samples.  The single nucleon knockout FSI processes also had distortions that were negligible.  The fix to the pion absorption by two nucleons reduces the number of back-to-back protons, and doesn't seem to strongly contribute elsewhere.   This last case is harder to probe unambiguously, even with generator samples, because it implies at least one other hadron is present and may experience its own FSI fate.

The narrowness of the inferred momentum $p_N$ of the struck nucleon distribution makes it ideal for testing a number of properties of the nuclear environment, including the predictions of the fixed implementation of hadron+nucleus scattering.  The distribution of nucleons in beyond-the-Fermi gas models, the energy to remove the nucleon, and Coulomb effects should be visible in this and related distributions.  The latest version of the current event generators default to not simulating this process.  The presence or absence of elastic hadron+nucleus scattering could play a role in describing the width and center of such a sharp distribution that is competitive with these other nuclear effects.

\begin{acknowledgments}

Support for these studies was granted by the United States National Science Foundation under Grant PHY-1607381.  Support, encouragement, and a careful reading of the manuscript were provided by many within the MINERvA collabortion.

\end{acknowledgments}

  \bibliographystyle{apsrev4-1title}
  \bibliography{elasticFSI}



\end{document}